\documentclass[fleqn,usenatbib]{mnras}

\usepackage{newtxtext,newtxmath}

\usepackage[T1]{fontenc}

\DeclareRobustCommand{\VAN}[3]{#2}
\let\VANthebibliography\thebibliography
\def\thebibliography{\DeclareRobustCommand{\VAN}[3]{##3}\VANthebibliography}

\usepackage{float}	
\usepackage{graphicx}	
\usepackage{amsmath}	

\newlength{\appfigheight}
\usepackage{subcaption}
\usepackage{dcolumn}
\usepackage{bm}
\usepackage{hyperref}
\usepackage[table]{xcolor} 
\usepackage{listings}

\definecolor{lightcyan}{rgb}{1,0,0}

\newcommand{\boundary}{ \cellcolor{yellow}}
\newcommand{\outlier}{ \cellcolor{cyan!30}}
\newcommand{\panel}[1]{ {\it #1}}
\newcommand{\IR}[2]{ {[#1]_{\|#2}}}
\newcommand{\Erf}{ {\rm Erf}}
\newcommand{\Fcal}{ {\cal F}}
\newcommand{\sspace}{\!\!}
\newcommand{\cosmoeft}{\texttt{CosmoEFT}}
\newcommand{\operon}{\texttt{Operon}}
\newcommand{\resumeft}{\texttt{ResumEFT}}
\newcommand{\class}{\texttt{CLASS}}

\newcommand{\classcosmoeft}{\texttt{CosmoEFT-Class}}

\newcommand{\pybird}{\texttt{Pybird}}
\newcommand{\classpt}{\texttt{CLASS-PT}}

\newcommand{\syren}{\texttt{Syren-New}}
\newcommand{\symeft}{\texttt{Sym-EFT}}

\newcommand{\Ast}{ \tilde{A}_s}
\newcommand{\kNL}{ k_{NL}}
\newcommand{\rhob}{ \bar{\rho}} 
\newcommand{\grad}{  \vec{\nabla}}
\newcommand{\thetat}{ \tilde{\theta}}
\newcommand{\adotoa}{  {\cal H}}
\newcommand{\hmpc}{ {h \rm Mpc}^{-1}} 
\lstdefinestyle{CStyle}{
    language=C,
    basicstyle=\ttfamily\small,        
    keywordstyle=\color{blue}\bfseries, 
    commentstyle=\color{green!50!black},
    stringstyle=\color{orange},         
    numberstyle=\tiny\color{gray},      
    numbers=left,                       
    stepnumber=1,                       
    tabsize=4,                          
    showstringspaces=false,             
    breaklines=true,                    
    frame=single,                       
    backgroundcolor=\color{gray!10},    
}

\title[Sym-EFT: Accelerating EFTofLSS with Symbolic Regression]{Sym-EFT: Accelerating Effective Field Theory of Large Scale Structure with Symbolic Regression}

\author[D. Farakou and C. Skordis]{
Despoina Farakou,$^{1,2}$\thanks{E-mail: farakou@fzu.cz}
Constantinos Skordis,$^{1,3}$\thanks{E-mail: skordis@fzu.cz}
\\
$^{1}$CEICO—FZU, Institute of Physics of the Czech Academy of Sciences, Na Slovance 1999/2, 182 00 Prague, Czechia \\
$^{2}$Institute of Theoretical Physics, Faculty of Mathematics and Physics, Charles University, V Holešovičkách 747/2, 180 00 Prague 8, Czechia \\
$^{3}$Department of Physics, University of Oxford, Denys Wilkinson Building, Keble Road, Oxford OX1 3RH, UK \\
}

\date{Accepted XXX. Received YYY; in original form ZZZ}

\pubyear{2026}

\begin{document}
\label{firstpage}
\pagerange{\pageref{firstpage}--\pageref{lastpage}}
\maketitle

\begin{abstract}
We present an emulator suite for the one- and two-loop cold dark matter power spectrum from
the Effective Field Theory of Large Scale Structures (EFTofLSS). Specifically, we emulate separately the various contributions to the 
one- and two-loop parts of the power spectrum, leaving out the possible counterterms which can be added as multiplicative prefactors. 
By leaving the time dependence of the counterterms unspecified at the emulation stage, our technique has the advantage of 
being extremely versatile in fitting any type of counterterm parametrisation to data, or to simulations, without having to change the emulator. 
We construct our emulators using the method of symbolic regression which results in functions that can be used directly in computer code,
while achieving errors of better than $0.5\%$ within the $k$-range of validity of EFT and maintaining ultra-fast computational evaluation of less than $\sim5\times10^{-4}s$ on a single core.
\end{abstract}

\begin{keywords}
dark matter -- large-scale structure of Universe
\end{keywords}


\section{Introduction}
\label{sec:Intro}
The  standard model of cosmology finds observational support from  several cosmological surveys.
Measurements of the Cosmic Microwave Background (CMB)  anisotropies~\citep{WMAP:2012fli,Planck:2018vyg,ACT:2025fju,SPT-3G:2025bzu},
 observations of the large-scale matter distribution~\citep{eBOSS:2020rpt,eBOSS:2021hod,DES:2025xii,Wright:2025xka} traced by the clustering and weak-lensing of galaxies,
and the determination of the expansion rate of the Universe~\citep{Pan-STARRS1:2017jku,Brout:2021mpj,Rubin:2023ovl}
have provided us with enhanced precision with which the model parameters are estimated.
Despite this success, several tensions between the measurement of some of the model parameters by different surveys 
have persisted~\citep{Abdalla:2022yfr},
 while recent data~\citep{DESI:2025zgx,DESI:2025fii,DESI:2025gwf} show possible hints of new physics.
As such, accurately constraining our standard model of cosmology remains particularly important.

Current surveys~\citep{Gebhardt:2021vfo,DESI:2025zgx,Euclid:2024yrr,LSSTDarkEnergyScience:2018jkl} 
are pushing observational precision to new levels, while upcoming surveys~\citep{Eifler:2020vvg,SimonsObservatory:2018koc,Zhao:2024alp} will offer a further increase in amount of data,
leading to even more precise measurements of the cosmological parameters.
Exploiting the data to their full potential requires computing the cosmological model predictions to smaller scales where 
linear perturbation theory breaks down and the non-linear evolution of dark matter must be taken into account.

N-body simulation methods are the standard and well tested way for precise modelling of observables to small enough scales so that data from surveys can be fully used. In the case 
of  dark-matter-only simulations, the power spectrum computed using different N-body codes has been tested to agree within $1\%$  for wavenumbers $k\lesssim 1 \hmpc$ and
  within $3\%$ for $k\lesssim 10 \hmpc$~\citep{Schneider:2015yka}, where $h$ is the dimensionless Hubble constant.
However, running N-body simulations is time-costly, and having to run them for several hundreds of thousands of times, as would be necessary when applying
 Markov chain Monte Carlo (MCMC) to parameter estimation, is prohibitive.

There are two ways to address this limitation. The first approach relies on the use of emulators~\citep{Heitmann:2006hr}, which are efficient interpolation methods for reproducing specific observables
within a chosen parameter range. Early emulators~\citep{Heitmann:2009cu} of wCDM could reach $1\%$ accuracy for the matter power spectrum for $k\lesssim 1 \hmpc$, later widened to $1\%$ for $k\lesssim 10\hmpc$\citep{Heitmann:2013bra} up to redshift $z=4$ using Gaussian process modelling.  Later emulators, also based on a Gaussian procces, provided various halo and galaxy clustering statistics~\citep{Nishimichi:2018etk,McClintock:2018uyf,Zhai:2018plk,McClintock:2019sfj}
while \cite{Giblin:2019iit}, \cite{Winther:2019mus} and~\cite{Saez-Casares:2023olw} provided matter power spectrum emulators in beyond-$\Lambda$CDM models, including extensions of general relativity.
The Euclid emulator II~\citep{Euclid:2020rfv} reaches $<1\%$ absolute accuracy in the non-linear power spectrum for $k\lesssim 1  \hmpc$ up to $z=3.5$ 
with the inclusion of massive neutrinos and parametrized dynamical dark energy.
Using neural networks,  the BACCO project provided accurate emulators for the dark matter-only power spectrum~\cite{Angulo:2020vky} and the baryon boost~\citep{Arico:2020lhq}. 
Finally,  neural networks were used to  emulate all CMB angular power spectra, namely the \emph{Capse.jl} package~\citep{Bonici_2024}, and CosmoPower~\citep{spuriomancini2021} which 
emulates in addition the matter power spectrum.

The second route is to use higher-order perturbation theory. Linear perturbation theory has been pivotal in the understanding of the cosmic microwave background (CMB) anisotropies, early Universe cosmology,
and establishing the concordance $\Lambda$CDM model~\citep{dodelson_2020_modern}. However, it is expected to become inaccurate around $k \gtrsim 0.1 \hmpc$ at $z=0$ due to the formation of non-linear structures.
By including higher-order terms~\citep{Jain:1993jh,Bernardeau:2001qr}, one can extend the validity of perturbation theory to smaller scales until the non-linear scale $\kNL\sim 0.5 - 1 \hmpc$, beyond which perturbation theory breaks down 
and using either N-body simulations, or emulators, seems inevitable. 
Yet, in this intermediate regime between large and small scales, this theoretical framework can provide additional insights into the Universe's evolution. 
The Effective Field Theory of Large Scale Structure (EFTofLSS)~\citep{Baumann:2010tm,Carrasco:2012cv,Hertzberg:2012qn, Carrasco:2012cv,Hertzberg:2012qn,Carrasco:2013sva,Carrasco:2013sva,Carrasco:2013mua,Pajer:2013jj,McQuinn:2015tva}, 
extends standard perturbation theory (SPT) by including the effects of non-linearities at smaller scales and becomes valuable in this context.
 In EFTofLSS, short-wavelength perturbations can influence large-scale physics through coupling with long-wavelength perturbations~\citep{Goroff:1986ep}.
Borrowing the analogy from~\cite{Senatore_Lectures}, similarly to how Maxwell's equations describe the behavior of dielectric materials, the large-scale structure of the Universe can be understood through an EFT up to the non-linear scale. 
This allows us to derive equations without needing intricate knowledge of the atomic-level structure, in this case galaxy formation, akin to electromagnetism.
The effect of integrating out the non-linearities is captured through counter-terms which are time-dependent functions that are otherwise not computable within the EFT. 

Given that EFTofLSS necessarily breaks down at the non-linear scale, and also, that with emulators ultra-fast percent-level accuracy can be achieved well beyond its regime of validity,
its use may seem an unnecessary step. However, there are good reasons for needing such an EFT.
The counter terms capture the short-distance  (ultraviolet or 'UV') physics that are to be found when running N-body simulations. 
Apart from providing the necessary consistency to the theory, 
it is these counter terms that give the EFTofLSS an edge that neither N-body simulations, nor their emulators, have. These terms can be fitted to N-body simulations (or emulators) and separately to observations, and there is no a priori reason that these fits should agree within a given model, unless it is a good model for our Universe. 
Indeed, fitting various EFTofLSS counterterms to simulations and generating informed priors on their range and redshift dependence has been employed, for example, in~\cite{Ivanov:2024hgq,Ivanov:2024xgb,Zhang:2024thl,DESI:2025wzd,Herle:2026klt}.
Hence, with the EFTofLSS one can test a wide range of models, and provide a robust way of parametrically  comparing $\Lambda$CDM with its various extensions beyond the regime of linear theory.

Nevertheless, while EFTofLSS offers a robust theoretical framework for modeling structure formation, its practical application can be computationally intensive, depending on the loop order used.
The \cosmoeft~code~\citep{Cataneo:2016suz} can easily take $\sim 15$~seconds for a single model computation at $1$-loop and $\sim 16$~minutes at $2$-loop, both when using $8$ cores in high-accuracy settings. While much faster than N-body simulations, it is still somewhat slow for efficient use in MCMC.
 Several codes have recently achieved ${\cal O}({\rm second})$ computation while maintaining excellent accuracy~\citep{Chudaykin:2020aoj,DAmico:2020kxu,Linde:2024uzr, Noriega_2022},
however, until recently this was only done for the $1$-loop power spectrum and the various $1$-loop counterterms relevant for biased tracers. These are currently sufficient for using the EFTofLSS with galaxy surveys.
However, it is desirable to have an ultra-fast computation of the $2$-loop power spectrum, as it contributes to CMB lensing at higher redshift and may provide complimentary information.
Two-loop accuracy is increasingly relevant to data from CMB experiments; the Atacama Cosmology Telescope (ACT)~\cite{louis2025atacamacosmologytelescopedr6}, South Pole Telescope (SPT)~\cite{2024arXiv241106000G}, 
and Simons Observatory (SO) \cite{SimonsObservatory:2018koc} demand high-precision theoretical predictions.

In this work, we build upon these advancements and introduce emulators for the various parts of the EFTofLSS $1$- and $2$-loop contributions by employing symbolic regression,
a machine learning technique that has only recently made its way to cosmology~\citep{Bartlett:2023cyr,Bartlett:2024jes,Sui:2024wob,Kammerer:2025dbi}. 
Our full emulator is differentiable and  achieves sub-millisecond computational speed while maintaining high accuracy, making it suitable for use in large parameter scans or MCMC pipelines.
Additionally, we provide a user-friendly interface with the Boltzmann solver \class, enabling seamless integration of our EFTofLSS emulator into existing cosmological analysis pipelines. 
This interface is designed to make the tool easily accessible to the broader cosmology community, facilitating the use of EFTofLSS-based predictions in a wide range of applications. 
We note that neural-network based emulators for the 1-loop EFTofLSS were created in~\cite{bonici2025effortfastdifferentiableemulator} and in~\cite{reeves2025pybirdjaxacceleratedinferencelargescale}.

This article is organised as follows.  In Section \ref{sec:EFTofLSS}, we describe the EFTofLSS framework at one- and two-loop order. In Section \ref{sec:emulator}, we introduce symbolic regression and describe the construction of our emulator, 
including dataset generation, and accuracy benchmarks. In Section \ref{sec:Error_comparison}, we evaluate the performance of our emulator relative to existing tools, and assess its accuracy
at higher redshift when applied to CMB lensing.
Finally, we conclude in section \ref{sec:Conclusion} with a discussion of future directions and applications. 
The full set of emulators and their validation is collected in appendix-\ref{sec:EmulatorSuite_1_loop} for one-loop and -\ref{sec:EmulatorSuite_2_loop} for two-loops.
Throughout this paper, we use a fiducial cosmology in our plots; for that, we chose the best-fit cosmology of \texttt{plikHM\_TTTEEE\_lowl\_lowE\_lensing} \citep{Planck:2018vyg}
as shown in Table~\ref{tab:fid}. 
\begin{table}
\centering
\begin{tabular}{|c|c|c|c|}
\hline
Parameter name &
Value 
&
Parameter name &
Value 
 \\
\hline
\hline
\rule{0pt}{2.3ex}
$\Omega_c$ & $0.2650$ & $\Ast \equiv 10^{9} A_{s}$ & $2.1005$  \\
\hline
$\Omega_b$ & $0.0494$ & $n_{s}$ & $0.9660$   \\
\hline
$h$ & $0.6732$ & $m_\nu$ & $0.06$  \\
\hline
\end{tabular}
\caption{Fiducial cosmological parameters used, see text.}
\label{tab:fid}
\end{table}

\section{Effective Field Theory of Large Scale Structure}
\label{sec:EFTofLSS}
\subsection{Standard perturbation Theory}
 We consider a late (spatially flat) Universe cosmology where the only
relevant components are that of cold dark matter (CDM), 
baryons, one species of massive neutrinos of fixed mass $m_{\nu} = 0.06eV$, and cosmological constant $\Lambda$.
We denote their relative densities at redshift $z=0$ as $\Omega_c$, $\Omega_b$, $\Omega_\nu$ and $\Omega_\Lambda$ respectively,
and set the Hubble constant $H_0 = 100 h$ $\rm{km/s/ Mpc}$, where $h$ is the dimensionless Hubble parameter. 
At $z<10$ the massive neutrino is already non-relativistic and we can collectively denote the total matter relative density today 
as $\Omega_m \equiv \Omega_{cb} + \Omega_\nu$, where $\Omega_{cb} \equiv \Omega_c + \Omega_b$ is the relative density of only CDM and baryons, that we use extensively in what follows.

We are interested in describing the mildly non-linear regime of structure formation using the EFTofLSS. 
At the lowest approximation the fluctuations of all matter species follow the adiabatic mode so that we may treat them collectively using the 
adiabatic matter density contrast $\delta$ and Eulerian velocity $u^i$. 
The effect of baryons has been treated in \cite{Lewandowski:2014rca} as an additional isocurvature mode and that of neutrinos in \cite{Senatore:2017hyk}. 
These make negligible difference at linear scales and can result to few percent differences at higher $k$.

The first step consists of linear and higher-order fluctuations on the FLRW background above~\citep{Bernardeau:2001qr,Jain:1993jh}.
We neglect vorticity such that  $u^i$ is given as a gradient of a scalar $\thetat$, i.e. $u_i  \equiv  \grad_i \thetat$.
On sub-horizon scales, the density contrast $\delta$  and $\thetat$ obey the continuity equation
    \begin{equation}
\dot{\delta} +  \grad_i \left[ (1+\delta)   \grad^i \thetat \right] = 0,
\label{2cont}
    \end{equation}
 and Euler equation 
      \begin{equation}
 \dot{\thetat} + \adotoa  \thetat + \frac{1}{2}  |\grad \thetat|^2 + \Phi = 0,
\label{2euler}
    \end{equation}
 where a dot denotes differentiation wrt conformal time $\tau$,
and  $\adotoa$ is the $z$-dependent conformal Hubble parameter. The 
gravitational potential $\Phi$ which sources \eqref{2euler} is determined from the total matter density via
 the Poisson equation $2\grad^2 \Phi = 8\pi G \rhob_m \delta$, where $\rhob_m$ is the background total matter density.

Passing to Fourier space~\footnote{We use the same symbols for position and Fourier space for brevity, since  the former
is not being used for the remainder of the article. The Fourier convention is that for any variable $A(\vec{x},t)$ its
Fourier space representation is $\int \frac{d^3x}{(2\pi)^3} e^{-i\vec{x}\cdot\vec{k}} A(\vec{x},t)$. },  
linearizing, and eliminating $\Phi$ leads to the two well-known linear continuity and Euler equations
   \begin{equation}
\dot{\delta}_1 +    \theta_1  = 0,
\label{2cont_lin}
    \end{equation}
 and Euler equation 
      \begin{equation}
 \dot{\theta}_1 + \adotoa  \theta_1  + 4\pi G \rhob_m \delta_1 = 0,
\label{2euler_lin}
    \end{equation}
where we have introduced the variable $\theta \equiv - k^2 \thetat$, and the subscript '1' is to mark these variables as 
corresponding to the linearized fluctuations. The system \eqref{2cont_lin} and \eqref{2euler_lin} has two linearly
independent solutions $D^+(\tau)$ and $D^-(\tau)$. In the case $\Omega_m=1$, these are $D^+ = (z+1)^{-1}$ and $D^- = (z+1)^{-3/2}$ 
while if $\Omega_m<1$, $D^+$ remains a growing mode and $D^-$ a decaying mode. Thus, 
 as it is commonly done, we choose only the growing mode $D^+ = D$ (dropping the '$+$'), 
such that $\delta_1 = D \delta_{in}(\vec{k})$ and $\theta_1 =   \adotoa \frac{z+1}{z} \frac{d\ln D}{d\ln z}  \delta_1$  are 
given in terms of the same initial condition $\delta_{in}(\vec{k})$.

We now  consider higher order terms in \eqref{2cont} and \eqref{2euler} in Fourier space, leading to
    \begin{align}
  \dot{\delta} + \theta  
=& -  \int  \int d^3k_1 \, d^3k_2 \, \delta^{(3)}\left(\vec{k} - \vec{k}_{12}\right) \frac{\vec{k}_1\cdot \vec{k}_{12}}{k_1^2} \theta(\vec{k}_1) \delta(\vec{k}_2),
\label{HOcont}
    \end{align}
and
    \begin{align}
  \dot{\theta} + \adotoa  \theta + \frac{3}{2}\adotoa^2 \Omega_m \delta 
=& -  \int d^3k_1  d^3k_2 \delta^{(3)}(\vec{k} -\vec{k}_{12})  
\nonumber
\\
& \times \frac{k_{12}^2 \vec{k}_1 \cdot \vec{k}_2  }{2 k_1^2k_2^2} \theta(\vec{k}_1) \theta(\vec{k}_2),
\label{HOEuler}
    \end{align}
respectively, 
where $\delta^{(3)}$ is the Dirac three-dimensional delta-function, $\vec{k}_{ij} \equiv \vec{k}_i + \vec{k}_j$ and where we 
have ommitted explicit time dependence in the arguments of $\delta$ and $\theta$ as well as explicit $\vec{k}$ dependence apart from
$\vec{k}_1$ and $\vec{k}_2$.

Equations \eqref{HOcont} and \eqref{HOEuler} form the basis for Eulerian perturbation theory to any order, given the initial condition $\delta_{in}(\vec{k})$ and assumption
of growing mode as discussed above. To a good approximation, higher order perturbations may be expanded as a series
\begin{align}
    \delta(\vec{k},z)=& \sum_n D^n \delta_{n}(\vec{k}),
\label{delta_exp}
\\
    \theta(\vec{k},z)=& \adotoa  \frac{z+1}{z} \frac{d\ln D}{d\ln z}  \sum_n D^n \theta_{n}(\vec{k}),
\label{theta_exp}
\end{align}
which enables separating out the time dependance from the $k$ dependence.  These are then used in \eqref{HOcont} and \eqref{HOEuler} to form the solution for $\delta$ and $\theta$ to any desired
order in perturbation theory, in terms of only one initial condition $\delta_1(\vec{k})$ and the growing mode $D(a)$. With this solution at hand,
 we may form the power spectrum $P(k,a)$ defined by
\begin{align} 
\langle \delta(\vec{k},z) \delta(\vec{k}',z)\rangle = P(k,z) \delta^{(3)}\left(\vec{k} + \vec{k}'\right).
\end{align}
Explicitly in terms of perturbation orders, we may split it into loop corrections, commonly called  Standard Perturbation Theory (SPT) terms, as
\begin{equation}
    P(k,z)= P_{11} + \underbrace{ P_{13} + P_{22}}_{1-loop}  +  \underbrace{ P_{51}+P_{42}+P_{33}}_{2-loop} + \ldots
\label{pk_expansion}
\end{equation}
where $P_{ij} \sim \langle \delta_i \delta_j\rangle$.  The terms on the RHS in \eqref{pk_expansion} are understood to have a $k$ and $z$ dependence, i.e.
 we have defined the linear power spectrum $P_{11}(k,z)$, the $1$-loop SPT power spectrum $ P_{{\rm 1-loop}}(k,z) \equiv  P_{13}(k,z) + P_{22}(k,z)$ 
and $2$-loop SPT power spectrum $ P_{2-loop}(k,z) \equiv P_{51}(k,z) +P_{33}(k,z) +P_{42}(k,z) $ respectively.
$P_{33}$ contains two contributions: $P_{33,I}$ which has one loop in each $\delta_{3}$, and $P_{33,II}$ in which each $\delta_1$ contracts with a $\delta_1$ from the other $\delta_3$.
\begin{figure}
\centering
\includegraphics[width=1\linewidth]{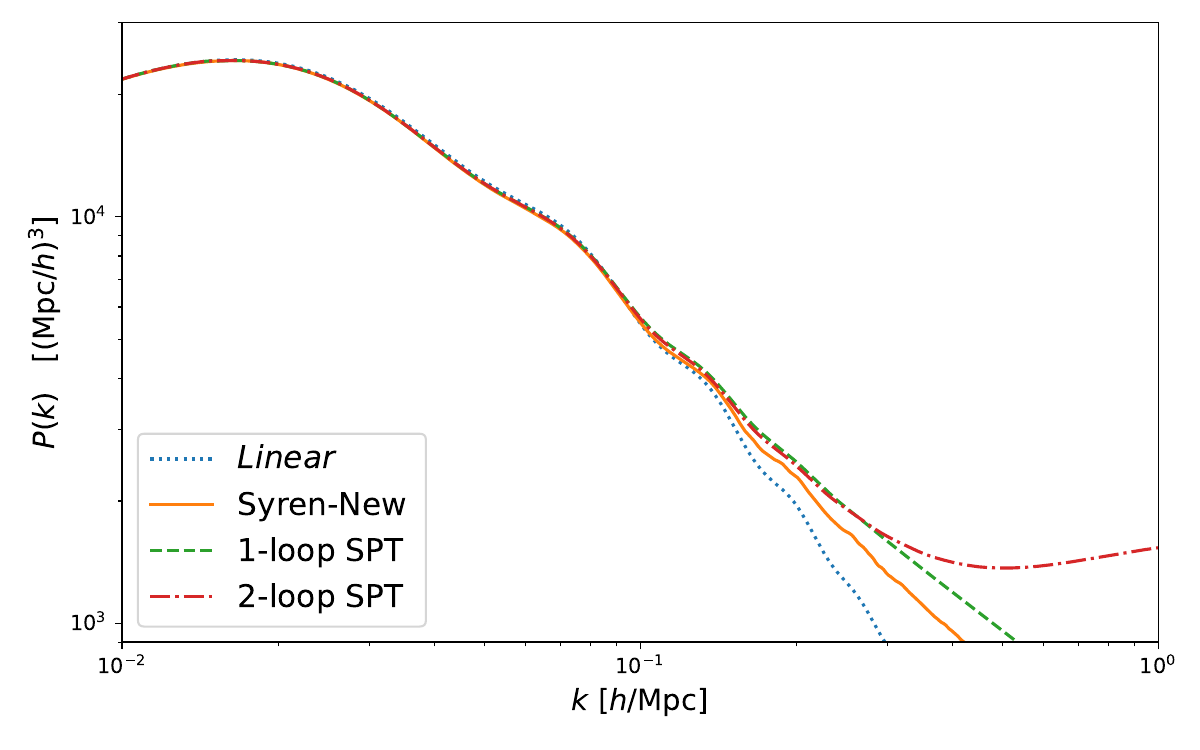}
\caption{The different contributions to the matter power spectrum. Shown is the  linear power spectrum (dotted, blue), 
1-loop SPT (dashed, green), 2-loop SPT (dash-dot, red) and the full nonlinear power spectrum from the \syren~emulator (solid, orange). }
\label{plot-SPT}
\end{figure}

Interestingly, owing to \eqref{delta_exp} and \eqref{theta_exp}, the late Universe power spectra factorise, so that we may write
\begin{subequations}
\begin{align}
 P_{11}(k,z) =& [D(z)]^2 P_{11}(k)
\\
 P_{{\rm 1-loop}}(k,z) =& [D(z)]^4 P_{{\rm 1-loop}}(k)
\\
 P_{{\rm 2-loop}}(k,z) =& [D(z)]^6 P_{{\rm 2-loop}}(k)
\end{align}
\label{D_dependence}
\end{subequations}
where from now on, we adopt the convention that power spectra without explicit time and $k$ dependence, or with only $k$ dependence, refer to redshift zero, that is,
$P_{11} = P_{11}(k) = P_{11}(k,z=0)$. Furthermore,  when time dependent power spectra are used, the $z$ dependence will be explicitly written.

We show the various contributions from  \eqref{pk_expansion} in Fig.\ref{plot-SPT} contrasted with 
the full non-linear result of the \syren~emulator~\citep{Sui:2024wob}. While the loop corrections start to be closer to the non-linear power spectrum, 
 over a wider range of scales, they eventually over-predict the power and diverge at smaller scales.

\subsection{Effective Field Theory}
In the previous subsection, we showed that at very small scales perturbation theory breaks down; see Fig. 1.
Thus, we focus on the intermediate scales that can still be described within perturbation theory.  
It is then useful to split the fields in the Boltzmann equations into short- and long-wavelength parts, 
 $\rho = \rho_l + \rho_s$ and $u^i = u^i _l+ u^i _s$, by applying a top-hat filter at a cut-off scale, ${\Lambda_{cut}}$.

Doing so results in a set of equations that involve the long wavelength fields, similar to equations (\ref{HOcont}) and (\ref{HOEuler}), plus an extra term that is entirely sourced by short modes. 
The continuity equation:
\begin{equation}
\label{continuity}
    \dot{\rho_l}+3H\rho_l+\frac{1}{a}\partial_i(\rho_lu^i_l)=0
\end{equation}
and the Euler equation:
\begin{equation}\label{euler}
    \dot{u^i_l}+Hu^i_l+\frac{1}{a}u^j_l(\partial_ju^i_l)+\frac{1}{a}\partial^i\phi_l=-\frac{1}{a\rho_l}\partial_j[\tau^{ij}]_{\Lambda_{cut}}
\end{equation}
where $[\tau^{ij}]_{\Lambda_{cut}}$ is the effective stress tensor
which 
 originates from the  smoothing out of  short-wavelength fluctuations ~\citep{Hertzberg:2012qn, Carrasco:2012cv, McQuinn:2015tva, Baumann:2010tm}. The effective stress tensor is a complicated function of all the terms that are allowed by General Relativity, and is non-local in time.
In practice, this introduces an imperfect fluid that cancels out the SPT divergences through counterterms, which provide time-dependent functions 
that further capture the smoothed-out short-scale physics.

It happens that certain infrared (IR) effects, particularly those arising from large-scale displacements, become non-negligible at relevant scales and can significantly degrade the accuracy of the results, especially concerning features like Baryon Acoustic Oscillations (BAO). For this reason, it is necessary to perform IR-resummation, which non-perturbatively incorporates these large-scale displacements, to obtain accurate predictions for the large-scale structure of the universe ~\citep{Senatore:2014via}.
The final resumed power spectrum will be given as a sum of the relevant loop terms, times a kernel that accounts for the order of resummation. 
Schematically, each term in the resulting expansion takes the form $\sim {\cal C}^n \IR{k^{2m} P^{(X_C)}_{{\rm L-loop}}}{N}$
where $N$ is the resummation order which depends on the kernel, $L$ is the loop order prior to resummation (with $P_{0-loop}$ being the linear power spectrum $P_{11}$), 
${\cal C}^n$ are possible counterterms to some power $n$ (if $n=0$ one is dealing with the SPT part),
and $(X_C)$ denotes possible counterterm-specific terms (and is omitted otherwise). We now consider the IR-resumed parts of the power spectrum, up to one and up to two loops.
 
\subsubsection{One-loop EFT}
At the one-loop level, the SPT power spectrum of equation (\ref{pk_expansion}) has a UV divergence that 
can be cancelled out by including the sound speed term of the effective fluid ~\citep{Hertzberg:2012qn, Baumann:2010tm, Pajer:2013jj, Foreman2015}

\begin{align}
P^{ \rm EFT}_{\rm 1-loop}(k, z) & = \underbrace{\IR{P_{11}(k,z)}{1}  + \IR{P_{\rm 1-loop}(k,z)}{0}}_{  P^{\rm SPT}_{\rm 1-loop }(k, z) } +  P_{c_s}(k,z)
\label{1loop_EFT}
\end{align}
where $ P_{c_s}(k,z)$ is the EFTofLSS counterterm contribution at $1$-loop given in terms of the sound speed $c^2_{s(1)}(z)$ as
\begin{equation}
P_{c_s}(k,z)=-2(2\pi)  [D(z)]^2  c^2_{s(1)}(z) \; \IR{k^2P_{11}}{0}.
\label{1loop_cs}
\end{equation}
where we have conventionally set $k_{NL}=1 \hmpc$.  The $z$-dependent power spectra in \eqref{1loop_EFT} are understood to be  given following our convention
explicitly defined through \eqref{D_dependence}.
The function  $c^2_{s(1)}(z)$  is not determined by the EFT but must be fitted to N-body simulations or to observations, up to 
a wavenumber $k_{fail}$ which stipulates the breakdown of $1$-loop EFT.

\subsubsection{Two-loop EFT}

The power spectrum up to two loops has a more complicated UV behavior and requires a different set of counterterms ~\citep{Carrasco:2013mua, Carrasco:2013sva, Foreman2015}.

\begin{align}
& P^{\rm EFT}_{\rm 2-loop }(k, z)=
  \underbrace{ \IR{P_{11}(k, z)}{2} + \IR{P_{\rm 1-loop}(k, z)}{1} + \IR{P_{ \rm 2-loop}(k, z)}{0} }_{P^{\rm SPT}_{\rm 2-loop}(k, z) }
\nonumber
\\
&
+ (2 \pi) c_{s(1)}^2 \left(  \IR{P_{\rm 1-loop}^{\rm (cs)}(k, z)}{0} -2 \IR{k^2 P_{11}(k, z)}{1} \right)
\nonumber
 \\
& 
-2(2 \pi) c_{s(2)}^2\IR{k^2 P_{11}(k, z)}{0}
+(2 \pi) c_1 \IR{P_{\rm 1-loop}^{\rm (quad)}(k, z)}{0}
\nonumber
\\
& +(2 \pi)^2 \left[ \left(c_{s(1)}^2\right)^2\left(1+\frac{2\zeta+5}{4\zeta+5}\right) +2  c_4 \right] \IR{k^4 P_{11}(k, z)}{0} 
\label{2loop_EFT}
\end{align}
where once more we have conventionally set $k_{NL}=1 \hmpc$, and $\zeta$ is a constant parameter.
  At $2$ loops new counterterms emerge in addition to $c^2_{s(1)}(z)$, and these are the
 time-dependent functions $\{c_{s(2)}^2(z),  \, c_1(z), \, c_4(z)\}$. These terms are in general non-local in time,
however,  local versions can be constructed approximately.

The factorisation of all terms appearing in \eqref{1loop_EFT} and \eqref{2loop_EFT} into separate functions  $z$ and $k$, see \eqref{D_dependence}, is what motivates our emulation strategy in
what follows. More importantly, the counter terms are left outside of the emulation scheme and are expected to be supplied as external functions by the user.
This makes \symeft~ more versatile as it allows different models for $z$ dependence of the counterterms to be used with the same emulator. We note that this
 decoupling strategy also (independently) employed in \citep{Donald_McCann_2022,Pezzotta:2025rxb,bonici2025effortfastdifferentiableemulator}.

\section{Symbolic representation of EFTofLSS}
\label{sec:emulator}

\subsection{Symbolic regression}
There is no faster way of approximating the output of a numerical algorithm, in our case the power spectrum, than by using an explicit functional form.
Furthermore, explicit functions can be inserted into any computer code quite easily, and offer an interpretable and fully differentiable way of interpolating, and in our case emulating the desired power spectra.

 Such explicit functional forms, in other words fitting functions,
 have traditionally been used in cosmology for this purpose. Old examples include, the CDM transfer function fit of~\cite{Bond:1984fp}, the CDM, WDM and masive neutrino transfer function fits of~\cite{Bardeen:1985tr},
 the non-linear matter power spectrum fit of~\cite{Peacock:1993xg} and the physics-informed improvement of \cite{Eisenstein:1997ik}. While these functions have served their purpose by providing sufficient accuracy at the time,
the very high precision that is necessary now and the shear number of terms that need interpolation makes this task humanly impossible.

Symbolic regression is a supervised machine learning method which generates explicit mathematical expressions that can be used to model a given dataset. 
In contrast with other regression methods where the functional form is fixed and optimisation takes place only on the free parameters, in symbolic regression  the structure of these functions 
is also unknown and is part of the  optimisation procedure. How well a functional form fits the data is encapsulated in a ``loss'' function, and the task of a symbolic regression algorithm
 is not only to minimise the loss function, but may also be
to select expressions of smaller length to avoid over-fitting and enable generalization. As such, one implements a multi-objective strategy.
\begin{figure}
\centering
\includegraphics[width=\linewidth]{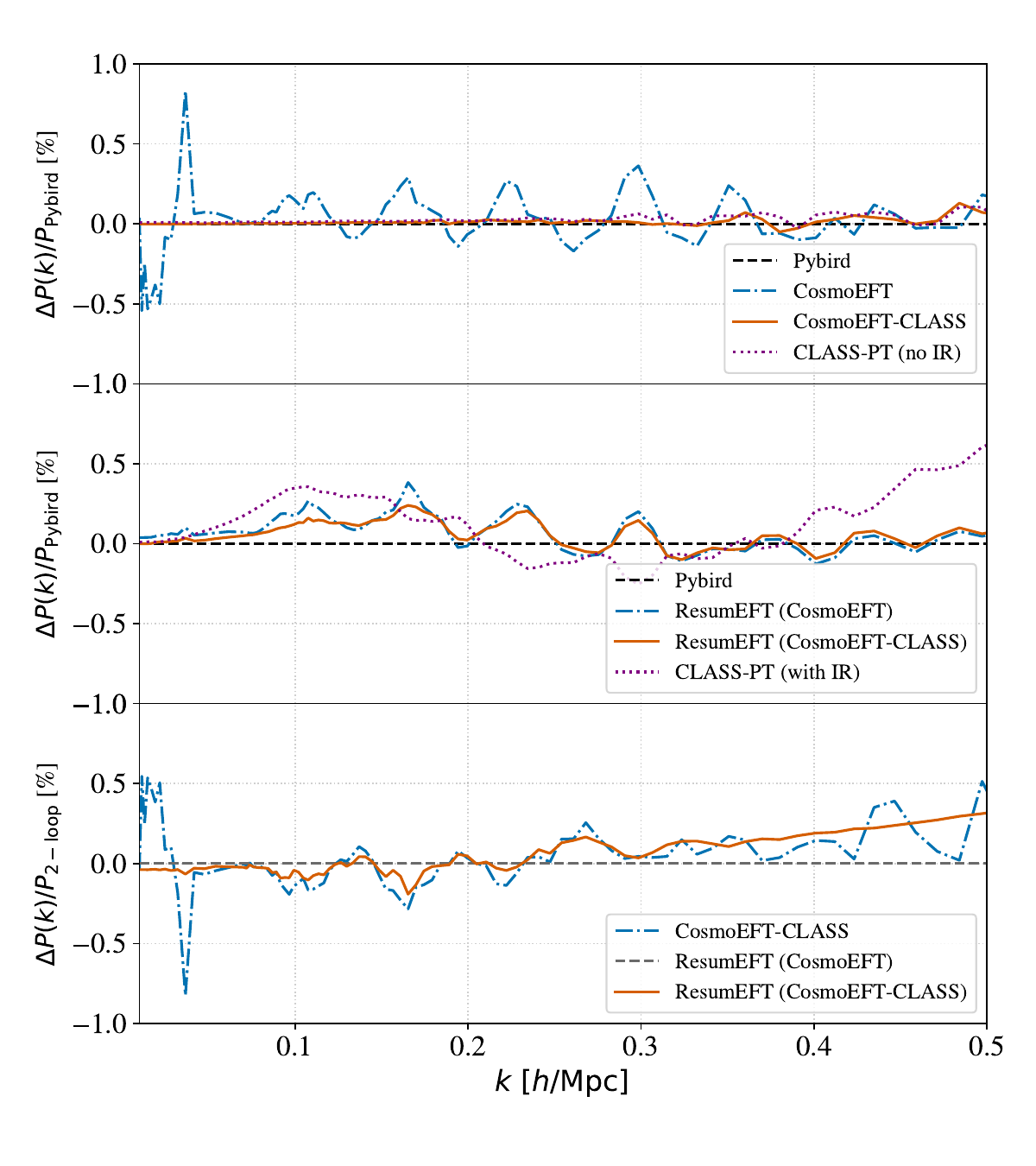}
 \caption{ \panel{Top:} $P^{\rm SPT}_{\rm 1-loop}(k)$ $\%$-relative difference  of \cosmoeft~(dash-dot, blue), \classcosmoeft~(solid, orange) and \classpt~(dotted, purple) 
compared respectively with \pybird (baseline, dashed black). \panel{Middle:} IR-resummed $P^{\rm SPT}_{\rm 1-loop}(k)$ $\%$-relative difference of \cosmoeft~(dash-dot, blue), \classcosmoeft~(solid, orange) and \classpt~ (dotted, purple) compared with \pybird(baseline). \panel{Bottom:} 
Comparison between the $2$-loop SPT power spectrum $P^{\rm SPT}_{\rm 2-loop}(k)$  
with and without IR resummation. Shown is the $\%$-relative difference between the baseline model (IR-resummed \cosmoeft)  and \cosmoeft~prior to resummation (dash-dot, blue),
and with \classcosmoeft(solid, orange) with resummation.}
\label{fig:compare_CosmoEFT_CosmoEFT-Class_Pybird_ClassPT}
\end{figure}

A commonly used approach to symbolic regression is by applying Genetic Programming~\citep{Koza_1992}, a method inspired by natural selection. In the context of symbolic regression, expressions are drawn from a pool and manipulated
through either cross-breeding or through mutation. The ones which fit better are then retained and after several generations, more accurate expressions of various lengths are produced. The two-dimensional plot of accuracy versus expression length is called the 
Pareto front, and it signifies the most optimal set of expressions found until the moment the algorithm was terminated.

Several implementations of symbolic regression exist. We chose the \operon~framework~\citep{OperonPaper} which implements symbolic regression using genetic programming. 
Written in C++ and publicly 
available~\footnote{The C++ implementation of the \operon~framework can be obtained from \url{https://github.com/heal-research/operon}.
For practical purposes, it is far easier to use its Python wrapper Pyoperon from \url{https://github.com/heal-research/pyoperon} or \url{https://pypi.org/project/pyoperon/}.}, it is based 
on a compact and efficient linear  tree encoding and indexing scheme for representing expressions, it internally uses dual numbers for auto differentiation, has low memory footprint and is highly parallel. 
It has been shown to scale well to evolving over $10^6$ individual expressions  depending on multi-dimensional parameter spaces. A comparison of several symbolic regression codes, including some which use a different method than genetic programming, 
may be found in \cite{LaCavaEtAl2021}. It is found that $4$ out of $5$ top performers are based on genetic programming and \operon~is at a sweetspot on the Pareto front of models in terms of simplicity and accuracy of produced expressions.
A benchmarking tool for symbolic regression codes was introduced in~\citep{Thing_2025} and demonstrated in cosmological and astrophysical setups.
Finally, in~\cite{RadwanEtAl2024} it was demonstrated that genetic programming is still superior compared to modern implementations of symbolic regression using deep generative neural networks.
~\footnote{\label{symbolic_footnote}Nevertheless, genetic programming is not by itself an efficient method for implementing symbolic regression~\citep{Kronberger_Inefficiency}, as a large portion of visited expressions are of low quality, and several of them 
are semantically equivalent to previously visited expressions. 
Thus, there is still a lot of room for improving the techniques, although we note that symbolic regression has been shown to be an NP-hard problem~\citep{virgolin2022symbolic}.  }

Symbolic regression has recently been used in cosmology to emulate the linear power spectrum~\citep{Bartlett:2023cyr}, 
non-linear dark matter power spectrum~\citep{Bartlett:2024jes,Sui:2024wob}, the baryon boost to the matter power spectrum~\citep{Kammerer:2025dbi},
as well as the growth rate and comoving distance~\citep{bartlett2025symbolicemulatorscosmologyaccelerating}.

\subsection{EFTofLSS codes: generating the dataset}
We computed the terms of the EFTofLSS power spectrum in \eqref{1loop_EFT} and \eqref{2loop_EFT} using  the \cosmoeft~C++ code~\citep{Cataneo:2016suz} and passed the result through
the \resumeft~code ~\citep{Cataneo:2016suz} for performing the IR resummation~\footnote{
These codes used to be publicly available at \url{http://web.stanford.edu/~senatore/} which no longer exists.}.
Since we were not concerned with speed when generating the datasets, we set $\Lambda_{IR}=0.1$ in  \resumeft~to increase accuracy.
  The \cosmoeft~code takes as input a set of cosmological parameters and the linear matter transfer function. 
For efficient calculation of the EFtofLSS terms, \cosmoeft~uses internaly the \texttt{COPTER} code~\citep{Carlson_2009} which provides IR-safe integrands~\citep{Carrasco:2013sva} and computes the loop integrals 
using Monte Carlo integration routines from the CUBA library~\citep{Hahn:2004fe}. 
 Moreover, it is set to use a fiducial cosmology that is computed once (by the code) and used to calculate the desired power spectra for other cosmologies which are close enough, accelerating the computation while maintaining fair precision. 
Since in our case, speed was not a concern at the computation stage, we instead bypassed the fiducial cosmology and let the code to always compute the power spectra from scratch. 
Moreover, we interfaced \texttt{CosmoEFT} with the Boltzmann code \class~\citep{Diego_Blas_2011} and passed it the exact numerically calculated background and linear cosmology directly. Thus,
the resulting  \classcosmoeft~code was able to provide more accurate spectra in order to have better precision when coupling it to \operon. After tuning its precision parameters, 
the code reported estimated errors  $\sim0.1\%$ which we have taken as the base value.

Other codes for computing the $1$-loop EFTofLSS power spectra~\footnote{
Pybird is publicly available at \url{https://github.com/pierrexyz/pybird} while the Class-PT code can be found from \url{https://github.com/Michalychforever/CLASS-PT}. Class-OneLoop is not yet public.}
include Class-PT~\citep{Chudaykin:2020aoj}, Pybird~\citep{DAmico:2020kxu} and Class-OneLoop~\citep{Linde:2024uzr}, which in addition to the $1$-loop matter power spectrum compute correlators for biased tracers. These codes are generally faster than CosmoEFT,
however, they do not provide the $2$-loop matter power spectrum needed here.

We compare $P^{\rm SPT}_{\rm 1-loop }$ from \eqref{1loop_EFT}  of the matter power spectrum  from \cosmoeft, \classcosmoeft, \pybird~ and \classpt~ in Fig.\ref{fig:compare_CosmoEFT_CosmoEFT-Class_Pybird_ClassPT},
without IR resummation (top panel) and with IR resummation (middle panel). 
We see that without IR resummation the difference between \pybird~,\classpt~and \classcosmoeft~is negligible until $k\sim 0.5\hmpc$ (within $0.1\%$) 
while \cosmoeft~differs at the $0.5\%$ level -- the difference is largely due to the use of the fiducial cosmology in \cosmoeft, and its internal use of \texttt{COPTER} which does not account for massive neutrinos.
We note that performing the comparison between \cosmoeft~and \classcosmoeft~ without massive neutrinos reduces the difference significantly. 

Switching on IR resummation increases the difference between  \pybird~and \classcosmoeft~to $\sim0.25\%$.
While we tried increasing the accuracy settings of both codes at the cost of speed, 
and also varied the $\Lambda_{IR}$ parameter, we were unable to reduce this difference~\footnote{We thank Pierre Zhang for discussions on this issue and help with Pybird.}. 
In the \classpt~case, we see larger differences reaching $\sim0.5\%$ and growing to over $2\%$ at $k\sim0.8\hmpc$. This larger difference in \classpt~is related to the IR resummation scheme. 
In doing both comparisons, with and without IR resummation, we had adjusted the 1-loop sound speed to a small non-zero value to align the power spectra, after having observed a difference proportional to $\IR{k^2P_{11}(k)}{0}$.
A more detailed comparison of all the EFT codes is left for a future, more dedicated study.

Although no other $2$-loop code apart from~\cosmoeft~was available to us, we performed an {\it internal} comparison between \cosmoeft~and \classcosmoeft~at the bottom panel 
  of Fig.\ref{fig:compare_CosmoEFT_CosmoEFT-Class_Pybird_ClassPT}, with  and without IR resummation. 
This shows that  the \cosmoeft~method of using a fiducial cosmology and then computing the desired spectra for cosmologies which are close enough to this fiducial 
cosmology can introduce an error of $\sim0.25\%$ at $k\lesssim 0.4\hmpc$, increasing to $0.5\%$ at larger $k$, with or without IR resummation. 

\subsection{The emulators}
\begin{figure*}
\centering
\includegraphics[width=0.9\linewidth]{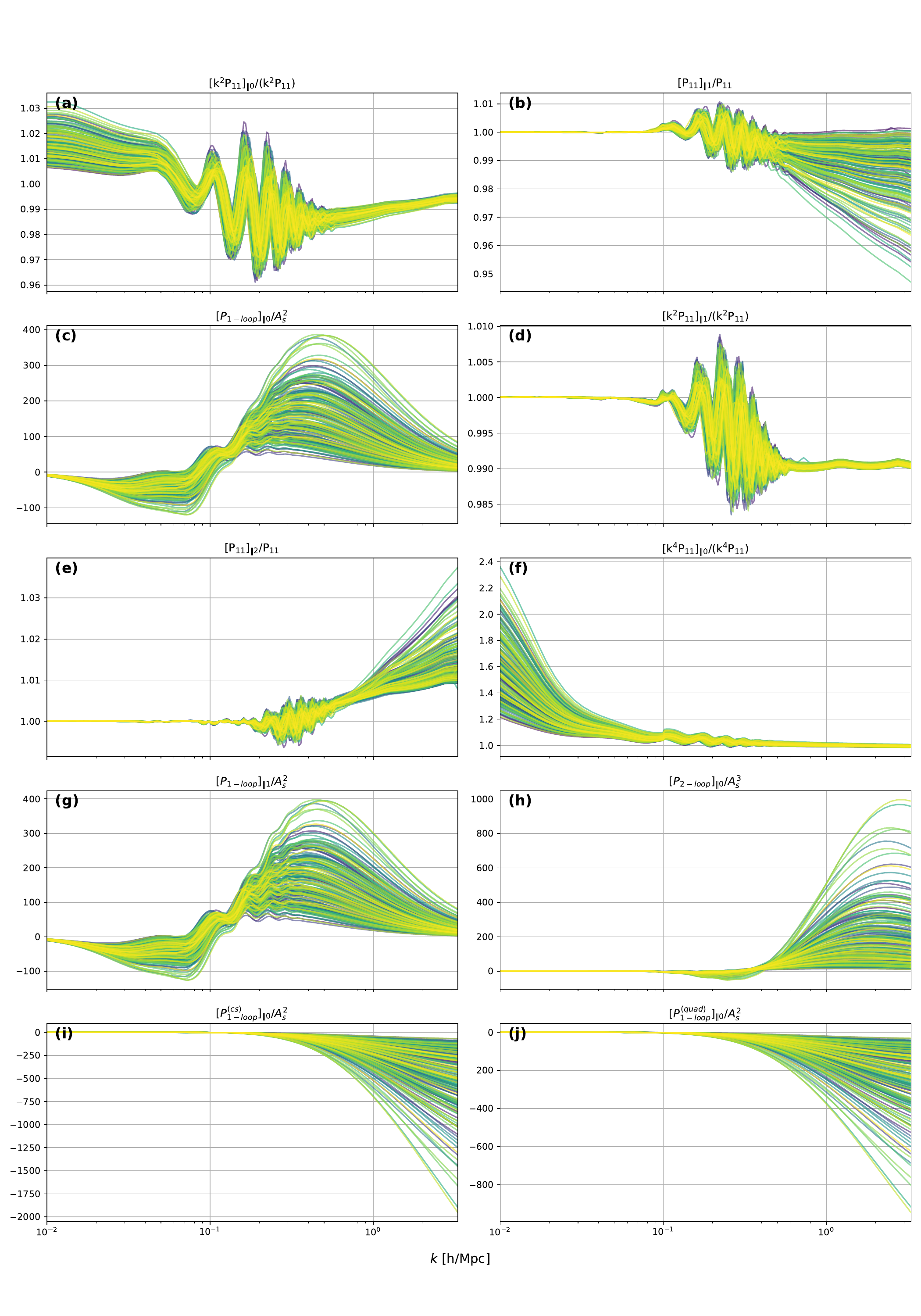}
\caption{Training Data for 200 cosmologies. Each line corresponds to a specific set of cosmological parameters, sampled with a Latin hypercube within the bounds in Table.~\ref{tab:cosmological_parameters}}
\label{fig:data}
\end{figure*}

We used \classcosmoeft~and \resumeft~to generate our training set of $200$ cosmologies drawn from the set of parameters $\{ \Ast \, ,\Omega_{cb} \, , \Omega_{b} \, ,h \, ,n_s\}$,
where $\Ast \equiv 10^9 A_s$.
We sampled these with a Latin hypercube within the same range as for \syren~\citep{Sui:2024wob} and the Euclid emulator II~\citep{Euclid:2020rfv},
 shown in  Table \ref{tab:cosmological_parameters}.
The codes return all the individual terms entering the $1$-loop \eqref{1loop_EFT} and $2$-loop \eqref{2loop_EFT} power spectra. The
full set is listed in  Table \ref{tab:emulated_functions}.
 We kept a fixed $114$ $k$-values for all models between $k=0.01\hmpc$ and $k_{max} = 3.3\hmpc$,  log-spaced, but with more points
 where EFTofLSS is relevant, which is $0.05<k<1 \hmpc$. For validation, we generated a further $100$ cosmologies from the same set,
however, at different points in parameter space and sampled on the same $k$ values as with the training set.
Since our $k$ values are densely populated  and given that functions generated by Symbolic regression are typically smooth, we expect fluctuations between $k$ values to be small.
 Furthermore, in several cases we found it more efficient to normalize the EFT functions tabulated in Table \ref{tab:emulated_functions}  
with some convenient known functional combination to achieve faster convergence of the emulator. The training set is shown collectively in Fig.\ref{fig:data}.
\begin{table}
\centering
\begin{tabular}{ccc}
\hline
Parameter & Lower bound & Upper bound
 \\
\hline
\hline
\noalign{\vskip 2pt} 
$\Ast \equiv 10^{9} A_{s}$ & $1.7$ & $2.5$ \\
\hline
$\Omega_m$ & $0.24$ & $0.4$ \\
\hline
$\Omega_b$ & $0.04$ & $0.06$  \\
\hline
$h$ & $0.61$ & $0.73$\\
\hline
$n_{\mathrm{s}}$ & $0.92$ & $1$  \\
\hline
\end{tabular}
\caption{Cosmological parameter ranges used for generating our training sets. We also kept two massless neutrinos and one massive with $m_{\nu}=0.06eV$.}
\label{tab:cosmological_parameters}
\end{table}

\begin{table}
\centering
\begin{tabular}{ccr}
1-loop  & 2-loop    & Notes
 \\
\hline
\hline
$\IR{k^2P_{11}(k)}{0}$ &
$\IR{k^2P_{11}(k)}{1}$
\\
\hline
$\IR{P_{11}(k)}{1}$ &  
 $\IR{P_{11}(k)}{2}$ &
\\
\hline
-&  $ \IR{k^4P_{11}(k)}{0}$ 
\\
\hline
$\IR{P_{\rm  1-loop }(k)}{0}$ & $ \IR{P_{\rm 1-loop}(k)}{1}$  & Three regions each
\\
\hline
- & 
$\IR{P_{\rm 2-loop}(k)}{0}$ & Two regions 
\\
\hline
- &  $\IR{P_{\rm 1-loop}^{\rm (cs)}(k)}{0} $ 
\\
\hline
- & $\IR{P_{\rm 1-loop}^{\rm (quad)}(k)}{0} $
\\
\hline
\end{tabular}
\caption{Emulated functions for 1-loop and 2-loop EFTofLSS terms.}
\label{tab:emulated_functions}
\end{table}

While the genetic programming algorithm of \operon~is completely deterministic, the output does depend on the initial random seed. 
Thus, it may happen that the generated expressions were not of sufficient quality, and this is not easy to control beforehand (cf. footnote \ref{symbolic_footnote}).
Therefore, we run \operon~for few different hyperparameters, namely,  $\epsilon=\{10^{-3}, 10^{-4}\}$, `maximum selection pressure'$=\{80,100\}$,
`population size'$=\{1000,1500\}$, `tournament size'$=\{5,10\}$, `optimizer iterations'$=\{8,9,10\}$ and `pool size'$=\{100,150\}$. Furthermore, we run $\sim10$ different
 initial random seed numbers keeping the same hyperparameters and collected the best expressions found after all these runs had finished.

Through experimentation we found that running Operon for hyperparameters within this range~\footnote{We thank Bogdan Burlacu and Gabriel Kronberger for advise on the choice of hyperparameters.} 
 generally produces better expressions. Once within this range, we did not observe substantial sensitivity as to the exact choice, while the choice of initial random seed led to 
wider variations in the resulting expressions.  Our experiments indicated that
around one third of runs led to better expressions and close to the best ones chosen.

\begin{figure*} 
\centering
  \includegraphics[width=0.48\linewidth,height=\appfigheight,keepaspectratio]{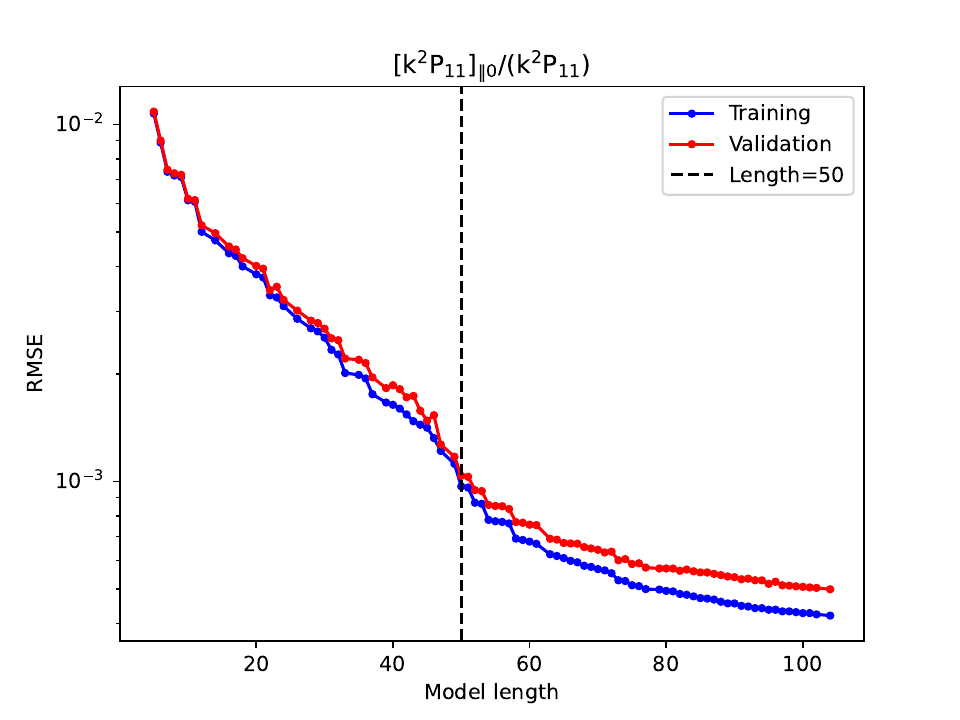}
   \includegraphics[width=0.48\linewidth,height=\appfigheight,keepaspectratio]{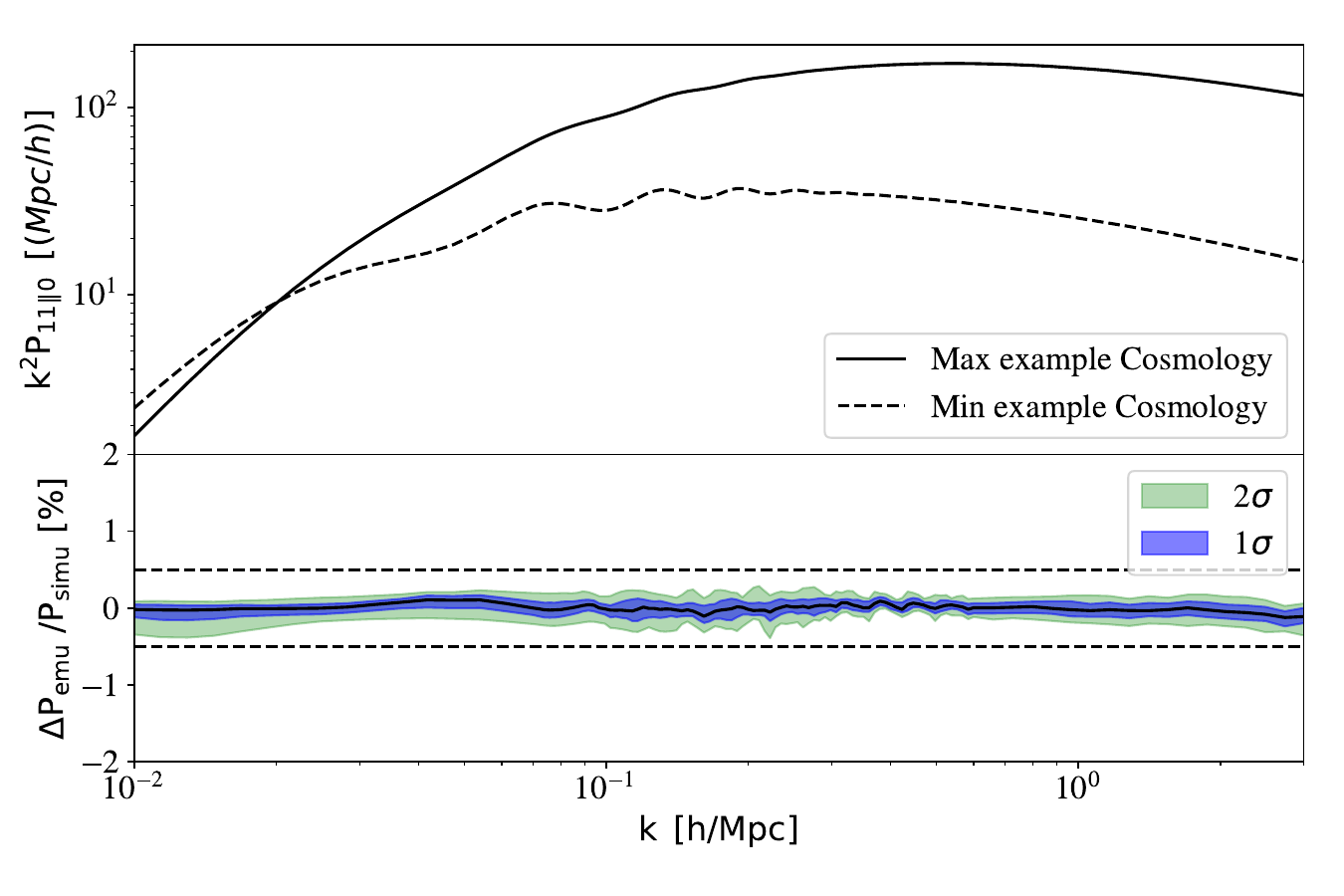}
\caption{ \panel{Left:} The Pareto front of RMSE vs model length for the $\frac{\IR{k^2 P_{11}}{0}}{k^2 P_{11} }$ emulator runs as generated by \operon, with blue marking the training and red the validation error, 
and with the chosen model of length 50 indicated by the vertical dashed line.  \panel{Right:} The top plot shows the  $\IR{k^2 P_{11}}{0}$ function for two extreme cases of cosmological parameters
while the bottom plot displays the resulting $1\sigma$ and $2\sigma$  emulator $\%$ error for all 300 cosmologies. The horizontal dashed lines mark the $0.5\%$ threshold.}
  \label{Pareto_and_error:k2p11_R0}
\end{figure*}

 Expressions generated by \operon~can very often be simplified further, for example, by combining constants together. 
Moreover, \operon~optimizes all the coefficients in the expressions it determines to machine precision by default. However, since our 
error tolerance is larger, we reduced the significant figures of each constant in our chosen expression up to the point that the error would 
start to change by an amount larger than $10^{-3}$. This process may result to slight reductions to the expression length compared to that reported by \operon~and 
displayed in the Pareto fronts.

To make the emulators more useful, we have tried to make the emulation error as small as possible. However, noticing that the differences between one-loop EFT codes are within 
$0.5\%$, as discussed above and seen in Fig.\ref{fig:compare_CosmoEFT_CosmoEFT-Class_Pybird_ClassPT}, it makes little sense to reduce the error to smaller values.
Thus we set our emulation error target to $\sim0.5\%$.

\subsubsection{The 1-loop emulators}
\label{1_loop_methodology}
At $1$-loop we have the following $3$ functions: $\IR{k^2P_{11}(k)}{0}$, $\IR{P_{11}(k)}{1}$ and $\IR{P_{\rm 1-loop }(k)}{0}$,
the first one being used for the $c_{s(1)}^2$ counterterm and the latter two forming $P^{\rm SPT}_{\rm 1-loop }$.

Consider first $\IR{k^2P_{11}(k)}{0}$. The functional form of this term is fairly close to the bare linear spectrum multiplied by $k^2$, that is, $k^2 P_{11}(k)$, upto $O(1)$ deviations. 
Thus it is far more efficient to emulate instead their ratio $E^{(1)}_{cs^2} \equiv \frac{ \IR{k^2 P_{11}}{0}}{k^2 P_{11}}$,
shown in panel `(a)' of Fig.\ref{fig:data}. The resulting Pareto front of the Root Mean Square Error (RMSE) vs expression length 
is shown on the left panel of Fig. \ref{Pareto_and_error:k2p11_R0}. We chose an expression of length 50, which after our reduction process described above, takes the form 
\begin{align}
E^{(1)}_{cs^2} =& 
C_{19} 
+ \Bigg\{  \Omega_{cb} \left[C_{12} h - \cos{\left(C_{13} \Ast \right)}\right] - C_{14} \Omega_{b}  \Bigg\} 
\nonumber
\\
&
\times \left(\frac{C_{15}}{\sqrt{C_{16} k^{2} + 1}} - \frac{C_{17}}{\sqrt{C_{18} k^{2} + 1}}\right)
\nonumber
\\
&
- \frac{\left( C_{0} \Omega_{b} - C_{1} k\right) \cos{\left(C_{6} k + \frac{k \left(C_{2} h + C_{3}\right)}{\sqrt{k^{2} + \left(C_{4} \Omega_{cb} + C_{5} \Omega_{b}\right)^{2}}} + \ln{\left(C_{7} k \right)} \right)}}{\sqrt{\left(C_{8} \Ast + \frac{\frac{C_{10} \Omega_{cb}}{k}  - C_{9} \Ast}{\sqrt{C_{11} k^{2} + 1}}\right)^{2} + 1}} 
\label{equation_k2P11_R0}
\end{align}
where the coefficients are 
\begin{align}
 C_0&= 0.709  & C_5 &=  10.322  & C_{10} &=  1.813   & C_{15}  &=  0.1188,
\nonumber
\\
 C_1 &= 0.059 & C_6 &=  55.1   & C_{11} &= 0.015   &  C_{16}  &=  249,
\nonumber
\\
 C_2 &= 52.48 & C_7 &=   6.125   & C_{12} &= 3.395    & C_{17}  &= 0.0609,
\label{constants_k2P11_R0}
\\
 C_3 &= 23.42 & C_8 &=  1839.487 & C_{13} &=   0.472   & C_{18}  &=  5.68,
\nonumber
\\
 C_4 &= 3.305 & C_9 &=   1840.901  & C_{14} &=  3.9   & C_{19}  &= 0.99508.
\nonumber
\end{align}
Our choice was informed by inspecting the final emulation error on the $1$-loop spectrum,
as smaller expressions were not reducing the emulation error within out target threshold. We return to this issue in the discussion and conclusion section below.

\begin{figure*}
    \centering
        \includegraphics[width=0.48\textwidth]{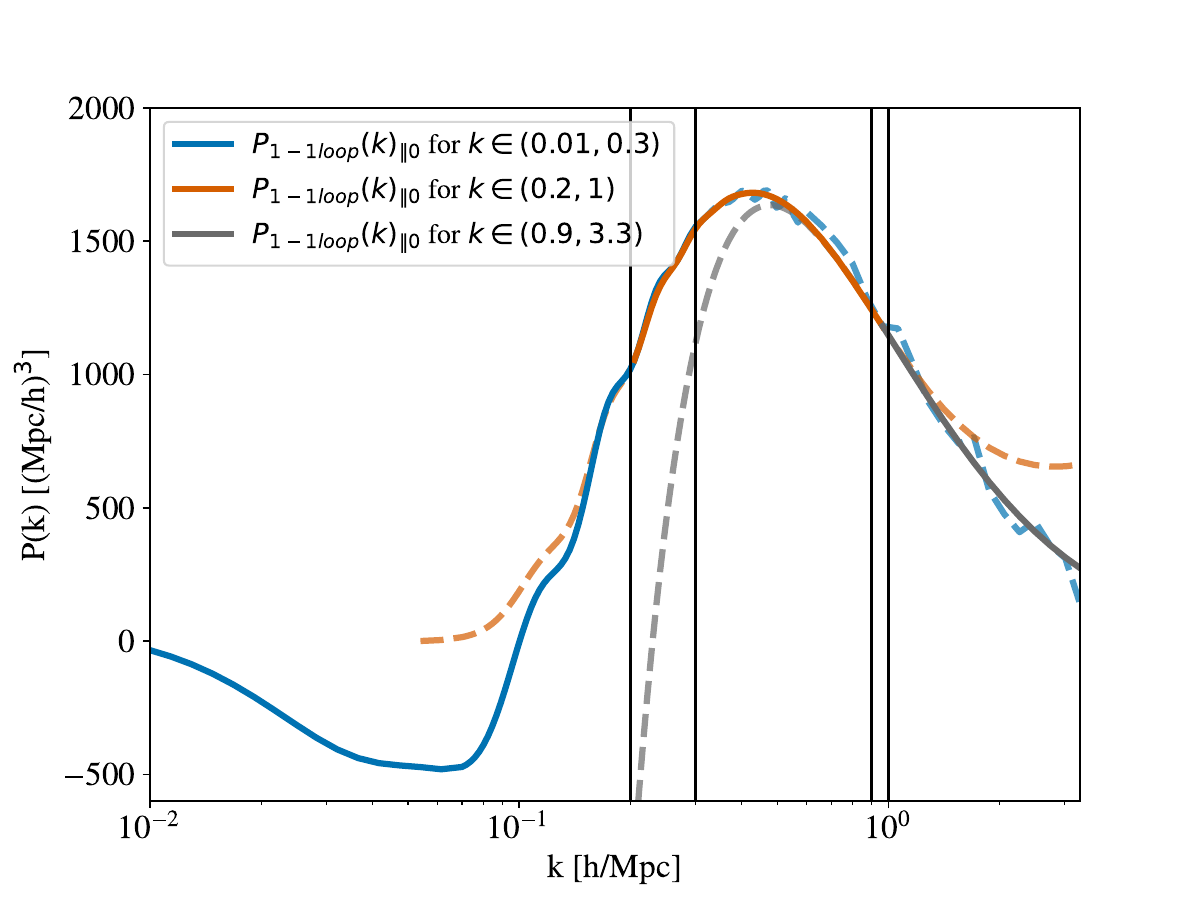}
        \includegraphics[width=0.48\textwidth]{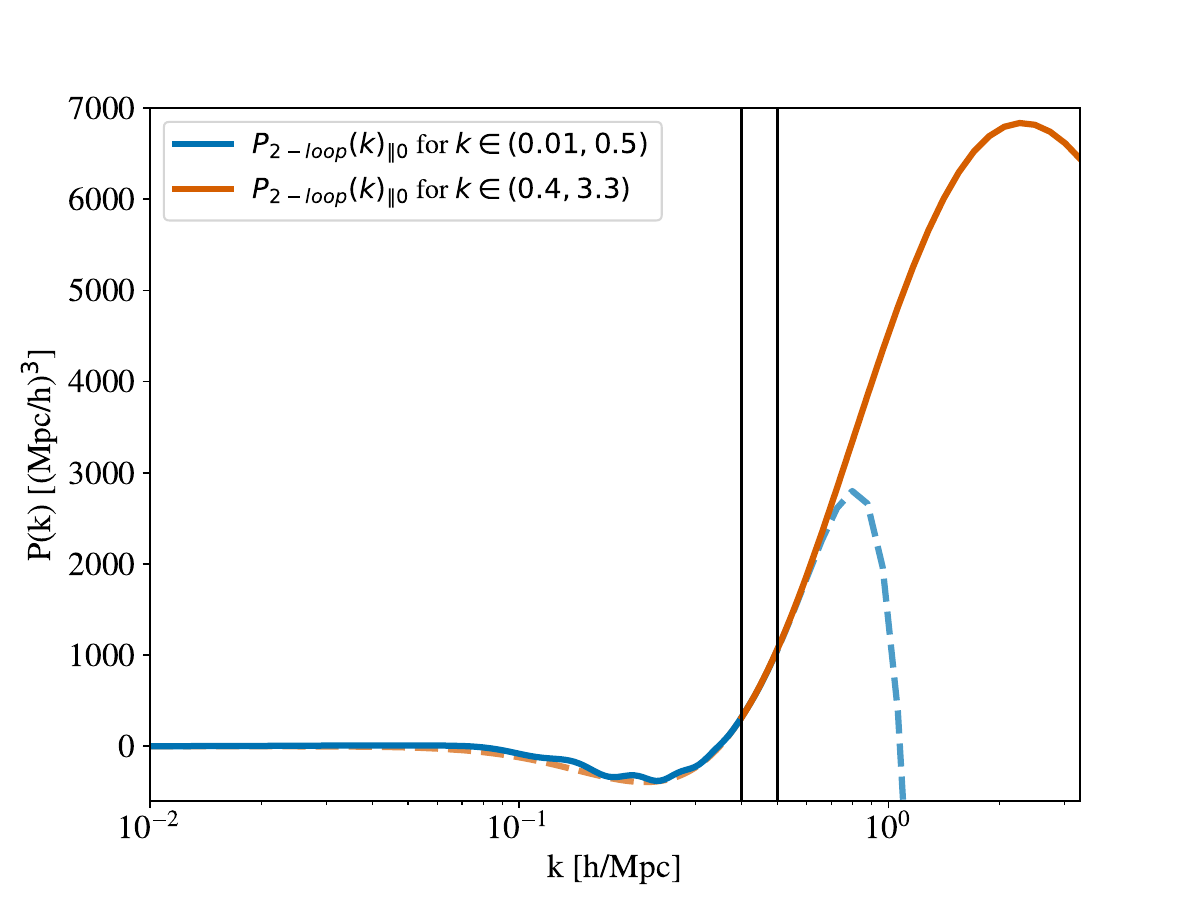}
    \caption{\panel{Left:} The $\IR{P_{\rm 1-loop}}{0}$ power spectrum for one set of cosmological parameters, 
split in three overlaping regions whose boundaries are marked with vertical lines.
We plot each emulated function within its region with a solid line, and outside its region of validity with a dashed line.
 \panel{Right:} The $\IR{P_{\rm 2-loop}}{0}$ power spectrum, split in two overlaping regions, whose boundaries are marked with vertical lines.
We use the same plotting conventions as the left panel.}
    \label{intervals}
\end{figure*}

We turn now to the $\IR{P_{11}}{1}$ part of the spectrum. In this case, it is better to emulate the ratio $\IR{P_{11}}{1} / P_{11} $, 
shown in panel  `(b)' of Fig.\ref{fig:data}, which removes the 
variation of $P_{11}$ over three orders of magnitude. 
The resulting emulator of length 50 and its coefficients as well as the Pareto front are shown in appendix \ref{sec:EmulatorSuite_1_loop}.

The final part of the $1$-loop spectrum is the function $\IR{ P_{1loop}}{0}$, normalized by dividing wth $\Ast^2$, motivated by the fact that $\IR{ P_{1loop}}{0}$
is proportional to  $\Ast^2$.
We note that the strategy of rescaling spectra by powers of  $\Ast$ has been also used in \cite{Bonici_2024,bonici2025effortfastdifferentiableemulator,derose2026lensingcounternarrativeeffective}.
This function has proven to be more difficult to emulate due its large variation from very negative to very large and positive values, 
in addition to the squeezed part around $k\sim 0.1 \hmpc$ which shows less variation between the different cosmologies. 
Thus, to achieve better precision we split the $\IR{ P_{1loop}}{0}$ function into three overlapping regions in $k$-space.
We checked various ways to split the regions and found that the best choice resulting in lower errors is the one described below.

Our choice was motivated as follows. On very small scales the dominant term in the power spectrum is $P_{11}$ (and its IR resummations), and so we defined the first region
as $k=[0.01, 0.3] \times \hmpc$. In the second region we require high precision in the EFTofLSS contribution which has to perform well to higher $k$ and so we chose
 $k=[0.2, 1.] \times \hmpc$, to allow overlap with region 1.  In the third region the integrals for all cosmology 
seem to have a similar decaying behavior, see panel `(c)' of Fig.\ref{fig:data}, and we chose this to be $k=[0.9, 3.3] \times \hmpc$. 
Even though the EFT is expected to fail already around $k_{\rm fail}\sim 0.4 \hmpc$ at redshift zero, 
 having the spectra out to higher $k$ is necessary as $k_{\rm fail}$ increases with increasing redshift. This does not mean, however, that the higher-loop contributions are more important at higher
redshift; rather, the higher-loop contributions are also multiplied by higher power of $D(z)$ and thus, also more suppressed at higher redshift relative to the linear spectrum.

\begin{figure}
\centering
 \includegraphics[width=\linewidth,keepaspectratio]{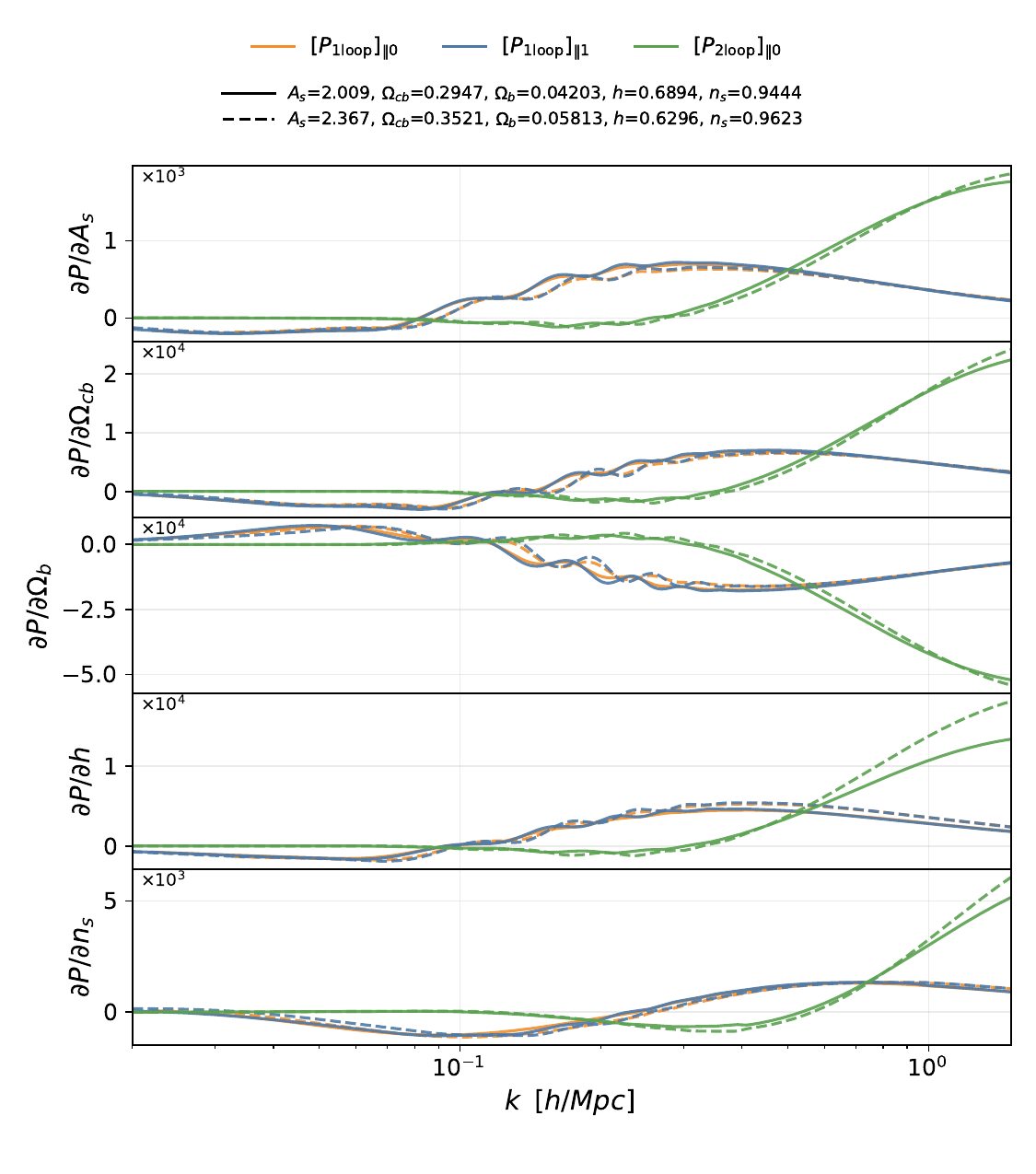}
\caption{The first derivative of  $\IR{ P_{1loop}}{0}$ (orange),  $\IR{ P_{1loop}}{1}$ (blue) and  $\IR{ P_{2loop}}{0}$ (green) 
   with respect to each cosmological parameter and for two example cosmologies each. 
}
  \label{derivatives}
\end{figure}
We created three emulators, one for each region, corresponding to $\IR{ P_{1loop}}{0}$. Our chosen functions have lengths 70, 83 and 50 respectively, and are 
shown in appendix~\ref{sec:EmulatorSuite_1_loop},  along with their corresponding Pareto fronts generated by \operon~in Fig. \ref{fig:combined-pareto-1loop}.
While we have tried to choose smaller and thus simpler expressions, our primary measure was accuracy, which meant that choosing smaller expressions 
would not lead to our $0.5\%$ error target.
To create the full emulator for  $\IR{ P_{1loop}}{0}$, we joined these three functions along their respective overlap intervals, using the error function $\Erf\left[ C_{n} (k_{n} - k)\right]$,
where $C_{n}$ and $k_{n}$ are constants indexed by the left region. Specifically for the joining of regions 1 and 2 we set $C_1 = 100$ and $k_1= 0.25$,
while for joining regions 2 and 3 we set  $C_2 = 40$, $k_2 =0.95$.  While one would expect that the joining would be dependent on the cosmological parameters, 
in practice this is not necessary, and this leads to a simpler implementation.
\begin{figure*}
\centering
  \includegraphics[width=0.48\linewidth,height=\appfigheight,keepaspectratio]{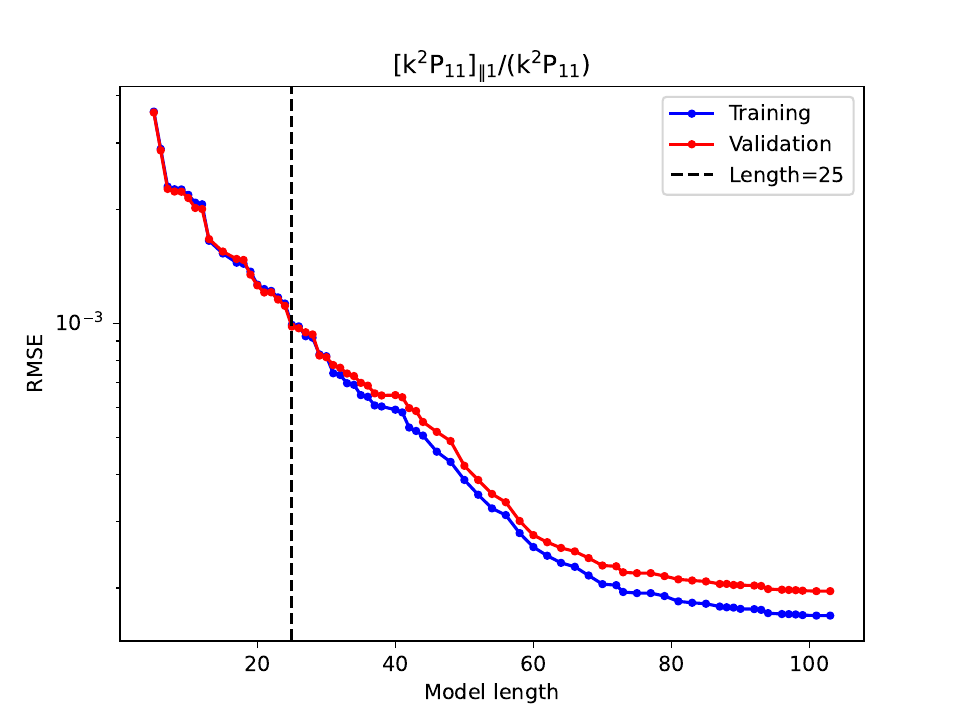}
  \includegraphics[width=0.48\linewidth,height=\appfigheight,keepaspectratio]{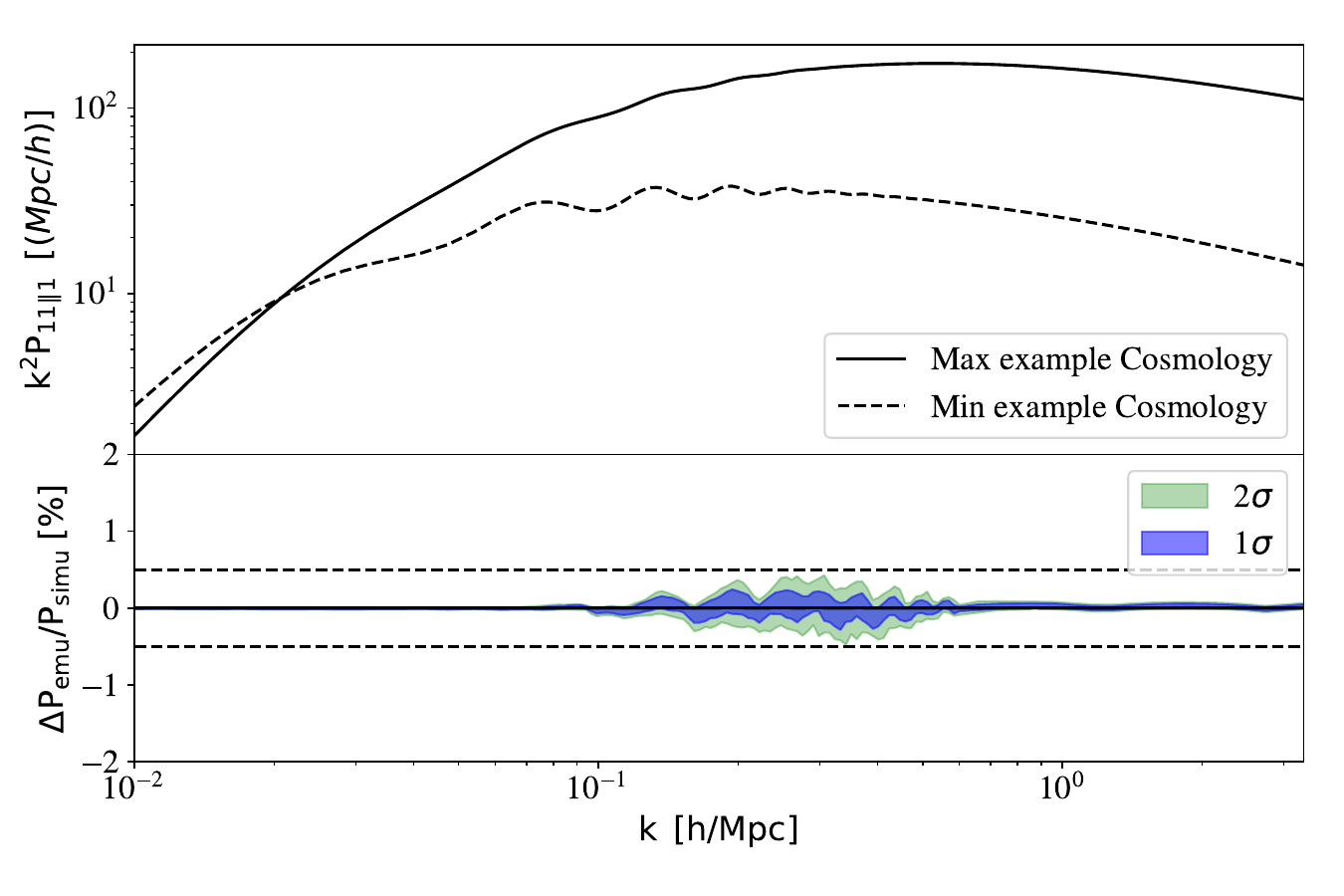}
\caption{\panel{Left:} The Pareto front of RMSE vs model length for the $\frac{\IR{k^2 P_{11}}{1}}{k^2 P_{11} }$ emulator runs as generated by \operon, with blue marking the training and red the validation error, 
and with the chosen model of length 25 indicated by the vertical line.  \panel{Right:} The top 
plot shows the  $\IR{k^2 P_{11}}{1}$ function for two extreme cases of cosmological parameters
while the bottom plot displays the resulting $1\sigma$ and $2\sigma$  emulator $\%$ error for all 300 cosmologies.
The horizontal dashed lines mark the $0.5\%$ threshold.}
  \label{k2p11_R1_combined}
\end{figure*}

On the left panel of Fig.\ref{intervals} we show the emulated $\IR{ P_{1loop}}{0}$ for one set of cosmological parameters, 
indicating  the three constituent emulators in blue, orange and grey (from large to small scales) and how they join within each overlapping region to produce the full emulated
function. The boundaries of the overlapping regions are marked by vertical lines. 
We use dashed lines to show how each emulator extrapolates outside their respective region, however, 
in practice we set the extrapolated part explicitly to zero. 

One may worry that joining together expressions may lead to non-smooth derivatives, posing problems for gradient-based inference methods, such as, Hamiltonian Monte Carlo. 
However, the transition from one region to another does not happen at a fixed point in $k$, but over a  continuous interval.
 Since within each region the functions are smooth (with smooth derivatives with respect to $k$ and all parameters), summing two such 
functions always results in another smooth function, including taking derivatives.
 We demonstrate this in practice in Fig.~\ref{derivatives}, showing the derivatives of $\IR{ P_{1loop}}{0}$, for 
each cosmological parameter, and for two example cosmologies.

\subsubsection{2-loop Symbolic-EFTofLSS}
\label{2_loop_methodology}
We now turn to the IR-resummed $2$-loop power spectra as they appear in \eqref{2loop_EFT}. There are $7$ functions to emulate, see Table. \ref{tab:emulated_functions}.
Consider first the  resummed $\IR{k^2 P_{11}}{1}$ function which is derived from the linear spectrum $P_{11}$ and is part of the $c_{s(1)}^2
$ counterterm contrubution to the  $2$-loop power spectrum. Just as in the $\IR{k^2 P_{11}}{0}$ case of the previous subsection, it is far more efficient to emulate the ratio
$ \frac{\IR{k^2 P_{11}}{1}}{k^2 P_{11} }$, as is easily seen from panel  `(d)' of Fig.\ref{fig:data}.
We show the RMSE vs model length Pareto front on the left panel of Fig.\ref{k2p11_R1_combined} with  blue marking the training and red the validation error. 
Even though \operon~found models resulting in 8 times smaller error, we chose a function with model length 25 as it was sufficient to pass our error targets.
The resulting emulator is given by 
\begin{align}
 \frac{\IR{k^2 P_{11}}{1}}{k^2 P_{11} }  =&
 C_{0} + \frac{1}{ \sqrt{C_{10} k^{C_{11}} + 1}} 
\Bigg[
 C_{8} \cos{\left(C_{9} k \right)}
\nonumber
\\
&
-\frac{ C_{1} \Omega_{b} \cos{\left(C_{2} \Omega_{cb} + C_{3} \Omega_{b} - C_{4} h - C_{5} k \right)}}{\sqrt{\frac{C_{6}}{k^{C_{7}}} + 1}} 
\Bigg]
\label{equation_k2P11_R1}
\end{align}
where the coefficients are 
\begin{align}
 C_0&= 0.99072  & C_4 &=   3.9        & C_{8} &=   9.218  \times 10^{-3}
\nonumber
\\
 C_1 &=  0.132   & C_5 &=   95.5      & C_{9} &=       3.44
\label{constants_k2P11_R1}
\\
 C_2 &= 19.23  & C_6 &=     10^{-6}   & C_{10} &=   5.6 \times 10^{4}
\nonumber
\\
 C_3 &= 63    &  C_7 &=  8.24         & C_{11} &=     9.9 
\nonumber
\end{align}
This was by far the simplest emulator we have constructed. Notice that there is no dependence on either the amplitude $\Ast$ nor the spectral index $n_s$, while 
the other three cosmologicical parameters, $\Omega_m$, $\Omega_b$ and $h$, appear only in the second cosine function, whose amplitude is modulated by $\Omega_b$. Interestingly, the double $k$-dependent
envelope given by the two square roots has no dependance on the cosmological parameters -- it is universal.

There are two further functions derived from the linear spectrum $P_{11}$. These are the IR-resummed $\IR{P_{11}}{2}$ which is part of $P^{\rm SPT}_{\rm 2-loop }$  
and $\IR{k^4 P_{11}}{0}$ which relates to the $c_4(z)$ counterterm. Just as the previous function,
it is more efficient to emulate the ratios $ \frac{\IR{P_{11}}{2}}{ P_{11} }$ and $ \frac{\IR{k^4 P_{11}}{0}}{k^4 P_{11} }$, shown 
in panels `(e)' and `(f)' of Fig.\ref{fig:data}, respectively. The resulting RMSE vs model length Pareto fronts are shown in appendix~\ref{sec:EmulatorSuite_2_loop}, 
left panel of Fig.\ref{Pareto_and_error:p11_R2} and Fig.\ref{Pareto_and_error:k4p11_R0}, respectively.
The chosen expressions have lengths 44 and 70, respectively, in order to be within our error threshold, as shown on the right panels of Fig.\ref{Pareto_and_error:p11_R2} and Fig.\ref{Pareto_and_error:k4p11_R0}.

We now turn to the $\IR{P_{\rm 1-loop}}{1}$ spectrum, displayed in panel  `(g)' of  Fig.\ref{fig:data}, normalized further by dividing with $\Ast^2$. This function has a striking resemblance to
 $\IR{P_{\rm 1-loop}}{0}$, as they are both IR resummations of the same underlying function $P_{\rm 1-loop}$, hence, we adopted the same emulation procedure as for the latter. 
To summarise, we split it into the same three regions in $k$-space as in Sec.\ref{1_loop_methodology} and emulated each region separately, joining the resulting functions as above. 
The Pareto fronts and chosen equations are displayed in appendix~\ref{sec:EmulatorSuite_P1loop_R1}.

Next in line is the $\IR{P_{\rm 2-loop}}{0}$ term, normalized by dividing with $\Ast^3$
 and shown in panel  `(h)' of Fig.\ref{fig:data}. While this function is markedly different from both the $\IR{ P_{\rm 1-loop}}{0}$  and $\IR{ P_{\rm 1-loop}}{1}$  cases, 
it exhibits similar difficulty in emulation because of its 3 orders of magnitude rise on scales smaller than $k\gtrsim 0.4\hmpc$ after first having gone through negative values on larger scales.
However, in this case having just two overlapping regions was sufficient. We chose the first region to be $k=[0.01, 0.5] \times \hmpc$
and the second as $k=[0.4, 3.3] \times \hmpc$, having tried different possibilities and finding that this combination 
was better at minimizing the error and capturing the features of the underlying $\IR{P_{\rm 2-loop}}{0}$  function.
The Pareto fronts resulting from the \operon~runs are shown in Fig.\ref{fig:p2loop_R0_pareto} of appendix~\ref{sec:EmulatorSuite_P2loop_R0}, along with the chosen equations of lengths  65 and 70 respectively.
\begin{figure*}
\centering
\includegraphics[width=0.45\linewidth]{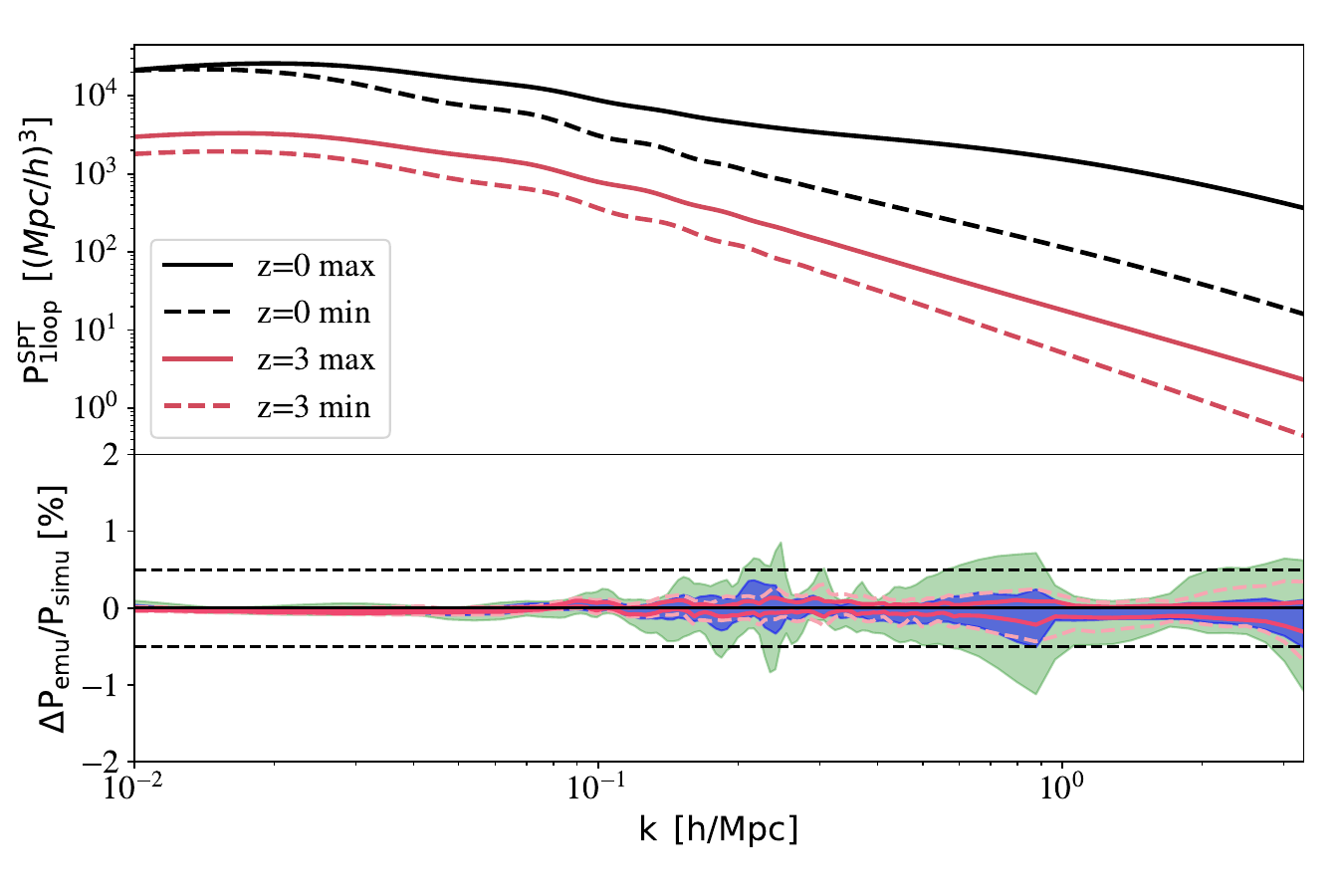}
\includegraphics[width=0.45\linewidth]{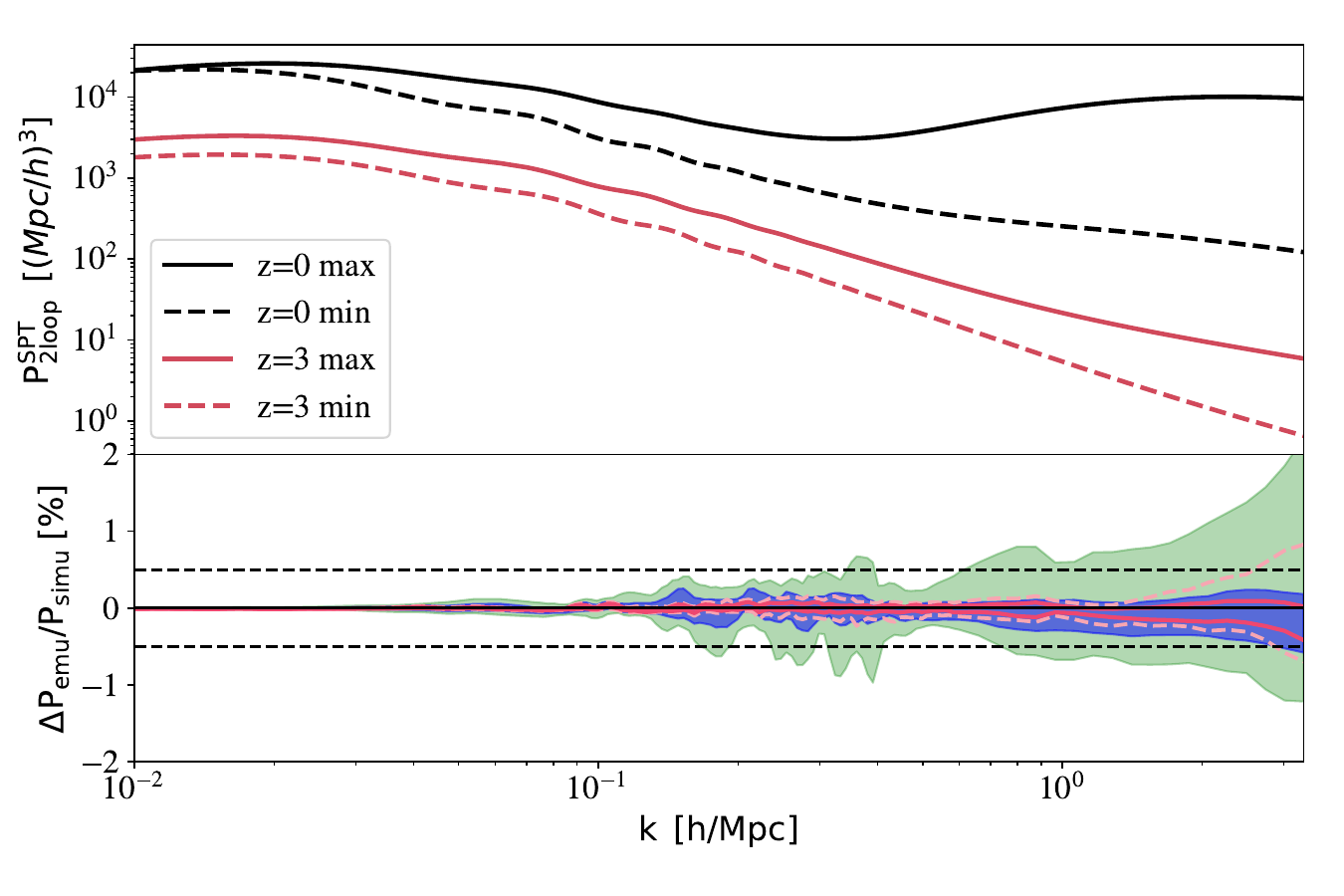}
\caption{\panel{Left:} The top panel shows the  $P^{SPT}_{1-loop}$ function for two extreme cases of cosmological parameters,
while the bottom panel displays the resulting $1\sigma$ and $2\sigma$  emulator $\%$ error.
 \panel{Right:}  The top panel shows the  $P^{SPT}_{2-loop}$ function for two extreme cases of cosmological parameters,
while the bottom panel displays the resulting $1\sigma$ and $2\sigma$  emulator $\%$ error. 
In both panels, we overlay the boundaries of the $1\sigma$(solid, red) and $2\sigma$ (dashed, red) emulation error at $z=3$, while the horizontal dashed lines mark the $0.5\%$ threshold.
}
\label{fig:1loop_and_2loopSPT}
\end{figure*}
Joining of the two regions was done using the error function as in Sec.\ref{1_loop_methodology}  but with different parameters set to $C_1 = 100$ and $k_1 = 0.4$.
The first derivatives of  $\IR{P_{\rm 1-loop}}{1}$ and $\IR{P_{\rm 2-loop}}{0}$ with respect to each cosmological parameter, for two example cosmologies, are shown in Fig.~\ref{derivatives}, demonstrating that they are smooth.

The final $2$-loop functions that we emulate are the IR resummed $\IR{P^{\rm (cs)}_{\rm 1-loop}}{0}$ and $\IR{P^{\rm (quad)}_{\rm 1-loop}}{0}$, shown in panels  `(i)'  and  `(j)' of Fig.~\ref{fig:data}, respectively. These display similar behaviour of $\sim3$ orders of magnitude 
rise on small scales with the former being around three times larger. Both display features on large scales which are not visible in the plot, and one  strategy would have been to split the $k$ interval in at least two regions, as we have done in 
the other cases above. However, we decided not to, given that for $k\lesssim 0.3 \hmpc$ their contribution to the total  $P^{\rm EFT}_{\rm 2-loop }$ is negligible. 
This is seen on the top panel of Fig.\ref{fig:p1loop_cs} and Fig.\ref{fig:p1loop_qd}.
Emulating the entire range and choosing models of length 78 and 72 respectively --see the left panels of Fig.\ref{fig:p1loop_cs} and Fig.\ref{fig:p1loop_qd} for the Pareto fronts-- 
leads to error less than $0.5\%$ within the relevant range.
We have checked that the large error for  $k\lesssim 0.3 \hmpc$ does not translate to large errors in the full $P^{\rm EFT}_{\rm 2-loop }$.

\subsection{Error comparison}
\label{sec:Error_comparison}
Having presented all our emulators, let us now assess the overall error and how the combined emulators fare against other existing codes. We first tested the error by keeping within the SPT part of either 1-loop or 2-loops, that is, setting 
all the counterterms to zero. We show the results in the left  (1-loop) and right (2-loop) panels of Fig.\ref{fig:1loop_and_2loopSPT}. The top part of each panel shows two extreme cases of how the total SPT function looks like.
While both extremes are practically intersecting around $k\sim 0.01\hmpc$, they are widely disparate on small scales upto a factor of $\sim10$ for the $1$-loop and $\sim100$ for $2$-loop spectrum. Yet, despite these large variations, the $\%$-error remains within our 
$0.5\%$ threshold for the whole range of $k$ at $1\sigma$, 
and with few deviations outside $0.5\%$, but still within $\sim 1\%$ at $2\sigma$ for $k\lesssim2\hmpc$. 

At larger $k$, the  $2\sigma$  1-loop error remains within $1\%$, dominated by the $\IR{P_{\rm 1-loop }(k)}{0}$ ,  while at $2$-loops it reaches $\sim 2\%$ at $k\sim 3\hmpc$,
dominated by the $\IR{P_{\rm 2-loop }(k)}{0}$ part.
 However, this is outside the regime of validity of EFT at $z=0$ while 
at high enough redshift when EFT is valid there, it gets suppressed by higher powers of  $D(z)$ and thus we do not expect this to be an issue.
To clarify, the linear $P_{11}(k)$ is proportional to $D^2(z)$, while each successive loop comes with an additional $D^2(z)$ contribution. Since $D(z)$ is normalized to
$1$ at $z=0$  then $D(z)<1$ at higher redshift, meaning that the higher loop contributions to the total power spectrum are suppressed. This in turn suppresses the total effective error
coming from the emulator.
We explicitly show how this process suppresses the effective emulation error at $z=3$ in Fig.~\ref{fig:1loop_and_2loopSPT}. 
A full investigation of the regime of validity of the EFT needs explicit 
specification  of the counterterms and will be done 
elsewhere.

\begin{figure}
\centering
\includegraphics[width=\linewidth]{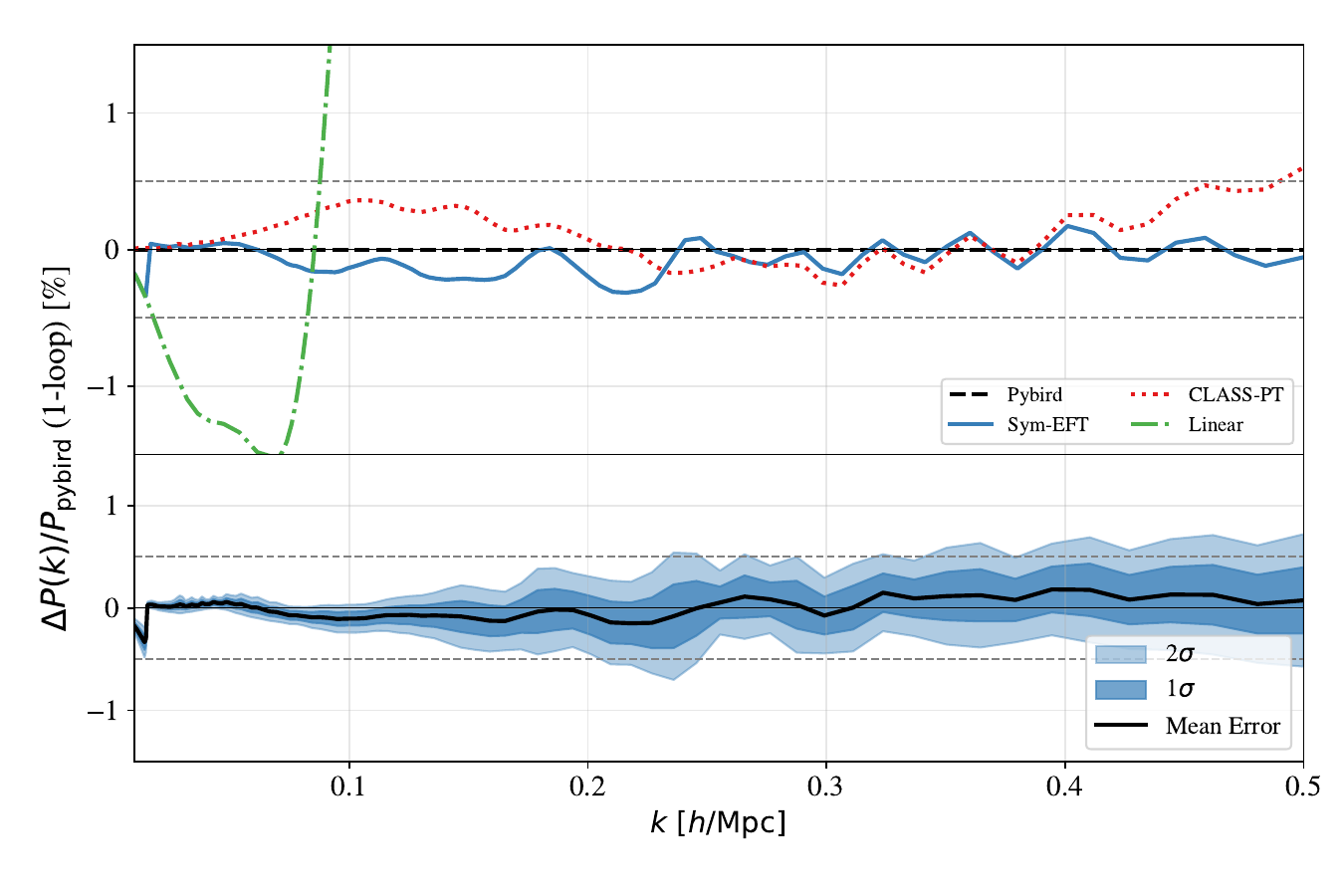}
\caption{\panel{Top:} IR-resummed $P^{\rm SPT}_{\rm 1-loop}(k)$ $\%$-relative difference  of \symeft~(solid blue), \classpt~(dotted red), 
and linear (dashed green) compare respectively 
with \pybird. \panel{Bottom:} IR-resummed $P^{\rm SPT}_{\rm 1-loop}(k)$ mean error (black), 1 $\sigma$ and $2 \sigma$ \% error for 1000 cosmologies, relative to \pybird. }
\label{1loop_pyb-1}
\end{figure}

Let us now compare our \symeft~emulator with other codes for several sets of cosmological parameters within our emulation bounds set in Table \ref{tab:cosmological_parameters}. 
We show the result of comparing one model with \pybird~(baseline - dashed black), \classpt~(dotted red) and \symeft~(solid blue) in the top panel of Fig~\ref{1loop_pyb-1}. 
The linear $P_{11}$ is shown as a guideline (green dot-dashed).  We see that our \symeft~emulator is closer to \pybird~than is \classpt~and 
all three are within $0.5\%$ of each other. Hence, our $0.5\%$ error threshold is within current deviations between codes and thus sufficient.
In the bottom panel of Fig~\ref{1loop_pyb-1}, we show a comparison between \symeft~and \pybird~for 1000 cosmologies sampled from a Latin hypercube within 
 the same bounds as in Table.\ref{tab:cosmological_parameters}. Our comparison shows excellent agreement within $0.5\%$ for
 $1\sigma$, and barely over for $2\sigma$. The dip around the lowest displayed $k$-values in both panels is due to our joining of \symeft~ to the linear spectrum
 for $k<0.01\hmpc$  using an error function. Let us note that these timings do not include the computation of $P_{11}$, for all codes used. To keep the total
computation time small, one can use an emulator for  $P_{11}$, preferably the symbolic emulator of \cite{Bartlett:2023cyr}.

We finally compare the mean running time of these codes to perform the EFT computations. These are shown in Table \ref{tab:timing}. 
CosmoEFT does not have the efficiency of \pybird~ and \classpt, however, \symeft~being based on pure function computations leads to $6000 - 24000$ speedup over the last two.
\texttt{PyBird-JAX}+ Emulator \citep{reeves2025pybirdjaxacceleratedinferencelargescale} is a JAX-compatible version of \pybird~ combined with a neural network emulator that significantly improves computational time.
At two loops we had only \cosmoeft~ to compare with, and the speedup increases to $\sim 2\times 10^6$.

We note that the \cosmoeft~computation was done using OpenMP on eight CPU cores, while \symeft~used only one CPU core, hence,
factoring this in, we have naive speedups around $\sim 10^6$ and $\sim 2\times 10^7$ compared to \cosmoeft~ at $1$- and $2$-loops respectively.
\begin{table}
\centering
\begin{tabular}{|c|c|c|}
\hline
\space &One-loop (sec) & Two-loop (sec) \begin{tabular}{l}
\end{tabular} 
 \\
\hline
\cosmoeft+\resumeft& 15.56 & 965.78  \\
\hline
\classpt& 1.46& -\\
\hline
\pybird& 0.59 & -\\
\hline
\texttt{PyBird-JAX} + Emulator&  $4.9 \times 10^{-4}$ & -\\
\hline
Sym-EFT (C)&   $1.3 \times 10^{-4}$ & $3.5 \times 10^{-4}$ \\
\hline
Sym-EFT (JAX)&   $1.6 \times 10^{-4}$ & $1.8 \times 10^{-4}$ \\
\hline
\end{tabular}
\caption{Running times for one and two loops of EFTofLSS codes. The \cosmoeft~used $8$ cores, while \symeft~only one. 
 The reported running times 
of \pybird~ and \classpt~ correspond to the high accuracy settings.}
\label{tab:timing}
\end{table}

\subsubsection{Redshift dependence}

We tested the accuracy of the redshift dependence of the  2-loop \symeft~emulator as it affects the CMB angular power spectra.
Specifically, the EFTofLSS matter power spectrum affects the CMB anisotropies through the integral of the potential $\Phi(k,z)$
over a redshift-dependent kernel, see \cite{Lewis2006}. Therefore any redshift variation of the EFTofLSS matter power spectrum will be directly translated to  $C_\ell^{\Phi\Phi}$,
and further into the (lensed) temperature $C_\ell^{TT}$, and polarization angular power spectra.
To do this properly it is necessary to make assumptions about the exact redshift dependence of the counterterms.
\begin{figure}
\centering
\includegraphics[width=\linewidth]{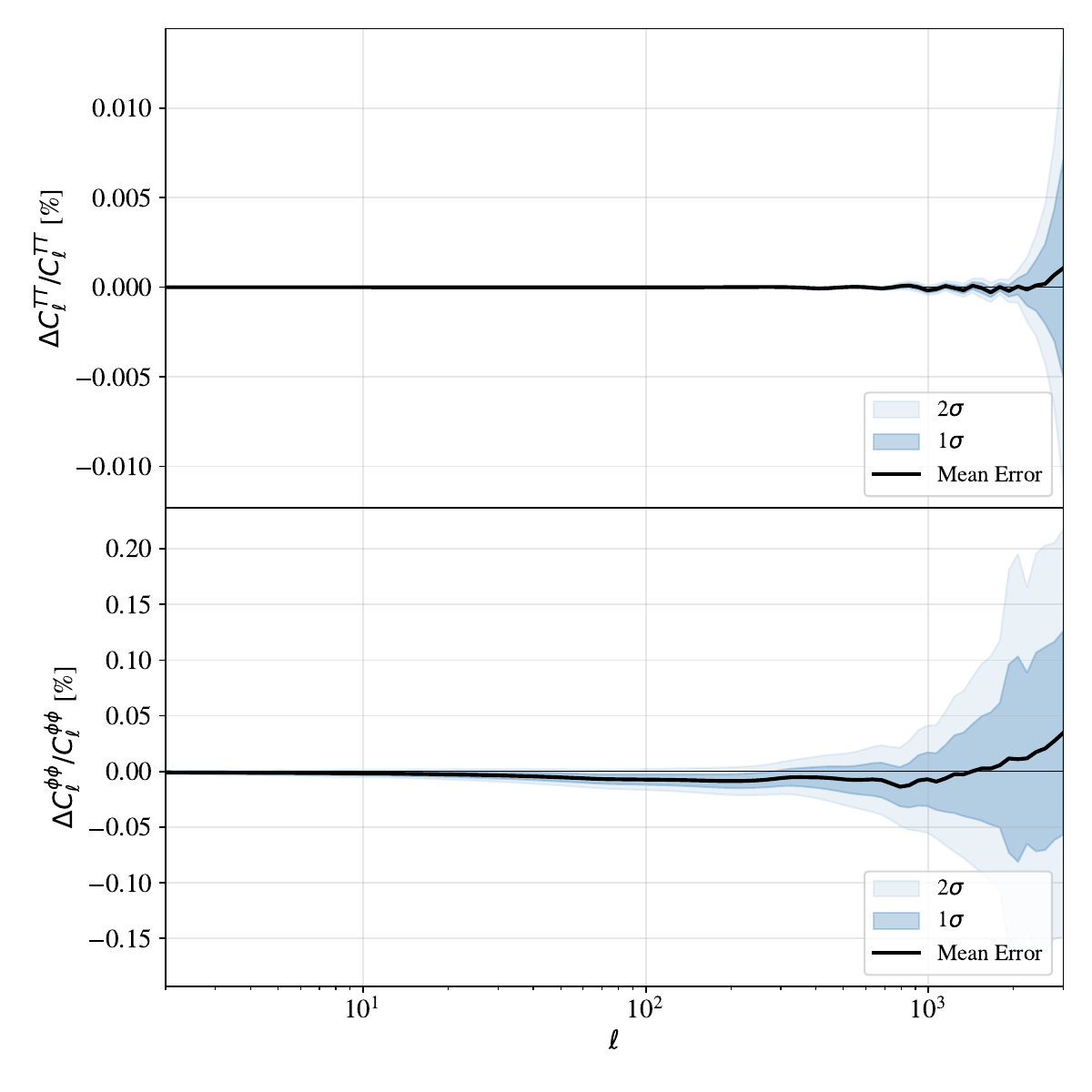}
\caption{\panel{Top:} The \%-relative error of 2-loop  \symeft~contrasted with \resumeft (\classcosmoeft) as they contribute to the
$C_l^{TT}$ lensed temperature angular power spectrum for our 200 training set cosmologies. 
Shown is the mean error (solid blue), $1\sigma$ band (blue shade) and $2\sigma$ \% band (light blue shade). 
 \panel{Bottom:} Same as the \panel{top} panel, but for the $C_l^{\Phi\Phi}$ lensing potential angular power spectrum.
}
\label{lensing-comparison}
\end{figure}
We chose to parametrise the counterterms as the sum of two powerlaws of the growth factor, following the proposal of~\cite{Foreman2015}. Specifically, 
we set
\begin{align}
    c_{s(1)}(z)&=A_{\rm cs}D(z)^{\alpha_s}+B_{\rm cs}D(z)^{\beta_s}~,
\nonumber
\\
    c_{1}(z)&=A_1D(z)^{\alpha_1}+B_1D(z)^{\beta_1}
\label{counter_z}
~,
\\
    c_{4}(z)&=A_4D(z)^{\alpha_4}+B_4D(z)^{\beta_4}\nonumber~,
\end{align}
with $A_{\rm cs}=0.0130$, $B_{\rm cs}=0.0220$, $a_{\rm cs}=8.9958$,  $b_{\rm cs}=-0.3957$, $A_1=-0.1059$,  $B_1=0.7644$, $a_1=0.5513$,  $b_1=1.2506$, $A_4=0.1061$,  $B_4=0.0824$, $a_4=2.5413$,  $b_4=4.7726$.
We have further set $c_{s(2)}=0$, as~\cite{Foreman2015} justified 
that this particular counterterm is very small and letting it vanish is a good approximation. 

We note that our chosen values for the parameters in \eqref{counter_z} are different from those in~\cite{Foreman2015},
 as we fitted the counterterms to the \syren~matter power spectrum and to the resulting $C_\ell^{\Phi\Phi}$ up to $\ell=1500$, taking into account a redshift dependent $k_{\rm fail}$
~\citep{Calderon_in_preparation}. In Fig.\ref{lensing-comparison} we show the error between \classcosmoeft~and \symeft~for the 200 cosmologies in our training set,
keeping the counterterms fixed as in \eqref{counter_z}. We find excellent agreement within $0.01\%$ at $2\sigma$ for $C_\ell^{TT}$ (top panel) 
up to $\ell = 2500$, and within $0.05\%$ at $2\sigma$ for $C_\ell^{\Phi\Phi}$ (bottom panel) up to $\ell = 1000$.

\section{Discussion and conclusion}
\label{sec:Conclusion}
We have presented an emulator suite for the EFTofLSS dark matter one- and two-loop power spectrum using the technique of symbolic regression.
The emulator suite consists of ten emulators for the different terms that appear in the EFTofLSS power spectra, see \eqref{1loop_EFT} and \eqref{2loop_EFT},
Three of these emulators were split in smaller overlaping regions in $k$ space, see Section \ref{1_loop_methodology} and \ref{2_loop_methodology},
as well as Table.\ref{tab:emulated_functions}.
Our emulators are pure fitting functions which depend on $k$ and the cosmological parameters, that can be inserted into any computer code, leading to $\sim 10^{-4} sec$ per model
computation time; see \eqref{equation_k2P11_R0}, \eqref{equation_k2P11_R1} and appendices \ref{sec:EmulatorSuite_1_loop} and \ref{sec:EmulatorSuite_2_loop}.
This method provides the fastest possible emulation technique. We have chosen as simple expressions as allowed to keep within a threshold of $\sim 0.5\%$ which in the case of $1$-loop
is comparable with the current differences between EFTofLSS codes. 

Apart from their use in ultra-fast MCMC based testing of $\Lambda$CDM with data in the mildly non-linear regime, one direct application is to use them in addition
with CMB data. It has been shown that the CMB lensing is sensitive to larger scales than the galaxy lensing surveys, 
and has a wider redshift range \cite{doux2025goings8fastinference}. 
Thus, we expect to see the complementary effect of two-loop EFTofLSS in the CMB lensing at different  redshift and $k$ ranges than 
cosmic shear.

While we targeted an emulation error of $\sim 0.5\%$, we are confident that our symbolic regression technique can reach better accuracy sacrificing neither model length nor
computational time. Given that current one-loop codes exhibit differences comparable with this error threshold, see Fig.\ref{1loop_pyb-1}, it seems that the current bottleneck is the accuracy
of single un-emulated models. To elaborate, it is possible that part of the `signal' that Operon is trying to fit contains some numerical noise 
 which might also, in principle, be correlated with the cosmological parameters and if absent could allow better expressions at fixed complexity. 
 It would thus be interesting to have a global comparison between all codes in order to reach sub-$0.1\%$ accuracy.

\begin{figure}
\centering
\includegraphics[width=0.48\textwidth]{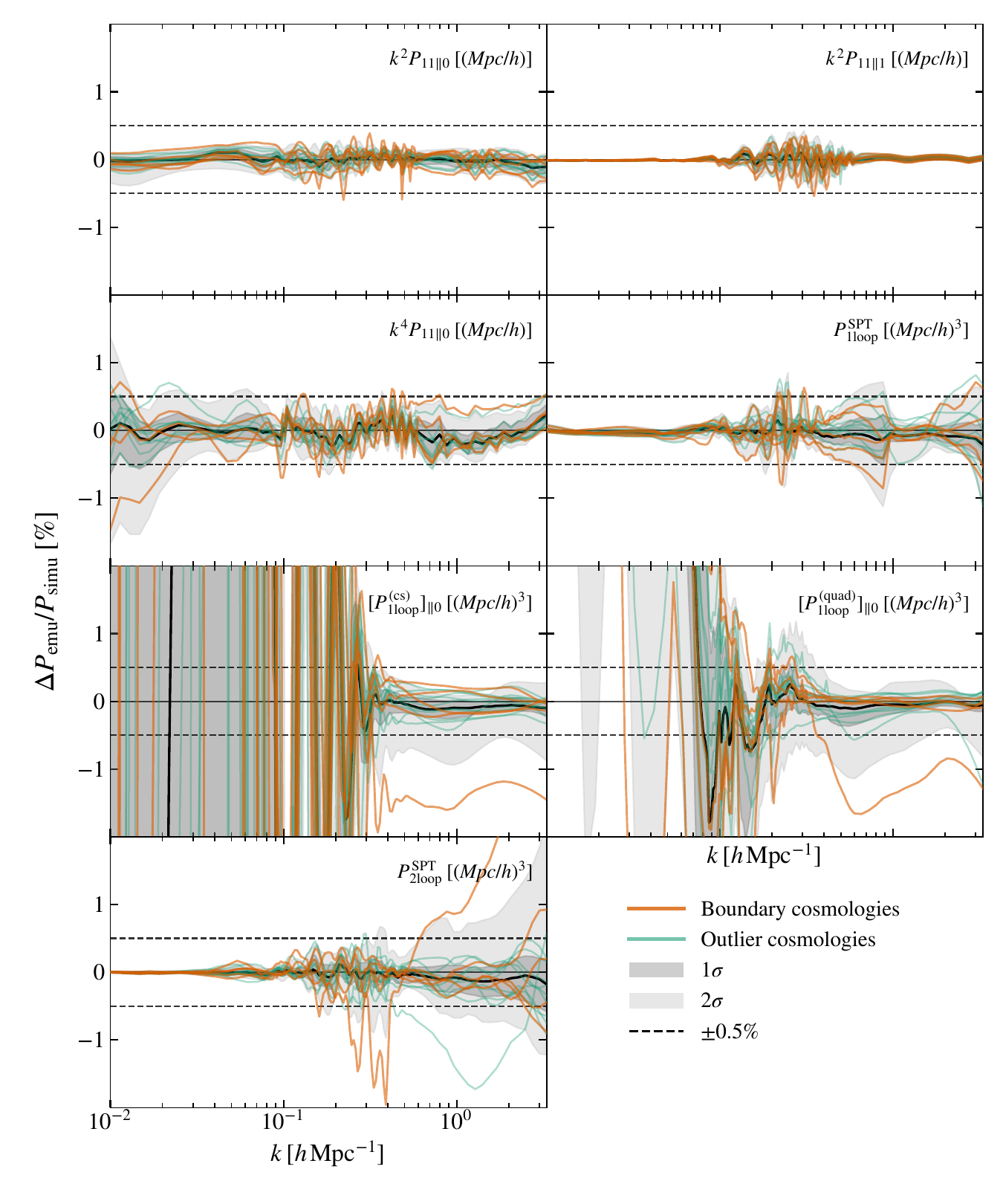}
\caption{The same error plots as in the lower panels of Figs.~\ref{Pareto_and_error:k2p11_R0},~\ref{k2p11_R1_combined},~\ref{fig:1loop_and_2loopSPT},~\ref{Pareto_and_error:k4p11_R0}, ~\ref{fig:p1loop_cs} and~\ref{fig:p1loop_qd}, here superimposed with errors from individual models from Table.\ref{tab:outliers}. Boundary models are in orange while outlier models are in blue. The one boundary model and one outlier model which 
have errors outside the $2\sigma$ band in the last three plots have $\Omega_m = 0.24$ and $\Omega_m = 0.232$ respectively (with corresponding $\Omega_b=0.06$ and $\Omega_b = 0.05$).}
\label{fig:outliers}
\end{figure}

An important question is that of generalisation, that is, whether these emulators can be used outside their training range, or, in $\Lambda$CDM extensions.  
We have checked the first case by considering a few models which have some of their parameters on the boundary of Table~\ref{tab:cosmological_parameters}
and also a few models with one parameter outside  the training range (outliers). The collected models are tabulated in Table~\ref{tab:outliers}. 
We find some degree of generalisation as long as the models are not too far ouside the boundary, although the full extend of how far has not been investigated.
Specifically, all outliers considered lie within the $2\sigma$ error band (sometimes within $1\sigma$) of the $\IR{k^2 P_{11}}{0}$,  $\IR{k^2 P_{11}}{1}$,  $\IR{k^4 P_{11}}{0}$ and $P^{SPT}_{1-loop}$ spectra, however, we found two models lying outside the $2\sigma$ band
in the   $\IR{P^{\rm (cs)}_{\rm 1-loop}}{0}$,  $\IR{P^{\rm (quad)}_{\rm 1-loop}}{0}$ and $P^{SPT}_{2-loop}$  cases. These two models have low $\Omega_m$ (on or outside the lower boundary) and high $\Omega_b$, meaning that the BAO are more pronounced.
 The results of this investigation are shown in Fig.~\ref{fig:outliers}.

To address the second case, we condidered two commonly used $\Lambda$CDM extensions: extending  $\Lambda$CDM with
dynamical dark energy, for instance, with the Chevallier-Polarski-Linder (CPL) parameterization, and extending  $\Lambda$CDM with neutrino mass.   
 While on sub-horizon scales CPL only introduces scale independent growth suggesting that this prescription should work,  super-horizon  scales introduce some small $k$ dependence in the full $P_{11}(k)$ which can affect the loop integrals. 
In the case of neutrino mass, letting $m_\nu$ to vary leads to scale-dependent growth throughout the entire $k$-range so that the loop integrals are expected to vary from the outset.
However, the common practice is to use the same underlying EFT with such extensions, modifying only the growth factors entering all expressions, but keeping the loop integrals intact. This is an approximation which only recently has started to be investigated more extensively, see
\cite{Donath:2020abv,Zhang:2021uyp,Hartmeier:2023brx,Fidler:2026wqg}, with positive results.
Nevertheless, emulating over approximated spectra may lead to additional biases, therefore, we suggest caution in using our emulators beyond $\Lambda$CDM and leave the full generalisation question for future work.

While this work was in its final stages, there have been two other articles on accelerating the 
two-loop EFTofLSS matter power spectrum,  \citep{bakx2025rapidcosmologicalinferencetwoloop} and \citep{Anastasiou:2025jsy}. These two articles use different techniques
and would be interesting to compare their output in a similar way as we have discussed above.

\section*{Acknowledgements}
We thank D. Bartlett for  insights in symbolic regression,  P. Ferreira and D. Vokrouhlicky for general discussions, G. Kronberger and B. Burlacu for providing \operon~ and insights on its
workings, L. Senatore for discussions on EFTofLSS and for sharing the \cosmoeft~ and \resumeft~ codes with us.
We particularly thank P. Zhang for discussions concerning \pybird.

The research leading to these results has received support from the European Structural and Investment Funds and the Czech Ministry of Education, Youth and Sports (project No. FORTE---CZ.02.01.01/00/22\_008/0004632). 
 CS acknowledges support from the Royal Society Wolfson Visiting Fellowship ``Testing the properties of dark matter with new statistical tools and cosmological data''. 
 DF thanks the Beecroft Institute for Particle Astrophysics and Cosmology, University of Oxford, for hospitality during which some of this work was completed.

\section*{Data Availability}
The raw power spectra of the cosmologies used in training and validating our emulators can be found on Zenodo; DOI:\url{https://doi.org/10.5281/zenodo.15649834}. 
The SymEFT emulator suite, including all emulator functions in C, JAX and Python can be accessed from \url{https://github.com/despf/symEFT.git}.

\begin{table}
\centering
%
\begin{tabular}{|c|c|c|c|c|}
\hline
\rule{0pt}{2.4ex}
$\Ast \equiv 10^{9}A_s$ &
$\Omega_m$ &
$\Omega_b$ &
$h$ &
$n_s$ \\
\hline
\hline
\boundary   1.70 & \boundary  0.400 & 0.050 & 0.670 & 0.960 \\
\hline
2.10 & \boundary 0.240 & \boundary 0.060 & 0.670 & 0.960 \\
\hline
2.10 & 0.320 & \boundary 0.040 & \boundary 0.730 & 0.960 \\
\hline
2.10 & 0.320 & 0.050 & \boundary 0.610 & \boundary 1.000 \\
\hline
\boundary  2.50 & 0.320 & 0.050 & 0.670 & \boundary 0.920 \\
\hline
\outlier  1.66 & 0.320 & 0.050 & 0.670 & 0.960 \\
\hline
\outlier 2.58 & 0.320 & 0.050 & 0.670 & 0.960 \\
\hline
2.10 & \outlier 0.232 & 0.050 & 0.670 & 0.960 \\
\hline
2.10 & \outlier 0.416 & 0.050 & 0.670 & 0.960 \\
\hline
2.10 & 0.320 & \outlier 0.039 & 0.670 & 0.960 \\
\hline
2.10 & 0.320 & \outlier 0.062 & 0.670 & 0.960 \\
\hline
2.10 & 0.320 & 0.050 & \outlier 0.604 & 0.960 \\
\hline
2.10 & 0.320 & 0.050 & \outlier 0.742 & 0.960 \\
\hline
2.10 & 0.320 & 0.050 & 0.670 & \outlier 0.916 \\
\hline
2.10 & 0.320 & 0.050 & 0.670 & \outlier 1.008 \\
\hline
\end{tabular}
\caption{The models used for testing generalisation. Boundary values are shaded in yellow, while outlier values are in blue. }
\label{tab:outliers}
\end{table}
%

\bibliographystyle{mnras}
\bibliography{Sym-EFT}


\appendix

\onecolumn

%
%

\section{The $1$-loop emulators}
\label{sec:EmulatorSuite_1_loop}
At $1$-loop the emulators concern the functions 
 $\IR{k^2P_{11}(k)}{0}$, $\IR{P_{11}(k)}{1}$ and $\IR{P_{\rm 1-loop }(k)}{0}$. The $\IR{k^2P_{11}(k)}{0}$ has been presented in the main part of the article,
see \eqref{equation_k2P11_R0} and \eqref{constants_k2P11_R0} as well as  Fig.\ref{Pareto_and_error:k2p11_R0} for the Pareto front and error plot respectively.
Here we present the remaining two emulators.

\subsection{The $\IR{P_{11}}{1}$ emulator}
\label{sec:EmulatorSuite_P11_R1}
\begin{figure}
\centering
  \includegraphics[width=0.48\linewidth,height=\appfigheight,keepaspectratio]{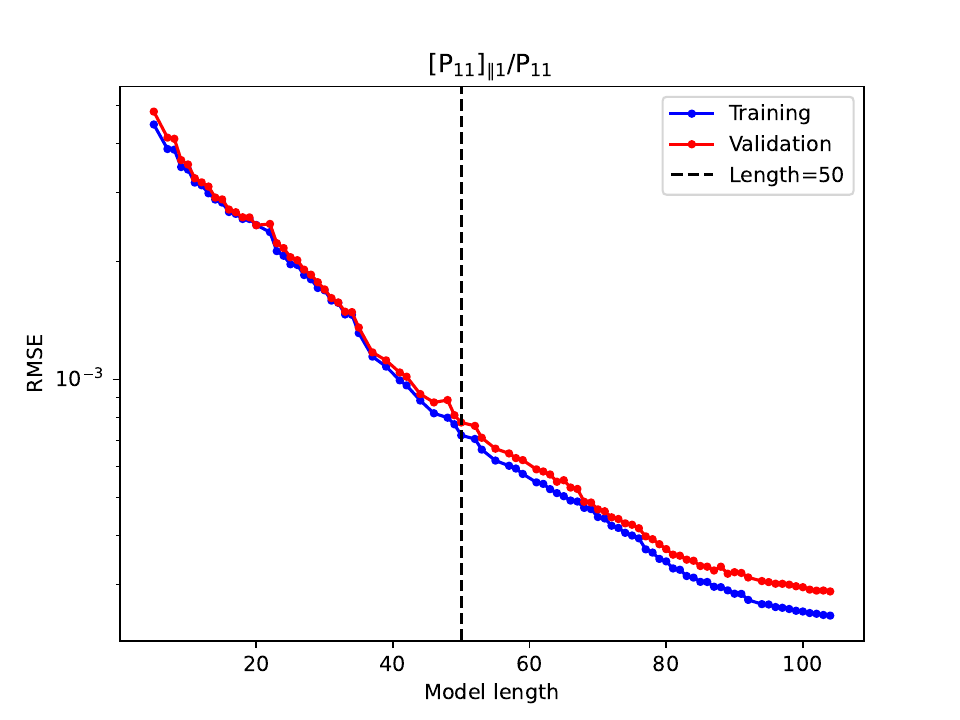}
   \includegraphics[width=0.48\linewidth,height=\appfigheight,keepaspectratio]{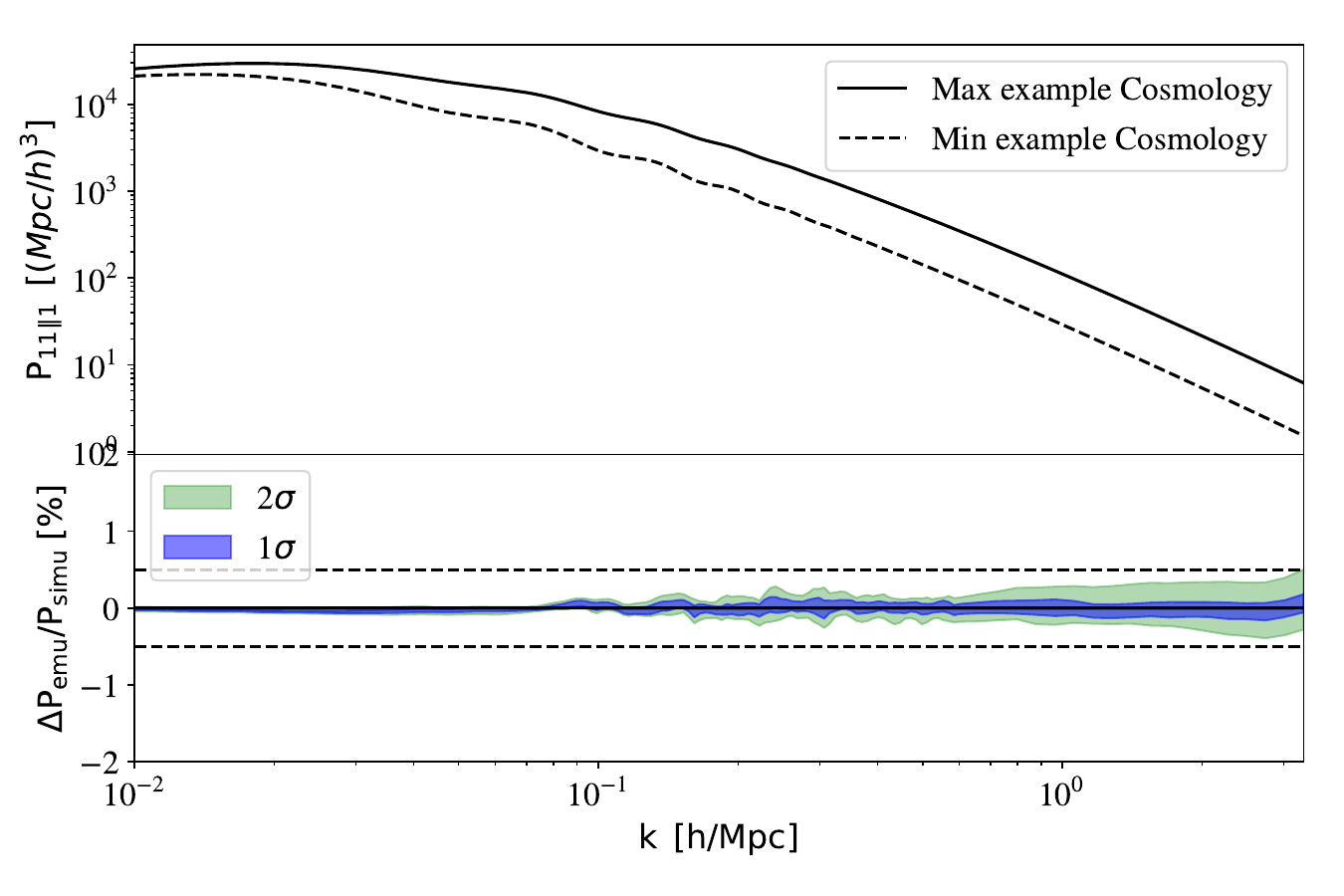}
\caption{ \panel{Left:} The Pareto front of RMSE vs model length for the $\frac{\IR{ P_{11}}{1}}{ P_{11} }$ emulator runs as generated by \operon, with blue marking the training and red the validation error, 
and with the chosen model of length 50 indicated by the vertical line. \panel{Right:} The 
top plot shows the  $\IR{P_{11}}{1}$ function for two extreme cases of cosmological parameters
while the bottom plot displays the resulting $1\sigma$ and $2\sigma$  emulator $\%$ error for all 300 cosmologies. The horizontal dashed lines mark the $0.5\%$ threshold.}
\label{p11_R1_combined}
\end{figure}

For the $\IR{ P_{11}}{1}$ emulator we chose a model of length 50.
We show the Pareto front of RMSE vs model length on the left of Fig.\ref{p11_R1_combined}.
 The form of the emulated function is 
\begin{align} 
\frac{ \IR{ P_{11}}{1}}{ P_{11} }   =& C_{21} 
+ k \left( C_{0} \Ast - C_{1} k\right) \left( C_{2} \Ast \left(C_{3} h\right)^{C_{4} \Omega_{cb}} - C_{5} \Omega_{b}\right) \cos{\left(C_{6} \Omega_{cb} + \frac{C_{9}}{\sqrt{C_{10} k^{2} + 1}} + \left(C_{7} h\right)^{C_{8} \Omega_{b}} \right)} 
\nonumber
\\
&
+ \frac{\left( C_{11} \Ast - \cos{\left(C_{12} k \right)}\right) \left( C_{13} \Omega_{cb} - C_{14} \Omega_{b}\right) \cos{\left(C_{15} k \left(C_{16} h\right)^{C_{17} \Omega_{cb} + C_{18} \Omega_{b}} \right)}}{\sqrt{\left( C_{19} \Omega_{b} - C_{20} k\right)^{2} + 1}}
\label{equation_P11_R1}
\end{align} 
where
\begin{align} 
C_{0} =& 0.859 & \sspace C_{1} =& 0.2675 & \sspace C_{2} =& 4.03 \times 10^{-3} & \sspace C_{3} =& 2.096 & \sspace C_{4} =& 5.96 & \sspace C_{5} =& 0.192 \nonumber \\
C_{6} =& 5.3 & \sspace C_{7} =& 1.2624 & \sspace C_{8} =& 145.7  & \sspace C_{9} =& 6.81 & \sspace C_{10} =& 121.6  &\sspace C_{11} =& 0.42 \nonumber \\
C_{12} =& 9.73 & \sspace C_{13} =& 0.0117 & \sspace C_{14} =& 0.1624 & \sspace C_{15} =& 138.55  & \sspace C_{16} =& 0.26537 & \sspace C_{17} =& 0.46944 \nonumber \\
C_{18} =& 1.487 & \sspace C_{19} =& 47.8 & \sspace C_{20} =& 11.62 & \sspace C_{21} =& 0.999883.
\label{constants_P11_R1}
\end{align} 
The resulting error is displayed on the right of Fig.\ref{p11_R1_combined}.

\subsection{The $\IR{P_{\rm 1-loop}}{0}$ emulator}
\label{sec:EmulatorSuite_P1loop_R0}
The IR resummed  $\IR{P_{\rm 1-loop}}{0}$  emulator is split into three regions in $k$ space as described in Section \ref{1_loop_methodology}, 
while all the Pareto fronts can be found in Fig.\ref{fig:combined-pareto-1loop}

\subsubsection{Region 1: $k = [0.01,0.3]\times \hmpc$}
\begin{figure*}
    \centering
        \includegraphics[width=0.33\textwidth]{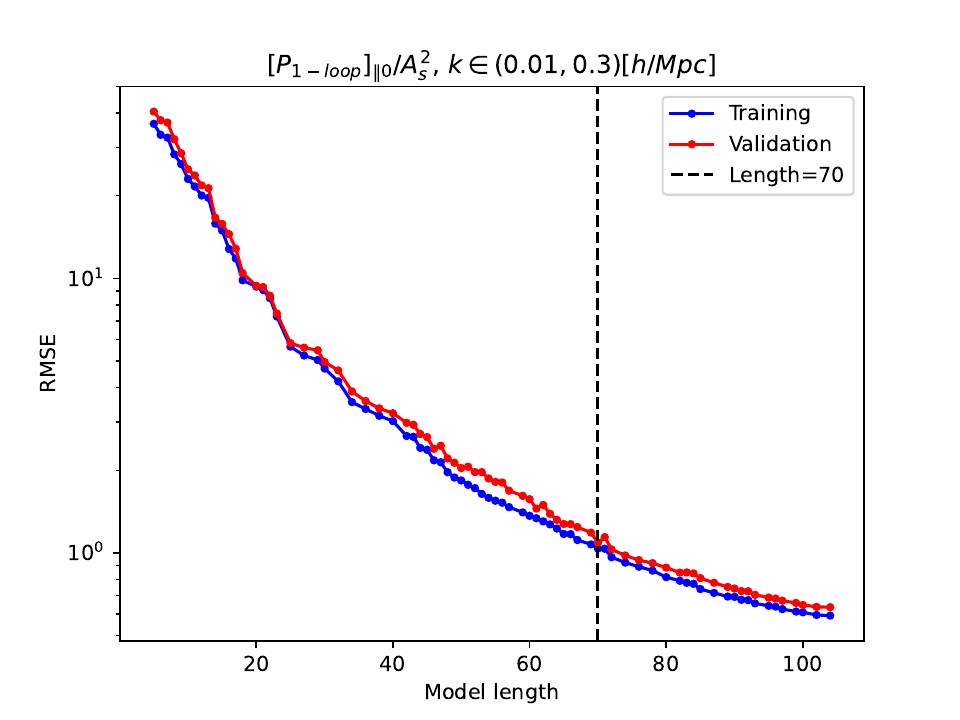}
        \includegraphics[width=0.33\textwidth]{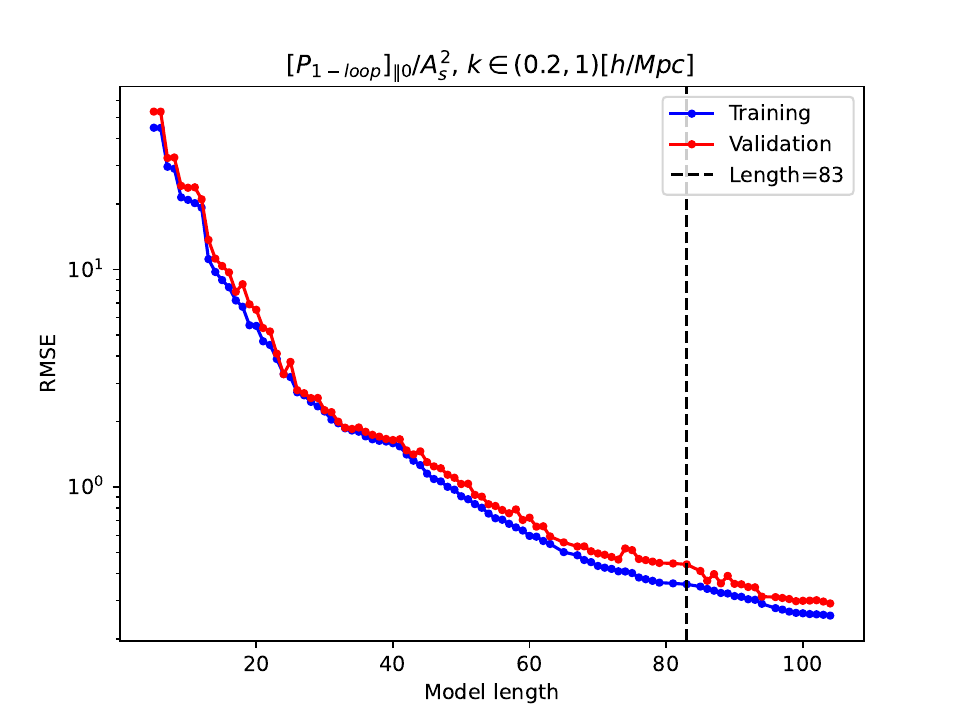}
        \includegraphics[width=0.33\textwidth]{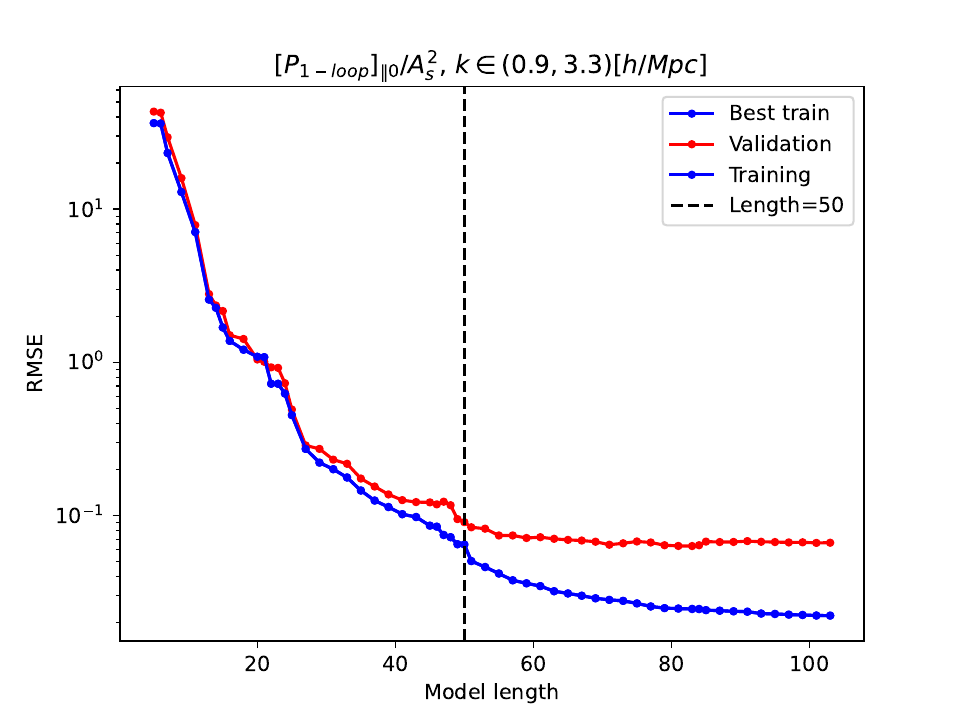}
    \caption{The Pareto fronts of RMSE vs model length for the three emulators of $\frac{\IR{ P_{{\rm 1-loop}}}{0}}{ \Ast^2 }$ runs as generated by \operon, with blue marking the training and red the validation error, and with the chosen models indicated by the vertical line. 
 }
    \label{fig:combined-pareto-1loop}
\end{figure*}

For the first region  of the $\IR{P_{\rm 1-loop}}{0}$  emulator we chose a model of length 70, see left panel of Fig.\ref{fig:combined-pareto-1loop}.
The resulting function is
\begin{align} 
\frac{ \IR{ P_{\rm 1-loop} }{0}^{(1)} }{ \Ast^2 }  =& - C_{0} h \left(C_{1} h\right)^{C_{2} \Omega_{b}} \cos{\left(\frac{C_{3} h + C_{4} k}{\sqrt{\left(C_{5} \Omega_{cb} + C_{6} \Omega_{b}\right)^{2} + 1}} \right)} \cos{\left(C_{7} \Omega_{cb} + \frac{C_{8}}{\sqrt{C_{9} k^{2} + 1}} \right)} -  C_{30}
\nonumber
\\
&
 - \frac{n_s \left(C_{10} \Omega_{cb} - C_{11} \Omega_{b}\right) \ln{\left(C_{12} k \right)}^{C_{13} h + \frac{C_{14} \Omega_{cb} - C_{15} \Omega_{b}}{\sqrt{\left(C_{16} n_s\right)^{- C_{17} \Omega_{cb}} + 1}}} \cos{\left(\frac{C_{18} \Omega_{cb} \cos{\left(C_{19} n_s \right)}}{\sqrt{C_{20} k^{2} + 1}} - C_{21} h - \frac{C_{22} n_s}{\sqrt{\frac{\left(C_{23} \Omega_{cb} + C_{24} k\right)^{2}}{\left(C_{25} \Omega_{cb} - C_{26} \Omega_{b}\right)^{2} + 1} + 1}} \right)}}{\sqrt{\left(C_{27} h\right)^{- C_{28} n_s} + 1} \sqrt{C_{29} \Omega_{cb}^{2} + 1}}
\label{equation_P1loop_R0_1}
\end{align} 
where
\begin{align} 
C_{0} =& 12.9 & \sspace C_{1} =& 2.355 & \sspace C_{2} =& 28 & \sspace C_{3} =& 3.89 & \sspace C_{4} =& 122.62  & \sspace C_{5} =& 1.5224 \nonumber \\
C_{6} =& 4.764 & \sspace C_{7} =& 1.656 & \sspace C_{8} =& 4.363 & \sspace C_{9} =& 188.7  & \sspace C_{10} =& 859.4  & \sspace C_{11} =& 1714  \nonumber \\
C_{12} =& 203.3  & \sspace C_{13} =& 0.9403 & \sspace C_{14} =& 7.62 & \sspace C_{15} =& 15.26 & \sspace C_{16} =& 0.555 & \sspace C_{17} =& 4.973 \nonumber \\
C_{18} =& 3.7 & \sspace C_{19} =& 4.742 & \sspace C_{20} =& 4200  & \sspace C_{21} =& 2.028 & \sspace C_{22} =& 50.37 & \sspace C_{23} =& 202.44  \nonumber \\
C_{24} =& 574.8 & \sspace C_{25} =& 27.634 & \sspace C_{26} =& 23.39 & \sspace C_{27} =& 0.84518 & \sspace C_{28} =& 6.498 & \sspace C_{29} =& 13.745 \nonumber \\
C_{30} =& 0.01 
\label{constants_P1loop_R0_1}
\end{align}

\subsubsection{Region 2: $k = [0.2,1]\times \hmpc$}
For the second region of the $\IR{P_{\rm 1-loop}}{0}$  emulator we chose a model of length 83, see the middle panel of Fig.\ref{fig:combined-pareto-1loop}.

\begin{align} 
\frac{ \IR{ P_{\rm 1-loop}}{0}^{(2)} }{ \Ast^2 }  =& 
\Fcal^{- C_{8} \Omega_{cb} + \ln{\left(- C_{10} \Omega_{b} + C_{11} n_s + C_{9} \Omega_{cb} + \left(C_{12} \Omega_{b}\right)^{- C_{13} h} \right)}
 - \frac{- C_{14} h + \frac{C_{15} k}{\sqrt{\left(- C_{16} \Omega_{b} + C_{17} k\right)^{2} + 1}} + \frac{C_{18} k}{\sqrt{C_{19} k^{2} + 1}}}{\sqrt{C_{20} n_s^{2} + 1}}} 
\nonumber
\\
&
\times 
\Bigg[C_{21} \Omega_{cb} + C_{22} h - C_{23} n_s + k \left(C_{24} - \frac{C_{25}}{\Omega_{b}}\right) + \left(C_{26} k\right)^{\frac{C_{27} \Omega_{b}}{\sqrt{\cos^{2}{\left(C_{28} \Omega_{cb} \right)} + 1}} - C_{29} k}
 \left(- C_{30} \Ast + C_{31} h + C_{32} n_s\right) 
\nonumber
\\
&
\times 
\cos{\left(C_{33} \Omega_{cb} + C_{34} \Omega_{b} - C_{35} h - C_{36} k \right)}\Bigg]
+ C_{37}
\label{equation_P1loop_R0_2}
\end{align} 
where
\begin{align} 
\Fcal =& - C_{0} \Ast + C_{1} \Omega_{cb} + C_{2} h + C_{3} n_s \left( C_{4} \Omega_{b} - C_{5} h + C_{6} k + \cos{\left(C_{7} \Omega_{cb} \right)}\right)
\end{align} 
and where
\begin{align} 
C_{0} =& 0.75 & \sspace C_{1} =& 91.73 & \sspace C_{2} =& 76.27 & \sspace C_{3} =& 57.16 & \sspace C_{4} =& 2.888 & \sspace C_{5} =& 2.6705 \nonumber \\
C_{6} =& 9.692 & \sspace C_{7} =& 5.12905 & \sspace C_{8} =& 1.1053 & \sspace C_{9} =& 44.189 & \sspace C_{10} =& 55.23 & \sspace C_{11} =& 1.3032 \nonumber \\
C_{12} =& 4.838 & \sspace C_{13} =& 0.5053 & \sspace C_{14} =& 1.00064 & \sspace C_{15} =& 3.2949 & \sspace C_{16} =& 0.97 & \sspace C_{17} =& 3.2 \nonumber \\
C_{18} =& 1.016 & \sspace C_{19} =& 0.515 & \sspace C_{20} =& 1.5425 & \sspace C_{21} =& 2.72 \times 10^{-3} & \sspace C_{22} =& 1.8803\times 10^{-2} & \sspace C_{23} =& 8.347 \times 10^{-3} \nonumber \\
C_{24} =& 2.42 \times 10^{-4} & \sspace C_{25} =& 2.67 \times 10^{-5} & \sspace C_{26} =& 54.5 & \sspace C_{27} =& 22.5 & \sspace C_{28} =& 8.5 & \sspace C_{29} =& 6.8 \nonumber \\
C_{30} =& 5.08 \times 10^{-4} & \sspace C_{31} =& 1.9795 \times 10^{-3} & \sspace C_{32} =& 8.1 \times 10^{-4} & \sspace C_{33} =& 17.18 & \sspace C_{34} =& 53.38 & \sspace C_{35} =& 7.601 \nonumber \\
C_{36} =& 93.4813 & \sspace C_{37} =& 8 \times 10^{-4} 
\label{constants_P1loop_R0_2}
\end{align}

\subsubsection{Region 3: $k = [0.9,3.3]\times \hmpc$}
For the third region  of the $\IR{P_{\rm 1-loop}}{0}$  emulator we chose a model of length 50, see the right panel of Fig.\ref{fig:combined-pareto-1loop}.
The resulting function  is
\begin{align} 
\frac{\IR{ P_{\rm 1-loop}}{0}^{(3)}}{ \Ast^2}  =& 
\left(C_{0} k\right)^{C_{1} \Ast + C_{2} n_s - C_{3}} \left(C_{4} h\right)^{- C_{5} \Omega_{b} + C_{6} n_s + C_{7}} \left(C_{9} + C_{8} \Omega_{cb}\right)^{\frac{C_{11} \Omega_{b} + C_{12} h}{\sqrt{C_{13} \Omega_{cb}^{2} + 1}} + \ln{\left(C_{10} k \right)}} 
\nonumber
\\
&
\times
\left( - C_{14} k
 + \ln{\left(C_{15} h \right)} + \frac{C_{16} \Omega_{cb} + C_{17} \Omega_{b} -  \frac{1}{\left(C_{18} k\right)^{C_{19} \Omega_{cb} \ln{\left(C_{20} h \right)}}}}{\sqrt{C_{21} k^{2} + 1}} 
\right) -  C_{22}
\label{equation_P1loop_R0_3}
\end{align} 
where
\begin{align} 
C_{0} =& 1.1086 & \sspace C_{1} =& 0.0154 & \sspace C_{2} =& 1.9261 & \sspace C_{3} =& 1.21245 & \sspace C_{4} =& 7.07172  &\sspace  C_{5} =& 6.339  \nonumber \\
 C_{6} =& 1.91047 & \sspace C_{7} =& 3.85764 & \sspace C_{8} =& 0.3437 & \sspace C_{9} =& 0.041539 & \sspace C_{10} =& 3.216055 & \sspace C_{11} =& 57.033  \nonumber \\
C_{12} =& 1.668 & \sspace C_{13} =& 171.47  & \sspace C_{14} =& 0.043 & \sspace C_{15} =& 2.8331 & \sspace C_{16} =& 4.8072 & \sspace  C_{17} =& 8.635 \nonumber \\
 C_{18} =& 0.3023 & \sspace C_{19} =& 1.2287 & \sspace C_{20} =& 3.4138 & \sspace C_{21} =& 0.4648 & \sspace C_{22} =& 8 \times 10^{-4} 
\label{constants_P1loop_R0_3}
\end{align}

%
%

\section{The $2$-loop emulators}
\label{sec:EmulatorSuite_2_loop}
At $2$-loop, the emulators concern the functions 
 $\IR{P_{11}(k)}{2}$, $\IR{k^2 P_{11}(k)}{1}$, $\IR{k^4 P_{11}(k)}{0}$,  $\IR{P_{\rm 1-loop }(k)}{1}$, $\IR{P_{\rm 2-loop }(k)}{0}$, 
 $\IR{P^{\rm (cs)}_{\rm 1-loop}}{0}$ and $\IR{P^{\rm (quad)}_{\rm 1-loop}}{0}$.
The $\IR{k^2P_{11}(k)}{1}$ has been presented in the main part of the article,
see \eqref{equation_k2P11_R0} and \eqref{constants_k2P11_R1} as well as  Fig.\ref{k2p11_R1_combined} for the Pareto front and error plot.
Here we present the remaining emulators.

\subsection{The $ \IR{P_{11}}{2}$ emulator}
\label{sec:EmulatorSuite_P11_R2}

\begin{figure*}
  \includegraphics[width=0.48\linewidth,height=\appfigheight,keepaspectratio]{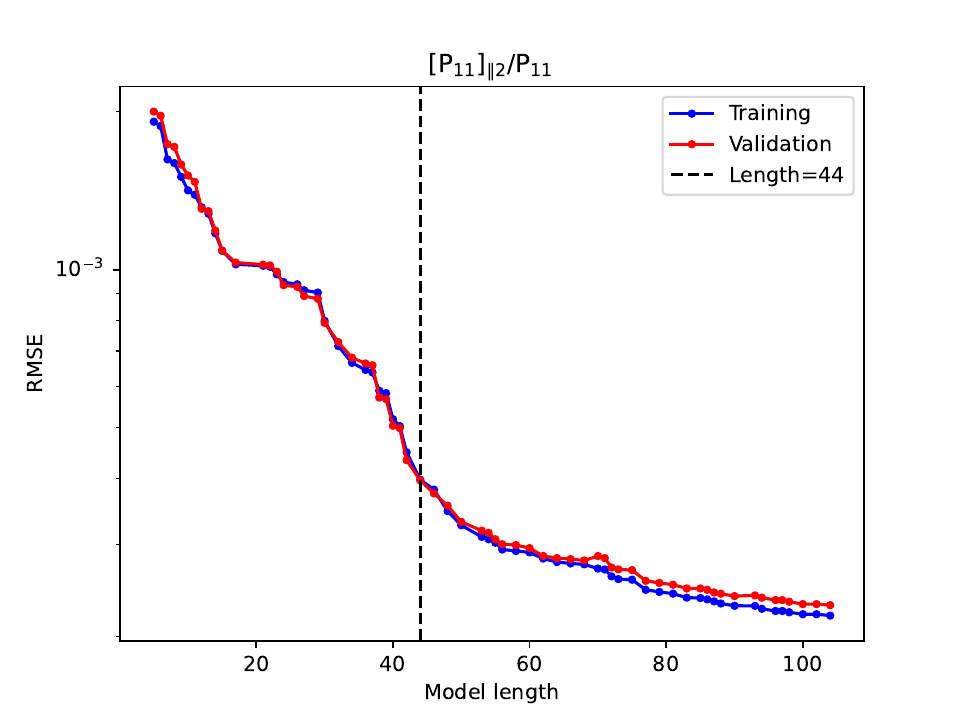}
   \includegraphics[width=0.48\linewidth,height=\appfigheight,keepaspectratio]{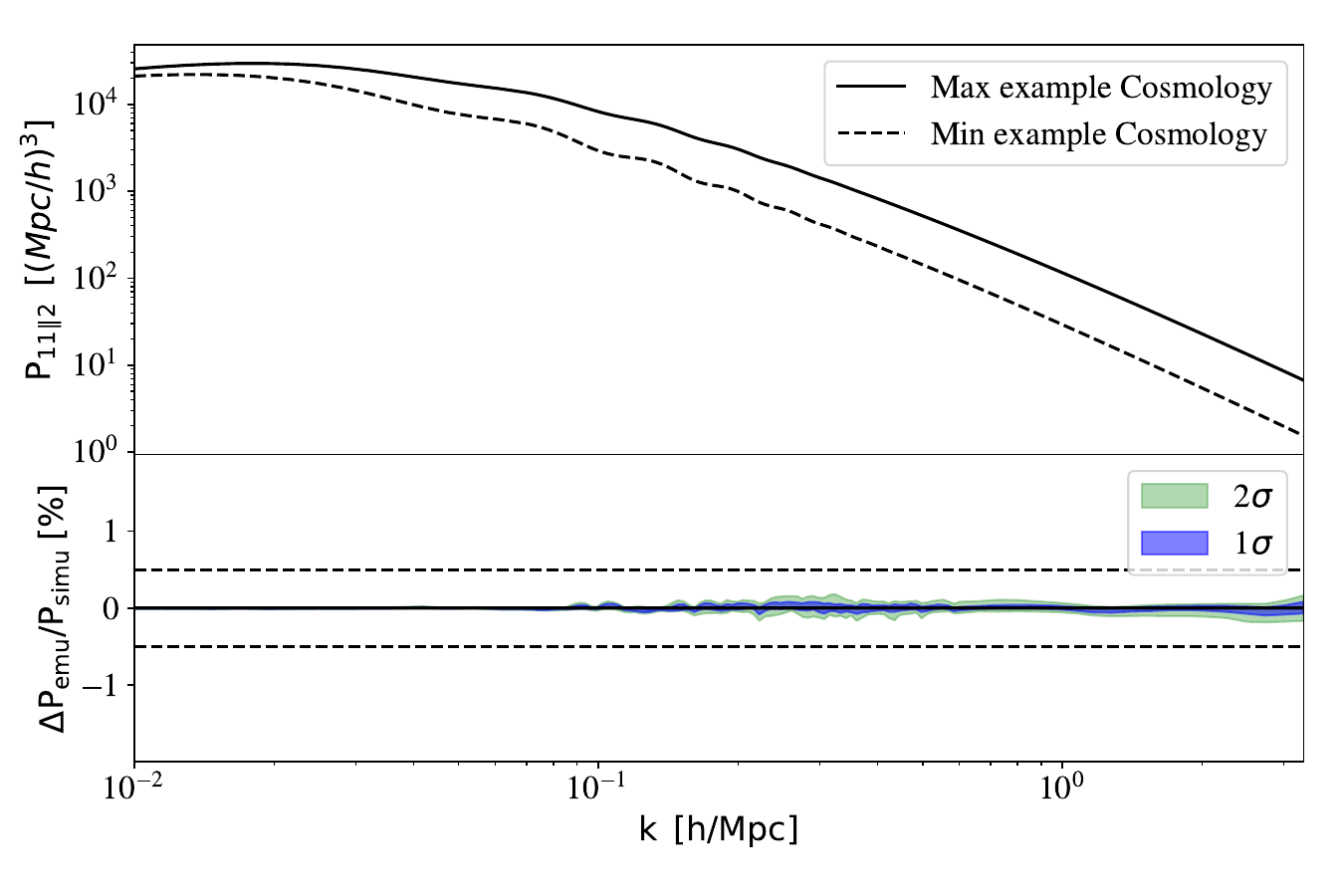}
\caption{{\bf Left:} The Pareto front of RMSE vs model length for the $\frac{\IR{P_{11}}{2}}{P_{11} }$ emulator runs as generated by \operon, with blue marking the training and red the validation error, 
and with the chosen model of length 44 indicated by the vertical line.  {\bf Right:} The top plot shows the  $\IR{P_{11}}{2}$ function for two extreme cases of cosmological parameters
while the bottom plot displays the resulting $1\sigma$ and $2\sigma$  emulator $\%$ error for all 300 cosmologies.}
  \label{Pareto_and_error:p11_R2}
\end{figure*}

For the $\IR{ P_{11}}{2}$ emulator we chose a model of length $44$ in order to be within our $P^{SPT}_{2-loop}$ error threshold.
We show the Pareto front of RMSE vs model length on the left of Fig.\ref{Pareto_and_error:p11_R2}.
 The form of the emulated function is 
\begin{align} 
\frac{ \IR{ P_{11}}{2}}{ P_{11} }   =& 
C_{21} - \frac{\left(C_{0} h\right)^{C_{1} \Ast} \left(C_{2} \Omega_{cb} - C_{3} \Omega_{b}\right) \left(C_{4} n_s - \left(C_{5} k\right)^{C_{6} \Omega_{cb}}\right)}{\sqrt{\left(C_{7} k\right)^{- C_{8} \Omega_{cb}} + 1}}
 - \frac{1}{\sqrt{\left(C_{19} k\right)^{C_{20} \Ast} + 1}}
\left[
\frac{C_{10} \cos{\left(\frac{C_{11} h + C_{12} k}{\sqrt{\left(C_{13} \Omega_{cb} + C_{14} \Omega_{b}\right)^{2} + 1}} \right)}}{\sqrt{\left(C_{15} \Omega_{b} - C_{16} k - \left(C_{17} k\right)^{- C_{18} n_s}\right)^{2} + 1}} + C_{9}
\right]
\label{equation_P11_R2}
\end{align} 
where
\begin{align} 
C_{0} =& 2.746 & \sspace C_{1} =& 2.343 & \sspace C_{2} =& 7.63 \times 10^{-3} & \sspace C_{3} =& 0.0265 & \sspace C_{4} =& 0.711 & \sspace C_{5} =& 0.408 \nonumber \\
C_{6} =& 0.8 & \sspace C_{7} =& 1.78 & \sspace C_{8} =& 13.6 & \sspace C_{9} =& 5.525 \times 10^{-3} & \sspace C_{10} =& 0.01016 & \sspace C_{11} =& 11.5 \nonumber \\
C_{12} =& 113.84 & \sspace C_{13} =& 1.41 & \sspace C_{14} =& 4.63 & \sspace C_{15} =& 154.6  & \sspace C_{16} =& 25.97 & \sspace C_{17} =& 2.444 \nonumber \\
C_{18} =& 4.067 & \sspace C_{19} =& 2.51 & \sspace C_{20} =& 3.48 & \sspace C_{21} =& 1.005462 
\label{constants_P11_R2}
\end{align} 
The resulting error is displayed on the right of Fig.\ref{Pareto_and_error:p11_R2}.

\subsection{The $\IR{k^4 P_{11}}{0}$ emulator}
\label{sec:EmulatorSuite_k4P11_R0}
\begin{figure*}
  \includegraphics[width=0.48\linewidth,height=\appfigheight,keepaspectratio]{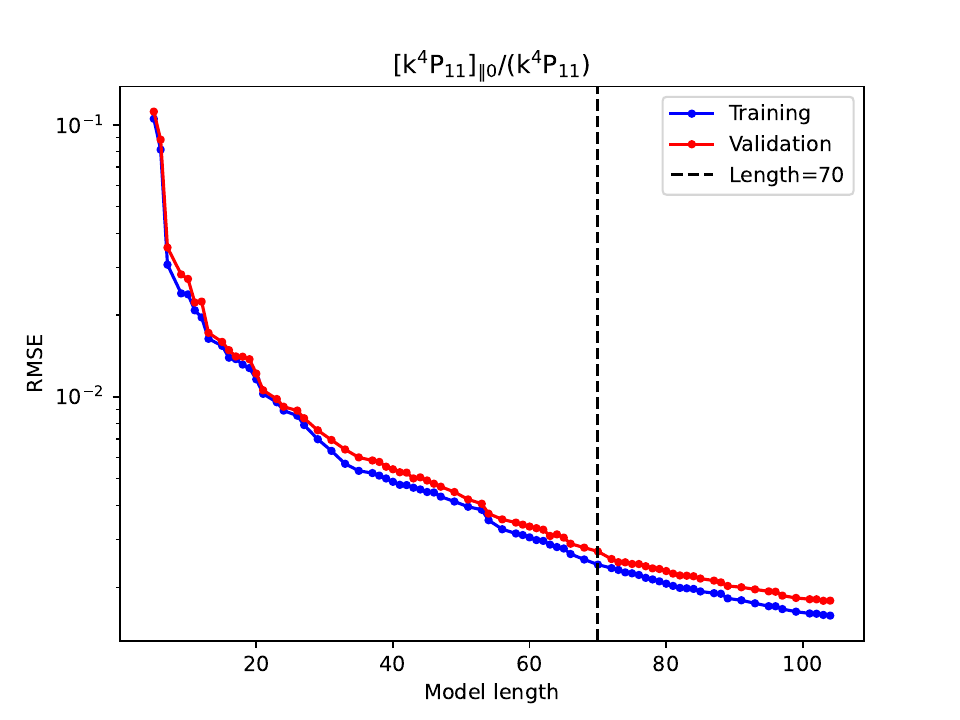}
   \includegraphics[width=0.48\linewidth,height=\appfigheight,keepaspectratio]{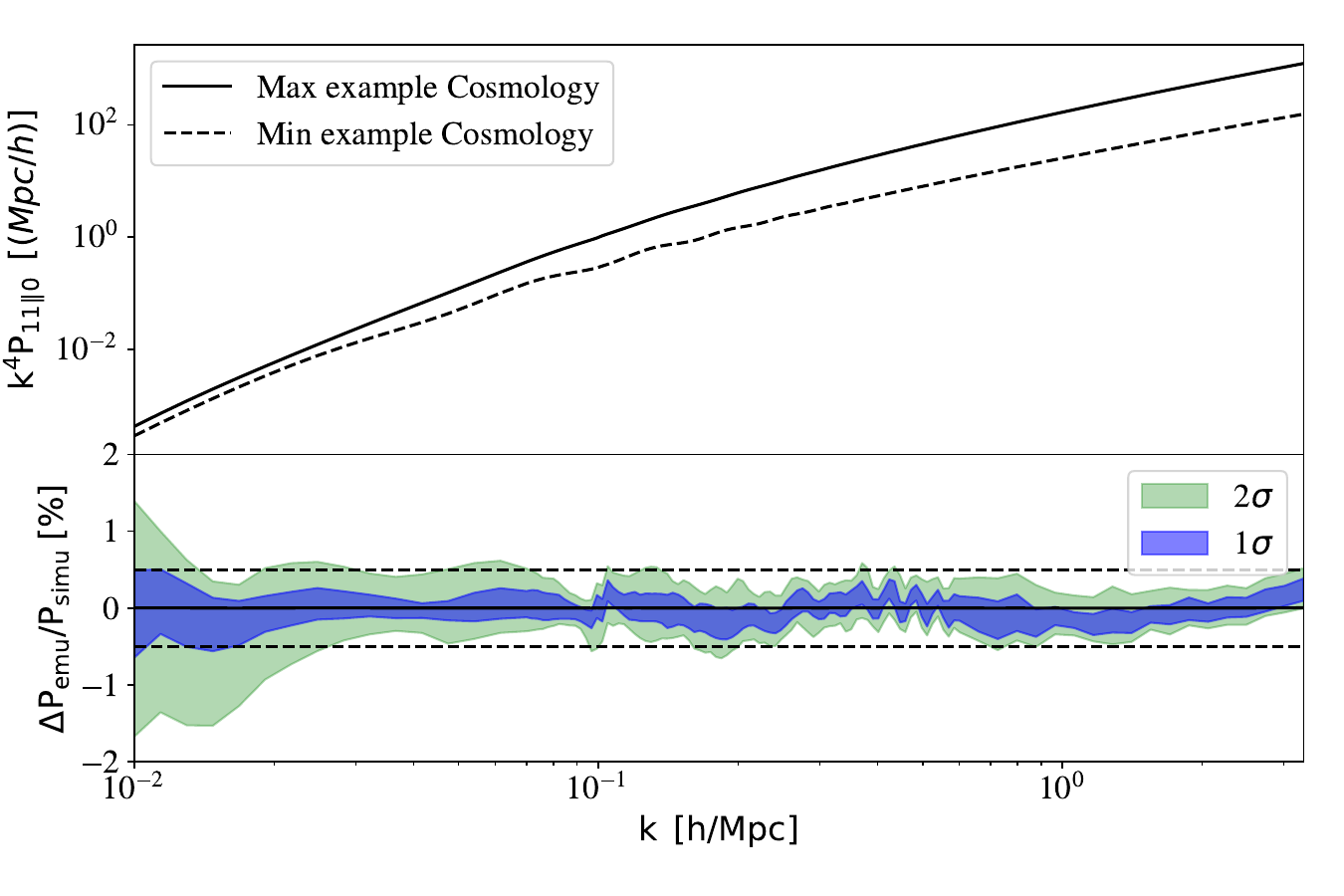}
\caption{{\bf Left:} The Pareto front of RMSE vs model length for the $\frac{\IR{k^4 P_{11}}{0}}{k^4 P_{11} }$ emulator runs as generated by \operon, with blue marking the training and red the validation error, 
and with the chosen model of length 70 indicated by the vertical line.  {\bf Right:} The top plot shows the  $\IR{k^4 P_{11}}{0}$ function for two extreme cases of cosmological parameters
while the bottom plot displays the resulting $1\sigma$ and $2\sigma$  emulator $\%$ error for all 300 cosmologies.}
  \label{Pareto_and_error:k4p11_R0}
\end{figure*}

For the $\IR{k^4 P_{11}}{0}$ emulator we chose a model of length $70$ in order to be within our $P^{SPT}_{2-loop}$ error threshold.
We show the Pareto front of RMSE vs model length on the left of Fig.\ref{Pareto_and_error:k4p11_R0}.
 The form of the emulated function is
\begin{align}
\frac{ \IR{k^4 P_{11}}{0}}{k^4 P_{11} }     =& C_{0} \Ast + C_{1} \Omega_{cb} + C_{18} 
+ \frac{C_{2} \left(C_{3} h\right)^{C_{4} \Omega_{cb} - C_{5} \Omega_{b}} 
\left(\frac{\Ast}{C_{13} k}\right)^{C_{14} \Omega_{cb}} 
}{
\sqrt{\left(C_{15} \Omega_{cb} - \frac{C_{16} k}{\sqrt{C_{17} k^{2} + 1}}\right)^{2} + 1}} 
\left(- C_{6} \Omega_{cb} + C_{7} n_s + \frac{C_{10} k + C_{9} \Omega_{b}}{\sqrt{\left(C_{11} k\right)^{- C_{12} k} + 1}} + \cos{\left(C_{8} k \right)}\right)
 \nonumber
\\
&
+ \frac{\left(- C_{19} \Ast + C_{20} \Omega_{cb} - C_{21} \Omega_{b} + C_{22} \cos{\left(C_{23} k \right)}\right) \cos{\left(\frac{C_{24} h + C_{25} k}{\sqrt{\left(C_{26} \Omega_{cb} + C_{27} \Omega_{b}\right)^{2} + 1}} \right)}}{\sqrt{\left( C_{28} \Omega_{b} - C_{29} k\right)^{2} + 1}} + \frac{- C_{30} n_s + C_{31} k + C_{32}}{\sqrt{C_{33} k^{2} + 1}}
\label{equation_k4P11_R0}
\end{align}
where
\begin{align} 
C_{0} =& 3.84 \times 10^{-3} & \sspace C_{1} =& 0.2188 & \sspace C_{2} =& 0.01 & \sspace C_{3} =& 5.586 & \sspace C_{4} =& 6.923 & \sspace C_{5} =& 8.222 \nonumber \\
C_{6} =& 34.7 & \sspace C_{7} =& 38.75 & \sspace C_{8} =& 117.27 & \sspace C_{9} =& 127  & \sspace C_{10} =& 18.87 & \sspace C_{11} =& 9.85 \nonumber \\
C_{12} =& 2000 & \sspace C_{13} =& 1297  & \sspace C_{14} =& 1.8743 & \sspace C_{15} =& 2.7745 & \sspace C_{16} =& 107.9  & \sspace C_{17} =& 613 \nonumber \\
C_{18} =& 2 \times 10^{-5} & \sspace C_{19} =& 7.02 \times 10^{-3} & \sspace C_{20} =& 0.074 & \sspace C_{21} =& 0.472 & \sspace C_{22} =& 0.01 & \sspace C_{23} =& 12.12 \nonumber \\
C_{24} =& 9.866 & \sspace C_{25} =& 116.5  & \sspace C_{26} =& 1.352 & \sspace C_{27} =& 4.704 & \sspace C_{28} =& 43.1 & \sspace C_{29} =& 14.4 \nonumber \\
C_{30} =& 0.3267 & \sspace C_{31} =& 50.062 & \sspace C_{32} =& 0.455 & \sspace C_{33} =& 3187.7
\label{constants_k4P11_R0}
\end{align} 
The resulting error is displayed on the right of Fig.\ref{Pareto_and_error:k4p11_R0}

\subsection{The $ \IR{P_{\rm 1-loop}}{1}$ emulator}
\label{sec:EmulatorSuite_P1loop_R1}

\begin{figure*}
    \centering
        \includegraphics[width=0.33\textwidth]{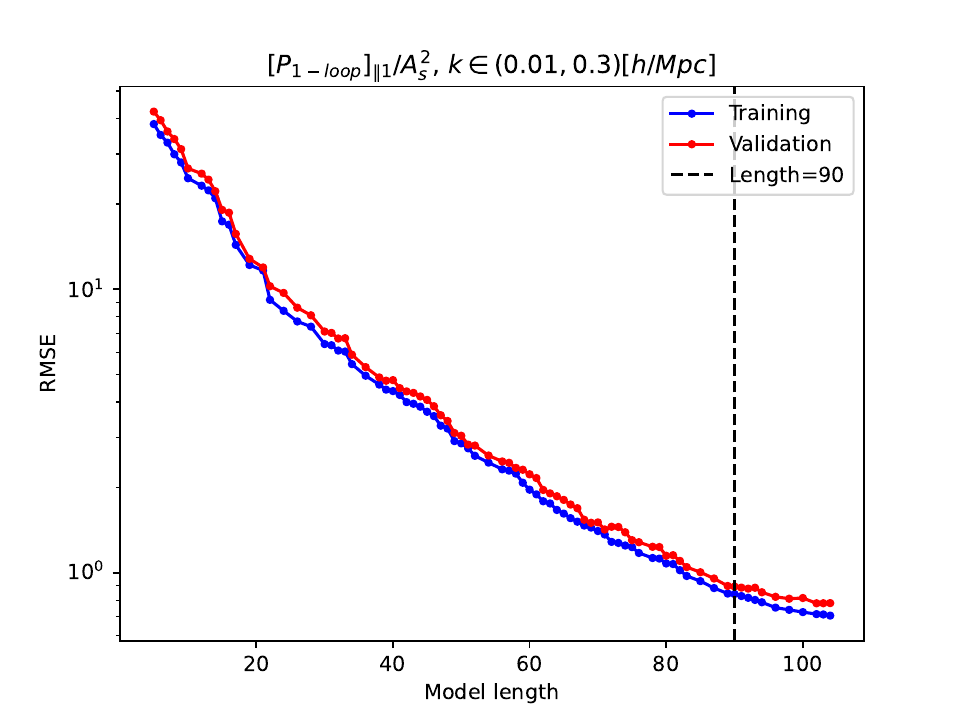}
        \includegraphics[width=0.33\textwidth]{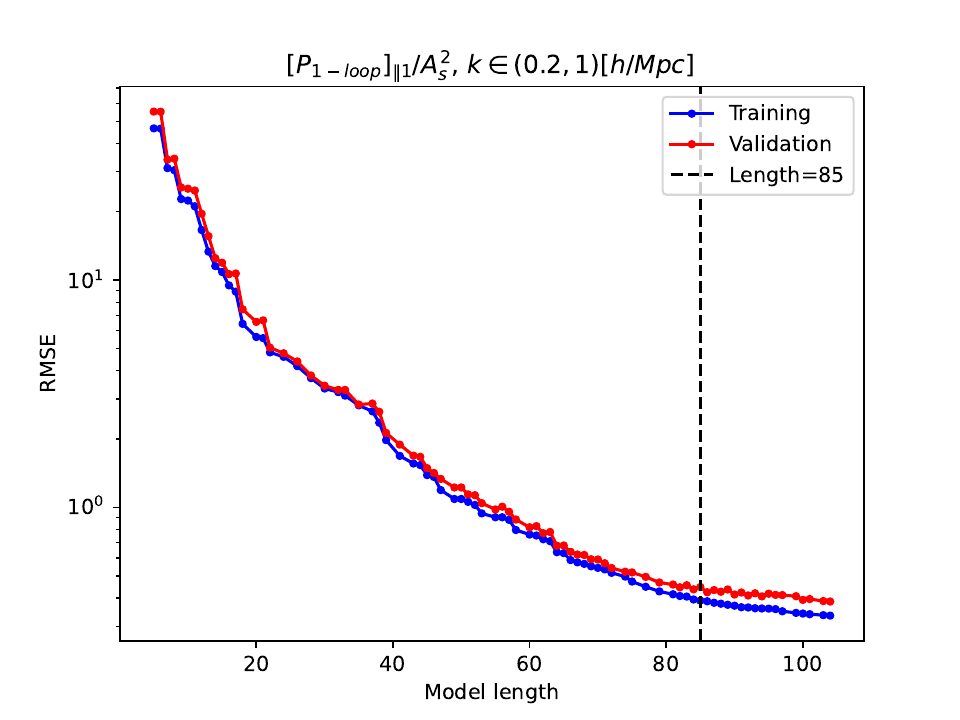}
        \includegraphics[width=0.33\textwidth]{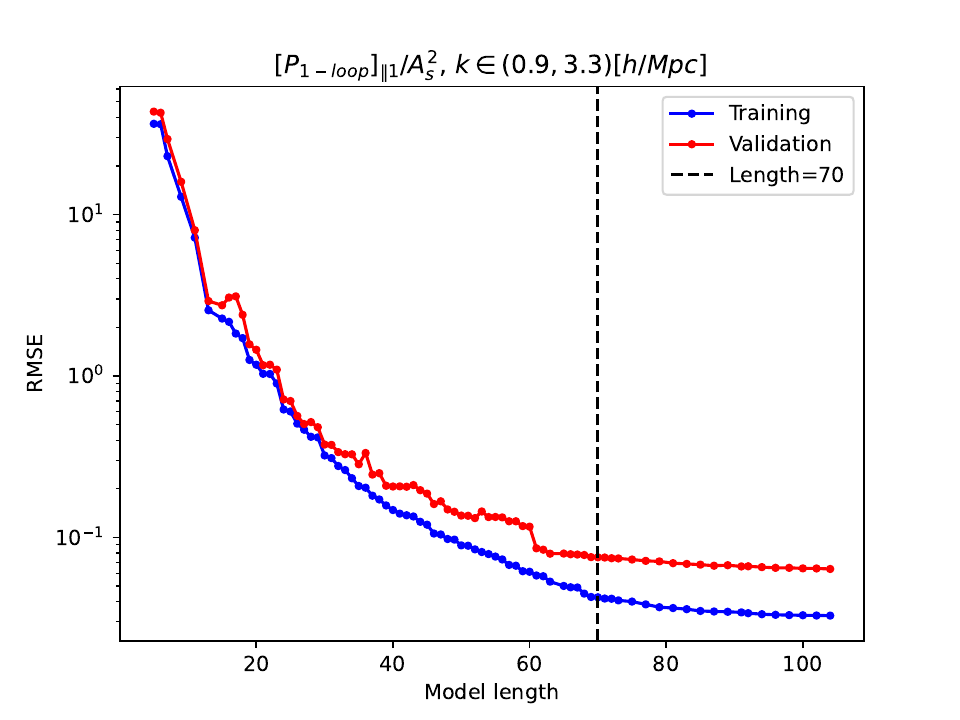}
    \caption{The Pareto fronts of RMSE vs model length for the three emulators of $\frac{\IR{ P_{{\rm 1-loop}}}{1}}{ \Ast^2 }$ runs as generated by \operon, with blue marking the training and red the validation error, and with the chosen models indicated by the vertical line. 
 }
    \label{fig:combined-pareto-2loop}
\end{figure*}

The IR resummed  $\IR{P_{\rm 1-loop}}{1}$  emulator is split into three regions in $k$ space as described in Section \ref{2_loop_methodology}, 
while all the Pareto fronts can be found in Fig.\ref{fig:combined-pareto-2loop}

\subsubsection{Region 1: $k = [0.01,0.3]\times \hmpc$}
For the first region  of the $\IR{P_{\rm 1-loop}}{1}$  emulator we chose a model of length 70, see left panel of Fig.\ref{fig:combined-pareto-2loop}.
The resulting function is
\begin{align} 
\frac{\IR{P_{\rm 1-loop}}{1}^{(1)}}{ \Ast^2 }  
=&
- C_{0} + 
\frac{
\left[ \left(\frac{C_{1} k}{\sqrt{C_{2} k^{2} + 1}}\right)^{C_{3} \Omega_{cb}} - \cos{\left(C_{4} n_s - \ln{\left(C_{5} k \right)} \right)}\right] 
\left[ C_{8} \Omega_{b} - \Omega_{cb} \left(C_{6} h - C_{7} n_s\right) - \cos{\left(C_{9} h - \frac{C_{10} k}{\sqrt{C_{11} \Omega_{cb}^{2} + 1}} \right)}\right]
}{\sqrt{\left(C_{35} k\right)^{- C_{36} \Omega_{cb} + \frac{C_{37} k}{\sqrt{C_{38} k^{2} + 1}}} + 1}}
\left\{
{}
- C_{12} \Omega_{cb}
\right.
\nonumber
\\
&
 + C_{13} k + \left(C_{14} h\right)^{C_{15} - C_{16} \Omega_{b}} \left(- C_{17} \Omega_{b} + \Omega_{cb}\right) \left[- C_{20} \Omega_{cb} - C_{21} k + \Omega_{cb} \left(C_{18} h + C_{19} n_s\right)\right]
 + \left(C_{33} k\right)^{- C_{34} \Omega_{cb}} 
\nonumber
\\
&
\left.
{}
+ \cos{\left(- C_{22} \Omega_{b} + C_{23} h + \frac{C_{24} k}{\sqrt{C_{25} \Omega_{cb}^{2} + 1}} \right)} + \cos{\left(C_{26} \Omega_{b} + C_{27} h + \frac{C_{28} k}{\sqrt{\left(C_{29} \Omega_{cb} + C_{30} \Omega_{b}\right)^{2} + 1}} + \frac{C_{31} k}{\sqrt{C_{32} k^{2} + 1}} \right)}
\right\}
\label{equation_P1loop_R1_1}
\end{align} 
where
\begin{align} 
C_{0} =& 0.01 & \sspace C_{1} =& 5.672 & \sspace C_{2} =& 23.847 & \sspace C_{3} =& 4.378  & \sspace C_{4} =& 0.81636 & \sspace C_{5} =& 3.336 \nonumber \\
C_{6} =& 430.88  & \sspace C_{7} =& 127.53  & \sspace C_{8} =& 276.55  & \sspace C_{9} =& 2.643 & \sspace C_{10} =& 119.07 & \sspace C_{11} =& 3.461 \nonumber \\
C_{12} =& 38.113 & \sspace C_{13} =& 23.725 & \sspace C_{14} =& 3.22496 & \sspace C_{15} =& 2.8447 & \sspace C_{16} =& 10.353 & \sspace C_{17} =& 2.1854 \nonumber \\
C_{18} =& 7.224 & \sspace C_{19} =& 49.34 & \sspace C_{20} =& 21.03 & \sspace C_{21} =& 75.24 & \sspace C_{22} =& 56.83 & \sspace C_{23} =& 3.291 \nonumber \\
C_{24} =& 116.47 & \sspace C_{25} =& 3.245 & \sspace C_{26} =& 27.45 & \sspace C_{27} =& 3.66 & \sspace C_{28} =& 120.39 & \sspace C_{29} =& 1.834 \nonumber \\
C_{30} =& 5.294 & \sspace C_{31} =& 22.9 & \sspace C_{32} =& 96 & \sspace C_{33} =& 2.922 & \sspace C_{34} =& 3.816 & \sspace C_{35} =& 5.834 \nonumber \\
 C_{36} =& 12.544 & \sspace C_{37} =& 31.5 & \sspace C_{38} =& 465 
\label{constants_P1loop_R1+1}
\end{align} 

\subsubsection{Region 2: $k = [0.2,1]\times \hmpc$}
For the second region  of the $\IR{P_{\rm 1-loop}}{1}$  emulator we chose a model of length 83, see the middle panel of Fig.\ref{fig:combined-pareto-2loop}.
\begin{align} 
\frac{ \IR{ P_{\rm 1-loop}}{1}^{(2)}}{ \Ast^2 }  =& 
\left(C_{0} h\right)^{C_{1} n_s} \Bigg\{ 
- C_{2} \Omega_{cb} \left(C_{3} \Omega_{b}\right)^{C_{4} k} \cos{\left(C_{5} k \right)} - C_{6} \Omega_{cb} + C_{7} \Omega_{b} - C_{8} n_s \left(C_{9} k\right)^{\frac{C_{10} \Omega_{b}}{\sqrt{C_{11} \Ast^{2} + 1}} - C_{12} k} 
\cos{\left(C_{13} \Omega_{cb} + C_{14} \Omega_{b} - C_{15} h - C_{16} k \right)} 
\nonumber
\\
&
\times 
+ \left(C_{17} k\right)^{- C_{18} k}
\bigg\}
 \left[ \left(C_{19} k\right)^{C_{20} k}\right]^{\left(C_{21} \Omega_{b}\right)^{C_{22} \Omega_{cb}}}
 \left\{ 
  \left[\left(C_{23} h\right)^{C_{24} \Omega_{cb} + C_{25} h}\right]^{\left(C_{26} k\right)^{C_{27} \Omega_{b} - C_{28} n_s}}\right\}^{\cos{\left(C_{29} \Omega_{b} + C_{30} n_s + \left(C_{31} k\right)^{- C_{32} \Omega_{cb}} \right)}} 
\nonumber
\\
&
\times 
\cos{\left(C_{33} n_s + \frac{\left(C_{34} \Omega_{cb}\right)^{C_{35} \Omega_{cb}}}{\sqrt{C_{36} k^{2} + 1}} + \left(C_{37} k\right)^{- C_{38} \Omega_{cb}} \right)} -  C_{39}
\label{equation_P1loop_R1_2}
\end{align} 
where
\begin{align} 
C_{0} =& 2.2735 & \sspace C_{1} =& 5.6863 & \sspace C_{2} =& 70 & \sspace C_{3} =& 2.4 & \sspace C_{4} =& 7.4 \nonumber \\
C_{5} =& 75.95 & \sspace C_{6} =& 291.08  & \sspace C_{7} =& 665.07  & \sspace C_{8} =& 5.256 & \sspace C_{9} =& 7.21 & \sspace C_{10} =& 66.65 \nonumber \\
C_{11} =& 0.185 & \sspace C_{12} =& 14.88 & \sspace C_{13} =& 18.6 & \sspace C_{14} =& 60.5 & \sspace C_{15} =& 8.716 & \sspace C_{16} =& 94.275 \nonumber \\
C_{17} =& 1.2283 & \sspace C_{18} =& 4.8575 & \sspace C_{19} =& 0.218604 & \sspace C_{20} =& 5.348 & \sspace C_{21} =& 0.7293 & \sspace C_{22} =& 1.59654 \nonumber \\
C_{23} =& 2.1905 & \sspace C_{24} =& 3.906 & \sspace C_{25} =& 6.2856 & \sspace C_{26} =& 2.5224 & \sspace C_{27} =& 7.78 & \sspace C_{28} =& 0.8806 \nonumber \\
C_{29} =& 0.891 & \sspace C_{30} =& 0.5633 & \sspace C_{31} =& 0.7723 & \sspace C_{32} =& 0.2949 & \sspace C_{33} =& 1.6517 & \sspace C_{34} =& 8.733 \nonumber \\
C_{35} =& 0.7744 & \sspace C_{36} =& 4.826 & \sspace C_{37} =& 4.963 & \sspace C_{38} =& 2.797 & \sspace C_{39} =& 0.014 
\label{constants_P1loop_R1_2}
\end{align}

\subsubsection{Region 3: $k = [0.9,3.3]\times \hmpc$}
For the third region  of the $\IR{P_{\rm 1-loop}}{1}$  emulator we chose a model of length 50, see the right panel of Fig.\ref{fig:combined-pareto-2loop}.
The resulting function  is
\begin{align} 
\frac{ \IR{ P_{\rm 1-loop}}{1}^{(3)} }{ \Ast^2 }  =& 
\frac{C_{0} \left(C_{1} n_s\right)^{C_{2} k + C_{3}} \left(C_{4} h - 1\right) \left(C_{5} \Omega_{cb} - C_{6} \Omega_{b} - 1\right) \left(C_{7} \Omega_{cb} - C_{8} \Omega_{b} - k\right) 
\left(- C_{10} h - C_{11} n_s + C_{9} \Ast + \left(C_{12} h + C_{13} k\right)^{C_{14} k}\right)}{
\sqrt{
\left[1 + \frac{\left( C_{15} \Omega_{cb} + C_{16} k - \ln{\left(C_{17} h \right)}\right)^{2}}{\left(C_{18} \Omega_{cb} - C_{19} \Omega_{b} + C_{20} h -  C_{21}\right)^{2} + 1} 
\right]\left(C_{22} k^{2} + 1\right)\left[\left(C_{23} \Omega_{b} + \left(C_{24} h\right)^{- C_{25} h}\right)^{2} + 1\right]}} 
\nonumber
\\
&
\times \frac{1}{\sqrt{  \left(C_{26} h + C_{27} n_s - C_{28} - \left(C_{29} k\right)^{- C_{30} \Omega_{cb}}\right)^{2} + 1}}
+ C_{31}
\label{equation_P1loop_R1_3}
\end{align} 
where
\begin{align} 
C_{0} =& 4032.6& \sspace C_{1} =& 0.30666 & \sspace C_{2} =& 0.38145 & \sspace C_{3} =& 2.0405 & \sspace C_{4} =& 2.7333 & \sspace C_{5} =& 12.0496 \nonumber \\
C_{6} =& 14.467 & \sspace C_{7} =& 80.831 & \sspace C_{8} =& 129.8  & \sspace C_{9} =& 1.1 \times 10^{-3} & \sspace C_{10} =& 0.3266 & \sspace C_{11} =& 0.6959 \nonumber \\
C_{12} =& 4.1534 & \sspace C_{13} =& 0.2649 & \sspace C_{14} =& 0.28992 & \sspace C_{15} =& 2.19 & \sspace C_{16} =& 2.9544 & \sspace C_{17} =& 3.225 \nonumber \\
C_{18} =& 8.357 & \sspace C_{19} =& 15.81 & \sspace C_{20} =& 0.5555 & \sspace C_{21} =& 0.9718 & \sspace C_{22} =& 5.666 & \sspace C_{23} =& 25.673 \nonumber \\
C_{24} =& 1.3491 & \sspace C_{25} =& 5.445 & \sspace C_{26} =& 3.6656 & \sspace C_{27} =& 5.7639 & \sspace C_{28} =& 9.1967 & \sspace C_{29} =& 0.841 \nonumber \\
C_{30} =& 1.256 & \sspace C_{31} =& 3 \times 10^{-3} 
\label{constants_P1loop_R1_3}
\end{align} 

\begin{figure*}
    \centering
        \includegraphics[width=0.48\textwidth]{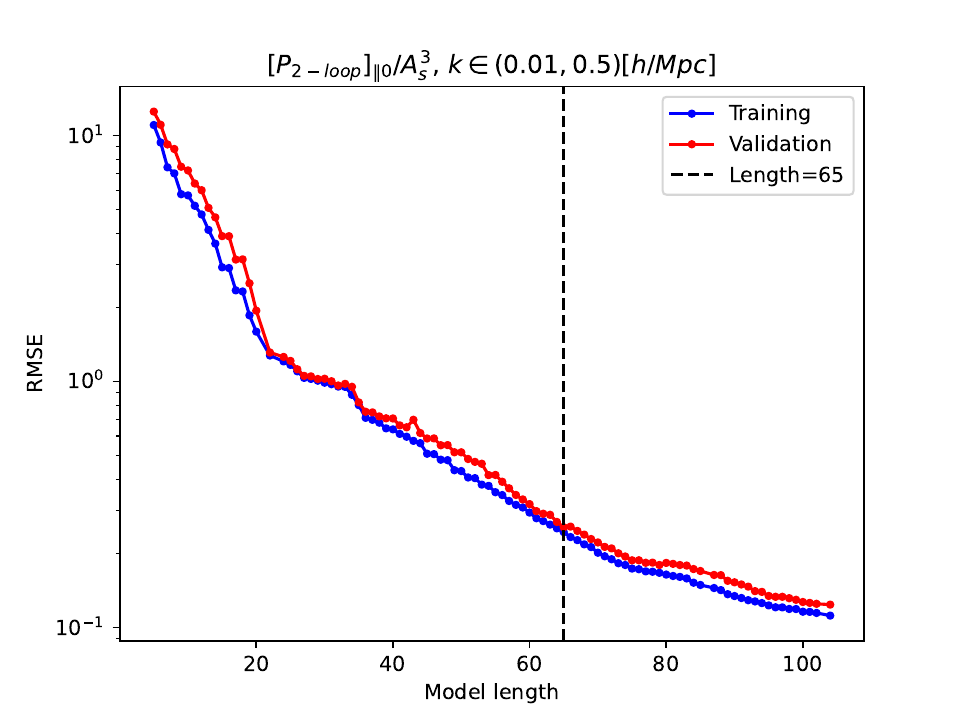}
        \includegraphics[width=0.48\textwidth]{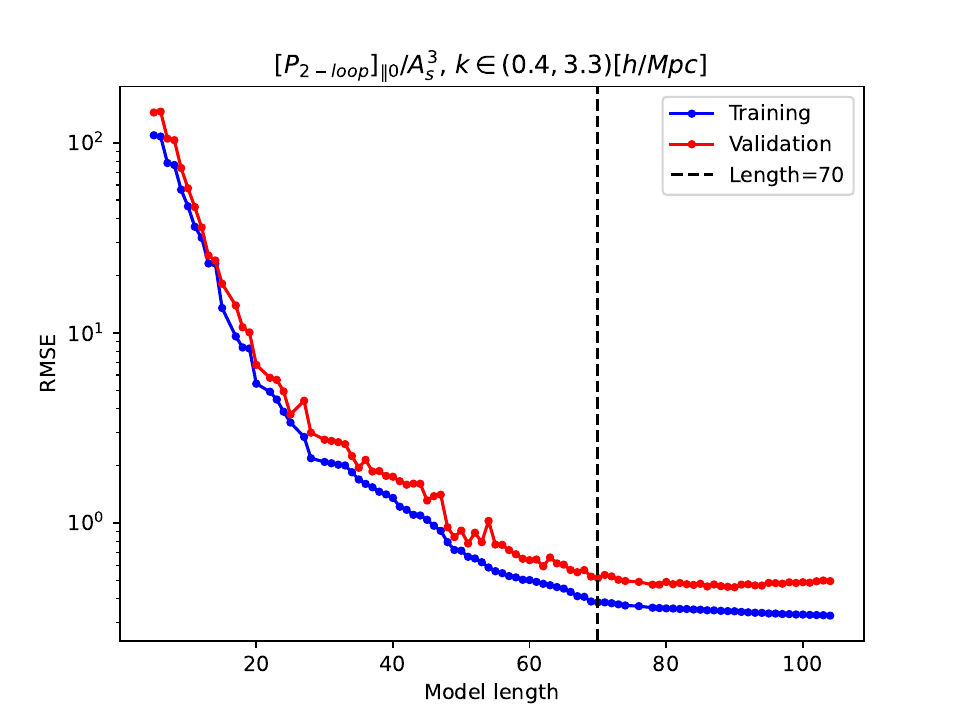}
    \caption{The Pareto fronts of RMSE vs model length for the three emulators of $\frac{\IR{ P_{2-loop}}{0}}{ \Ast^3 }$ runs as generated by \operon, with blue marking the training and red the validation error, and with the chosen models indicated by the vertical line. }
    \label{fig:p2loop_R0_pareto}
\end{figure*}

\subsection{The $ \IR{ P_{\rm 2-loop}}{0}$ emulator}
\label{sec:EmulatorSuite_P2loop_R0}
The IR resummed  $\IR{P_{\rm 2-loop}}{0}$  emulator is split into two overlaping regions in $k$ space as described in Section \ref{2_loop_methodology}, 
while all the Pareto fronts can be found in Fig.\ref{fig:p2loop_R0_pareto}

\subsubsection{Region 1: $k = [0.01,0.5]\times \hmpc$}
For the first region of the $\IR{P_{\rm 2-loop}}{0}$  emulator we chose a model of length 65, see the right panel of Fig.\ref{fig:p2loop_R0_pareto}.
The resulting function  is
\begin{align} 
\frac{ \IR{P_{\rm 2-loop}}{0}^{(1)}}{ \Ast^3 }  =& 
\frac{\left(C_{0} h\right)^{C_{1} \Omega_{cb} - \frac{C_{2} \Omega_{cb}}{\sqrt{C_{3} k^{2} + 1}} + C_{4} n_s}
}{\sqrt{\left(C_{27} k\right)^{- C_{28} n_s} + 1}}  
 \left(\frac{C_{5} \Omega_{cb}}{\sqrt{C_{6} \Omega_{b}^{2} + 1}} -  \frac{1}{\left(C_{7} k\right)^{C_{8} k}}\right) 
\Bigg[
C_{9} h \left(- C_{10} \Omega_{b} + \Omega_{cb}\right) 
\cos{\left(\frac{\left(C_{11} h\right)^{C_{12} \Omega_{cb}} \left(- C_{13} \Omega_{b} + C_{14} n_s\right)}{\sqrt{\left(C_{15} \Ast + C_{16} n_s + C_{17} k\right)^{2} + 1}} \right)} 
\nonumber
\\
&
- C_{18} k + C_{26} - \left(C_{19} k\right)^{C_{20} \Omega_{b} - C_{21} k} \cos{\left(- C_{22} \Omega_{b} + C_{23} h + \frac{C_{24} k}{\sqrt{C_{25} \Omega_{cb}^{2} + 1}} \right)}
\Bigg]
-  C_{29}
\label{equation_P2loop_R0_1}
\end{align} 
where
\begin{align} 
C_{0} =& 3.867 & \sspace C_{1} =& 2.187 & \sspace C_{2} =& 8 & \sspace C_{3} =& 146  & \sspace C_{4} =& 5.023 \nonumber \\
C_{5} =& 25.41 & \sspace C_{6} =& 698  & \sspace C_{7} =& 0.964 & \sspace C_{8} =& 2.316 & \sspace C_{9} =& 32.5 & \sspace C_{10} =& 1.588 \nonumber \\
C_{11} =& 5.62 & \sspace C_{12} =& 1.394 & \sspace C_{13} =& 19.27 & \sspace C_{14} =& 8.556 & \sspace C_{15} =& 0.084 & \sspace C_{16} =& 1.542 \nonumber \\
C_{17} =& 22.48 & \sspace C_{18} =& 7.585 & \sspace C_{19} =& 19 & \sspace C_{20} =& 23.5 & \sspace C_{21} =& 8.96 & \sspace C_{22} =& 38.4 \nonumber \\
C_{23} =& 6.76 & \sspace C_{24} =& 113  & \sspace C_{25} =& 3.55 & \sspace C_{26} =& 2.451 & \sspace C_{27} =& 0.454 & \sspace C_{28} =& 3.365\nonumber \\
C_{30} =& 0.02 
\label{constants_P2loop_R0_1}
\end{align} 

\subsubsection{Region 2: $k = [0.4,3.3]\times \hmpc$}
For the second region  of the $\IR{P_{\rm 2-loop}}{0}$ emulator, we chose a model of length 70, see the right panel of Fig.\ref{fig:p2loop_R0_pareto}
\begin{align} 
\frac{ \IR{ P_{\rm 2-loop}}{0}^{(2)}}{ \Ast^3 }  =& 
C_{0} \left(C_{1} h\right)^{C_{2} \Omega_{cb} - C_{3} \Omega_{b} + C_{4} n_s} 
\left\{ 
C_{5} k + \left(C_{6} k - \ln{\left(C_{7} h \right)}\right) \left[C_{8} n_s \left(- C_{10} \Omega_{b} + C_{9} \Omega_{cb} + k\right) - k\right]\right\}
\nonumber
\\
&
 \times
\left\{ 
C_{11} k + \left(C_{12} h\right)^{\ln{\left(C_{13} \Omega_{cb} \right)}} 
- \cos{\left[C_{14} \Omega_{b} - \left(C_{15} \Omega_{cb} + C_{16} \Omega_{b}\right) \left(C_{17} \Omega_{b} + h\right) \right]}
\right\}
\nonumber
\\
&
 \times
\ln{\left(\frac{C_{18} n_s}{\sqrt{\left(- C_{19} \Ast + C_{20} \Omega_{cb}\right)^{2} + 1}} + C_{21} k \right)}^{C_{22} h + C_{23} n_s + \ln{\left(\frac{C_{24} \Omega_{cb} \left(C_{25} \Omega_{cb} - \Omega_{b}\right)}{k} \right)}} -  C_{26}
\label{equation_P2loop_R0_2}
\end{align} 
where
\begin{align} 
C_{0} =& 6.656 & \sspace C_{1} =& 3.13 & \sspace C_{2} =& 6.826 & \sspace C_{3} =& 34.34 & \sspace C_{4} =& 4.031 & \sspace C_{5} =& 0.7532 \nonumber \\
C_{6} =& 0.2073 & \sspace C_{7} =& 3.982 & \sspace C_{8} =& 0.924413 & \sspace C_{9} =& 1.21654 & \sspace C_{10} =& 1.9249 & \sspace C_{11} =& 0.038 \nonumber \\
C_{12} =& 3.134 & \sspace C_{13} =& 5.859 & \sspace C_{14} =& 40.75 & \sspace C_{15} =& 10.27 & \sspace C_{16} =& 6.92 & \sspace C_{17} =& 4.885 \nonumber \\
C_{18} =& 1.14 & \sspace C_{19} =& 0.35 & \sspace C_{20} =& 3.1 & \sspace C_{21} =& 2.82 & \sspace C_{22} =& 3.02 & \sspace C_{23} =& 4.79 \nonumber \\
C_{24} =& 0.01946 & \sspace C_{25} =& 0.512 & \sspace C_{26} =& 0.02 
\label{constants_P2loop_R0_2}
\end{align} 

\subsection{The $\IR{P^{\rm (cs)}_{\rm 1-loop}}{0}$ emulator}
\label{sec:EmulatorSuite_P1loop_cs}
\begin{figure*}
    \centering
        \includegraphics[width=0.48\linewidth]{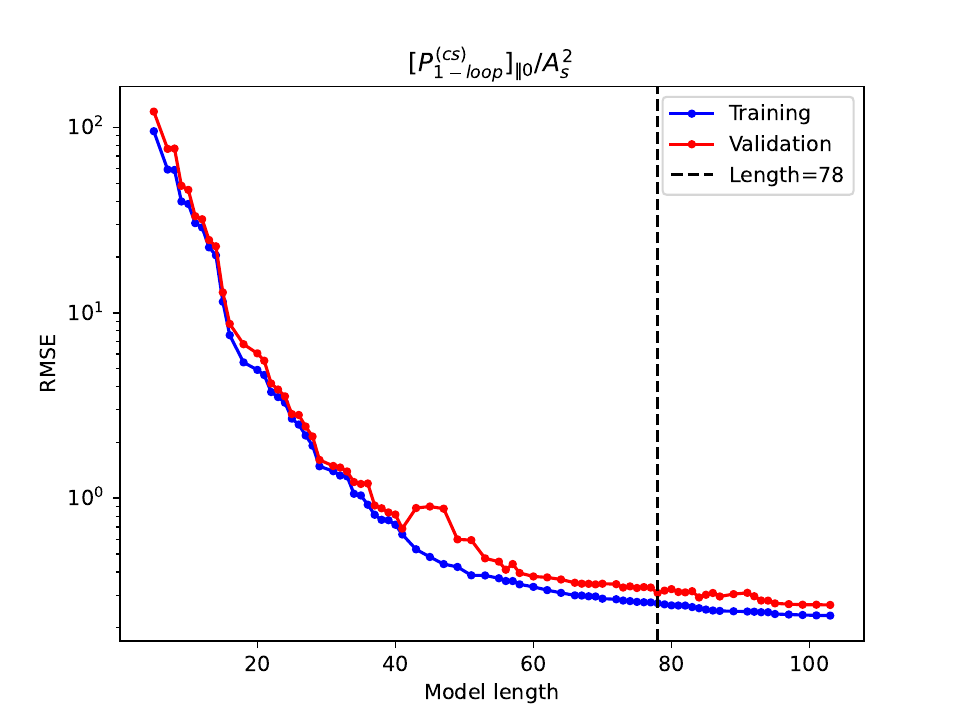}
        \includegraphics[width=0.48\linewidth]{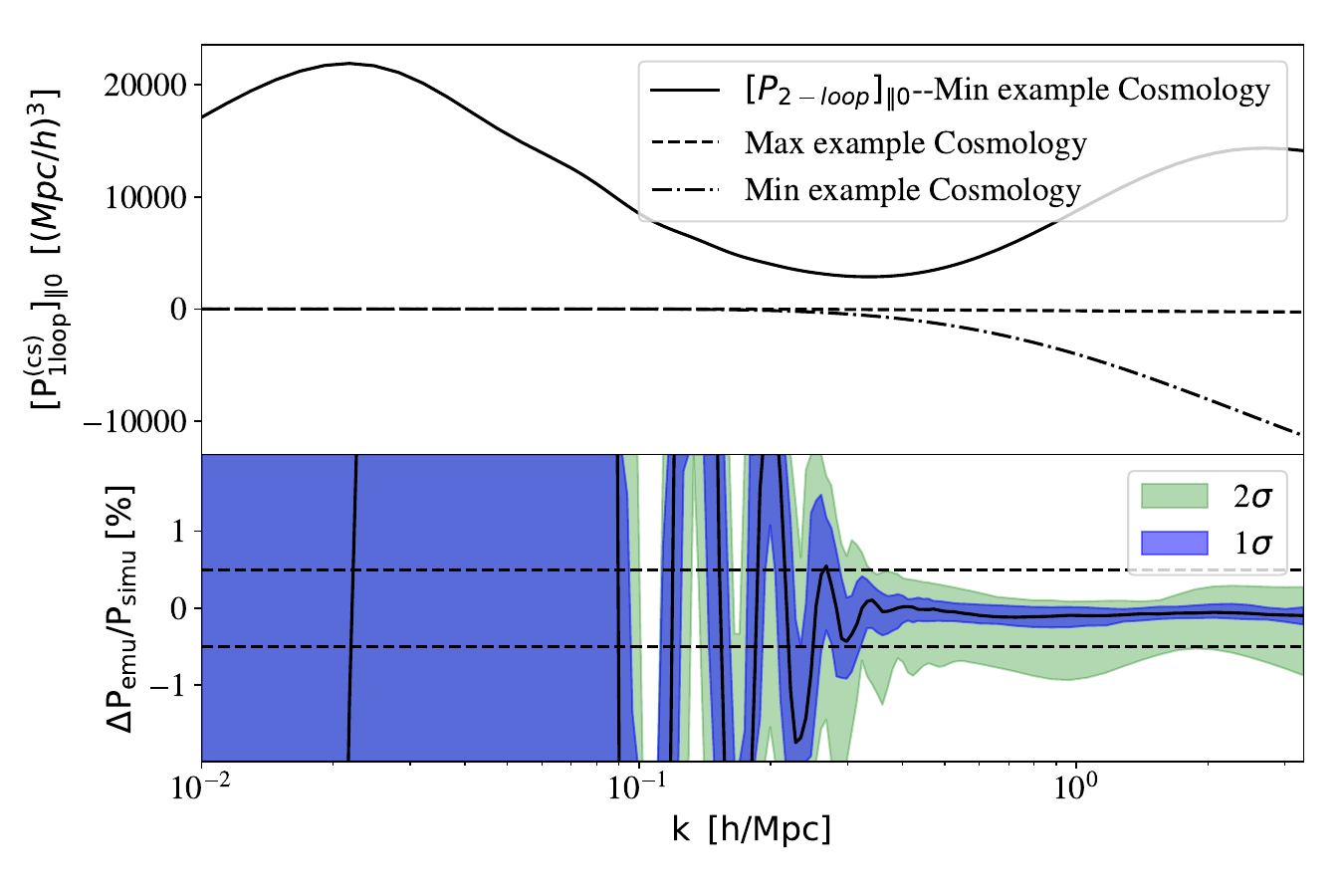}
\caption{\panel{Left:} The Pareto front of RMSE vs model length for the $\frac{\IR{ P^{(cs)}_{1loop}}{0}}{\Ast^2 }$ emulator runs as generated by \operon, with blue marking the training and red the validation error, 
and with the chosen model of length 78 indicated by the vertical line.  \panel{Right:} 
The top plot shows the  $\IR{P^{\rm (cs)}_{1-loop}}{0}$ function for two extreme cases of cosmological parameters
while the bottom plot displays the resulting $1\sigma$ and $2\sigma$  emulator $\%$ error for all 300 cosmologies. 
The seemingly large errors for  $k<0.3 \hmpc$ are inconsequential as  $\IR{ P_{\rm 2-loop}}{0}$ dominates there; see discussion in \ref{2_loop_methodology}. }
    \label{fig:p1loop_cs}
\end{figure*}

For the $\IR{P^{\rm (cs)}_{\rm 1-loop}}{0}$ emulator we chose a model of length $78$.  We show the Pareto front of RMSE vs model length on the left of Fig.\ref{fig:p1loop_cs}.
\begin{align} 
\frac{\IR{P^{\rm (cs)}_{\rm 1-loop}}{0}}{ \Ast^2 }    =& 
 \left(C_{0} h\right)^{- C_{1} \Omega_{b} + C_{2} n_s + \frac{C_{3} k}{\sqrt{C_{4} k^{2} + 1}}} \left(- C_{9} h + \left(C_{10} \Omega_{cb} - C_{11} n_s\right)^{\left(C_{12} k\right)^{- C_{13} \Omega_{cb}}}\right) \left(C_{5} \Omega_{cb} - C_{6} \Omega_{b}\right)^{\left(C_{7} k\right)^{C_{8} h}}
\nonumber
\\
&
\times 
\left(C_{18} n_s + \ln{\left(C_{19} k \right)} + \frac{C_{20} \Ast + C_{21} \Omega_{b} - C_{22} h + C_{23} k}{
\sqrt{C_{24} k^{2} + 1} \sqrt{\left( C_{25} n_s + \frac{C_{26} k}{\sqrt{\left(- C_{27} \Ast + C_{28} h\right)^{2} + 1}} - \left(C_{29} \Omega_{cb}\right)^{- C_{30} \Omega_{cb}}\right)^{2} + 1}}\right) 
\nonumber
\\
&
\times 
 \left[- C_{14} \Omega_{cb} + C_{15} \Omega_{b} - \left(C_{16} \Omega_{cb}\right)^{C_{17} \Omega_{b}}\right] 
\left(- C_{31} h - C_{32} n_s + C_{33} k + \left(C_{34} \Omega_{cb}\right)^{C_{35} \Omega_{cb} - C_{36} h}\right) - C_{37}
\label{equation_P1loop_cs}
\end{align} 
where
\begin{align} 
C_{0} =& 8.716 & \sspace C_{1} =& 10.523 & \sspace C_{2} =& 3.5605 & \sspace C_{3} =& 39.75 & \sspace C_{4} =& 490.06  \nonumber \\
C_{5} =& 0.98835 & \sspace C_{6} =& 1.84825 & \sspace C_{7} =& 0.2374 & \sspace C_{8} =& 0.1992 & \sspace C_{9} =& 8.79 \times 10^{-4} & \sspace C_{10} =& 0.058302 \nonumber \\
C_{11} =& 5.307 \times 10^{-3} & \sspace C_{12} =& 0.8495 & \sspace C_{13} =& 0.5929 & \sspace C_{14} =& 7.371 & \sspace C_{15} =& 15.45 & \sspace C_{16} =& 3.395 \nonumber \\
C_{17} =& 24.97 & \sspace C_{18} =& 3.916 & \sspace C_{19} =& 8.514 \times 10^{-4} & \sspace C_{20} =& 0.698 & \sspace C_{21} =& 292.8  & \sspace C_{22} =& 12.674 \nonumber \\
C_{23} =& 163.45  & \sspace C_{24} =& 15.317 & \sspace C_{25} =& 4.3412 & \sspace C_{26} =& 150.78  & \sspace C_{27} =& 0.339 & \sspace C_{28} =& 128.14  \nonumber \\
C_{29} =& 0.94006 & \sspace C_{30} =& 3.2281 & \sspace C_{31} =& 0.2662 & \sspace C_{32} =& 0.41405 & \sspace C_{33} =& 0.4143 & \sspace C_{34} =& 1.092744 \nonumber \\
C_{35} =& 1.73468 & \sspace C_{36} =& 0.16392 & \sspace C_{37} =& 5.2 \times 10^{-3} 
\label{constants_P1loop_cs}
\end{align} 

\subsection{The $\IR{P^{\rm (quad)}_{\rm 1-loop}}{0}$ emulator}
\label{sec:EmulatorSuite_P1loop_qd}
\begin{figure*}
    \centering
        \includegraphics[width=0.48\textwidth]{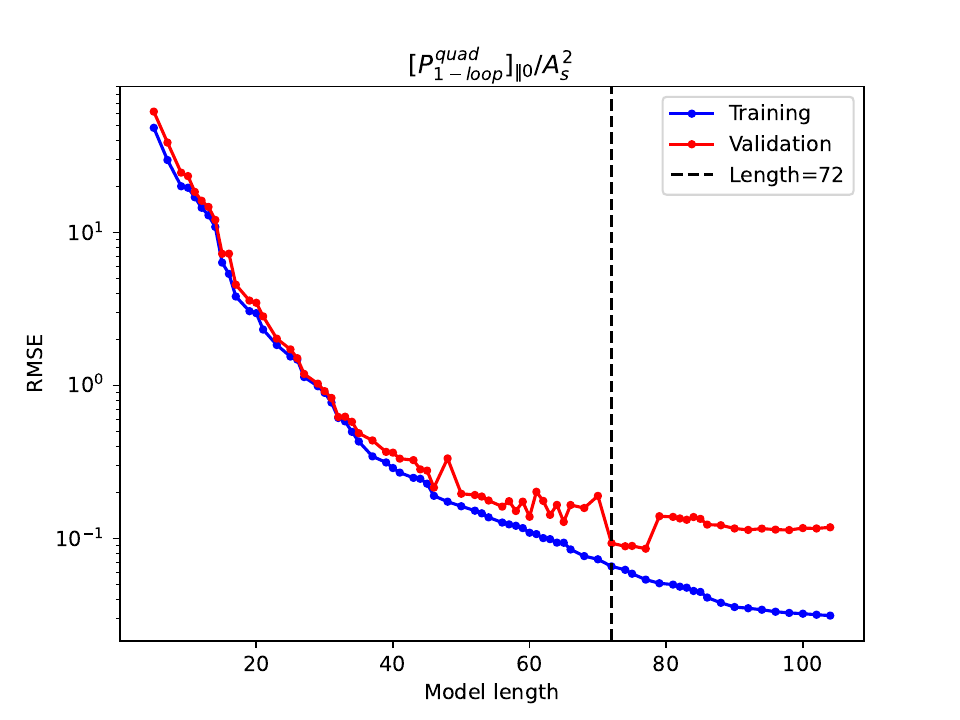}
        \includegraphics[width=0.48\linewidth]{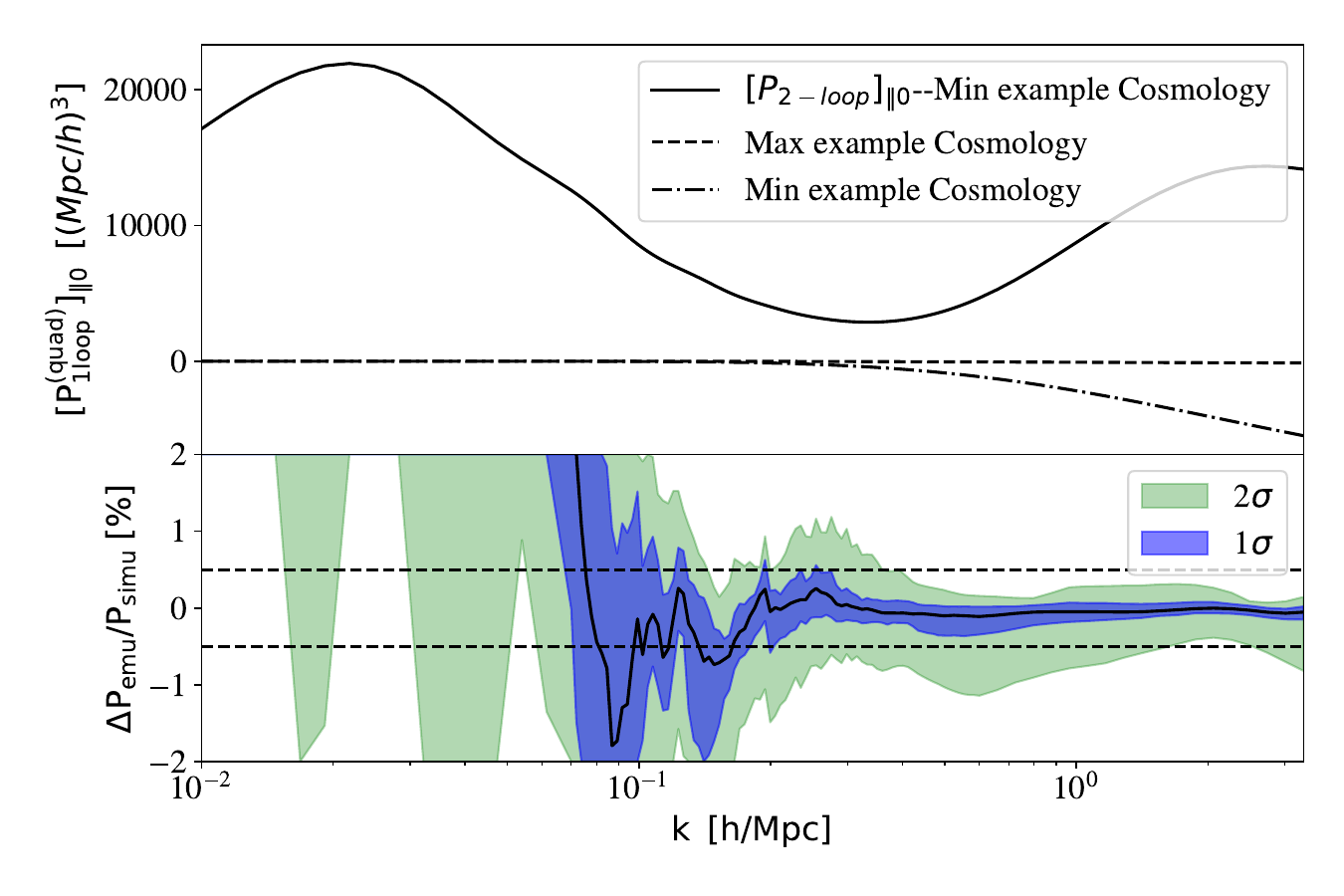}
\caption{\panel{Left:} The Pareto front of RMSE vs model length for the $\frac{\IR{ P^{(quad)}_{1loop}}{0}}{\Ast^2 }$ emulator runs as generated by \operon, with blue marking the training and red the validation error, 
and with the chosen model of length 72 indicated by the vertical line.  \panel{Right:} 
The top plot shows the  $\IR{P^{\rm (quad)}_{1-loop}}{0}$ function for two extreme cases of cosmological parameters
while the bottom plot displays the resulting $1\sigma$ and $2\sigma$  emulator $\%$ error for all 300 cosmologies.
The seemingly large errors for  $k<0.3 \hmpc$ are inconsequential as  $\IR{ P_{\rm 2-loop}}{0}$ dominates there; see discussion in \ref{2_loop_methodology}. }
    \label{fig:p1loop_qd}
\end{figure*}

For the $\IR{P^{\rm (quad)}_{\rm 1-loop}}{0}$ emulator we chose a model of length $72$.  We show the Pareto front of RMSE vs model length on the left of Fig.\ref{fig:p1loop_qd}.
The resulting function  is
\begin{align} 
\frac{ \IR{ P^{\rm (quad)}_{\rm 1-loop}}{0}}{ \Ast^2 }  =& 
\left[
\cos{\left(C_{18} \Omega_{b} 
- \frac{C_{19} h}{\sqrt{\left(C_{20} \Omega_{b} + C_{21} k\right)^{2} + 1} \sqrt{\cos^{2}{\left(\frac{C_{22} \Omega_{cb}}{\sqrt{C_{23} \Ast^{2} + 1}} \right)} + 1}} \right)} 
\right.
\left. {}
- \frac{C_{24} \Omega_{cb} + C_{25} k + \cos{\left(C_{26} \Omega_{cb} - C_{27} \Omega_{b} - C_{28} n_s - C_{29} \right)}}{\sqrt{\left(C_{30} \Omega_{b} + C_{31} k\right)^{2} + 1}}
\right]
\nonumber
\\
&
\times
\frac{\left(C_{0} h\right)^{- C_{1} \Omega_{b} - \frac{C_{5}}{\sqrt{C_{6} k^{2} + 1}} + \frac{C_{2} h + C_{3} n_s}{\sqrt{C_{4} h^{2} + 1}}}}{\sqrt{C_{32} k^{2} + 1}} 
\left(C_{7} k\right)^{\frac{C_{9} \Omega_{cb}}{\sqrt{\left(C_{10} \Omega_{cb} + C_{11} k\right)^{2} + 1}} + \frac{C_{12} \Omega_{cb}}{\sqrt{C_{13} k^{2} + 1}} - C_{14} \Omega_{b} + C_{15} h + C_{16} n_s + C_{17} k + C_{8} \Omega_{cb}}
-  C_{33}
\label{equation_P1loop_qd}
\end{align} 
where
\begin{align} 
C_{0} =& 1.0114 & \sspace C_{1} =& 9.199 & \sspace C_{2} =& 12.811 & \sspace C_{3} =& 4.371 & \sspace C_{4} =& 11.248 \nonumber \\
C_{5} =& 1.9858 & \sspace C_{6} =& 29.86 & \sspace C_{7} =& 12.475 & \sspace C_{8} =& 1.6152 & \sspace C_{9} =& 6.5046 & \sspace C_{10} =& 3.873 \nonumber \\
C_{11} =& 11.89 & \sspace C_{12} =& 0.5958 & \sspace C_{13} =& 0.0837 & \sspace C_{14} =& 5.748 & \sspace C_{15} =& 0.4183 & \sspace C_{16} =& 1.7877 \nonumber \\
C_{17} =& 0.01105 & \sspace C_{18} =& 5.0913 & \sspace C_{19} =& 3.467 & \sspace C_{20} =& 16.01 & \sspace C_{21} =& 4.183 & \sspace C_{22} =& 6.48 \nonumber \\
C_{23} =& 0.05 & \sspace C_{24} =& 17.557 & \sspace C_{25} =& 1.23 & \sspace C_{26} =& 13.345 & \sspace C_{27} =& 19.93 & \sspace C_{28} =& 0.334 \nonumber \\
C_{29} =& 4.051 & \sspace C_{30} =& 24.483 & \sspace C_{31} =& 1.3545 & \sspace C_{32} =& 2.85 & \sspace C_{33} =& 8 \times 10^{-4} 
\label{constants_P1loop_qd}
\end{align} 

\bsp	
\label{lastpage}
\end{document}